%% file: arxiv_v2.tex
\documentclass[sigconf]{acmart}

\AtBeginDocument{%
  }

\setcopyright{none}
\renewcommand\footnotetextcopyrightpermission[1]{}

\input{mydefs}

\tcbuselibrary{breakable}

\begin{document}

\title{\sysname: Holistic Evaluation of Watermark for Dataset Copyright Auditing}

\author{Xiao Ren}
\authornote{The first two authors contributed equally to this work.}
\affiliation{%
  \institution{Zhejiang University}
  \city{Hangzhou}
  \country{China}
}
\email{renx@zju.edu.cn}

\author{Xinyi Yu}
\authornotemark[1]
\affiliation{%
  \institution{Xi'an Jiaotong University}
  \city{Xi'an}
  \country{China}
}
\email{xinyiyu@stu.xjtu.edu.cn}

\author{Linkang Du}
\authornote{The corresponding authors.}
\affiliation{%
  \institution{Xi'an Jiaotong University}
  \city{Xi'an}
  \country{China}
}
\email{linkangd@xjtu.edu.cn}

\author{Min Chen}
\affiliation{%
  \institution{Vrije Universiteit Amsterdam}
  \city{Amsterdam}
  \country{The Netherlands}
}
\email{m.chen2@vu.nl}

\author{Yuanchao Shu}
\affiliation{%
  \institution{Zhejiang University}
  \city{Hangzhou}
  \country{China}
}
\email{ycshu@zju.edu.cn}

\author{Zhou Su}
\affiliation{%
  \institution{Xi'an Jiaotong University}
  \city{Xi'an}
  \country{China}
}
\email{zhousu@ieee.org}

\author{Yunjun Gao}
\affiliation{%
  \institution{Zhejiang University}
  \city{Hangzhou}
  \country{China}
}
\email{gaoyj@zju.edu.cn}

\author{Zhikun Zhang}
\authornotemark[2]
\affiliation{%
  \institution{Zhejiang University}
  \city{Hangzhou}
  \country{China}
}
\email{zhikun@zju.edu.cn}

\begin{abstract}
The surging demand for large-scale datasets in deep learning has heightened the need for effective copyright protection, given the risks of unauthorized use to data owners.
Although the dataset watermark technique holds promise for auditing and verifying usage, existing methods are hindered by inconsistent evaluations, which impede fair comparisons and assessments of real-world viability.
To this end, we organize existing methods according to two key dimensions, implementation and verification, to support a consistent analysis and evaluation pipeline across tasks.
Based on this framework, we develop \sysname, a unified benchmark and open-source toolkit for systematically evaluating image dataset watermark techniques in classification and generation tasks.
Using \sysname, we assess 25 representative methods under standardized conditions, perturbation-based robustness tests, multi-watermark coexistence, and multi-user interference.
To enable accurate and reproducible benchmarking, we use TPR@5\%FPR for unified sample-level comparison and introduce the verification success rate (VSR) for dataset-level auditing.
Key findings reveal that standard single-watermark evaluations tend to overestimate practical auditability.
Methods that verify reliably in isolation often suffer from performance degradation at low watermarked-sample ratios, while yielding ambiguous ownership evidence in complex multi-user and multi-watermark settings.
We hope that \sysname can facilitate advances in watermark reliability and practicality, thus strengthening copyright safeguards in the face of widespread AI-driven data exploitation.
\end{abstract}


\keywords{Dataset Copyright Auditing, Dataset Watermark, Deep Learning}

\maketitle
\pagestyle{plain}

\section{Introduction}
\begin{figure}[!ht]
\centering
\includegraphics[width=0.9\hsize]{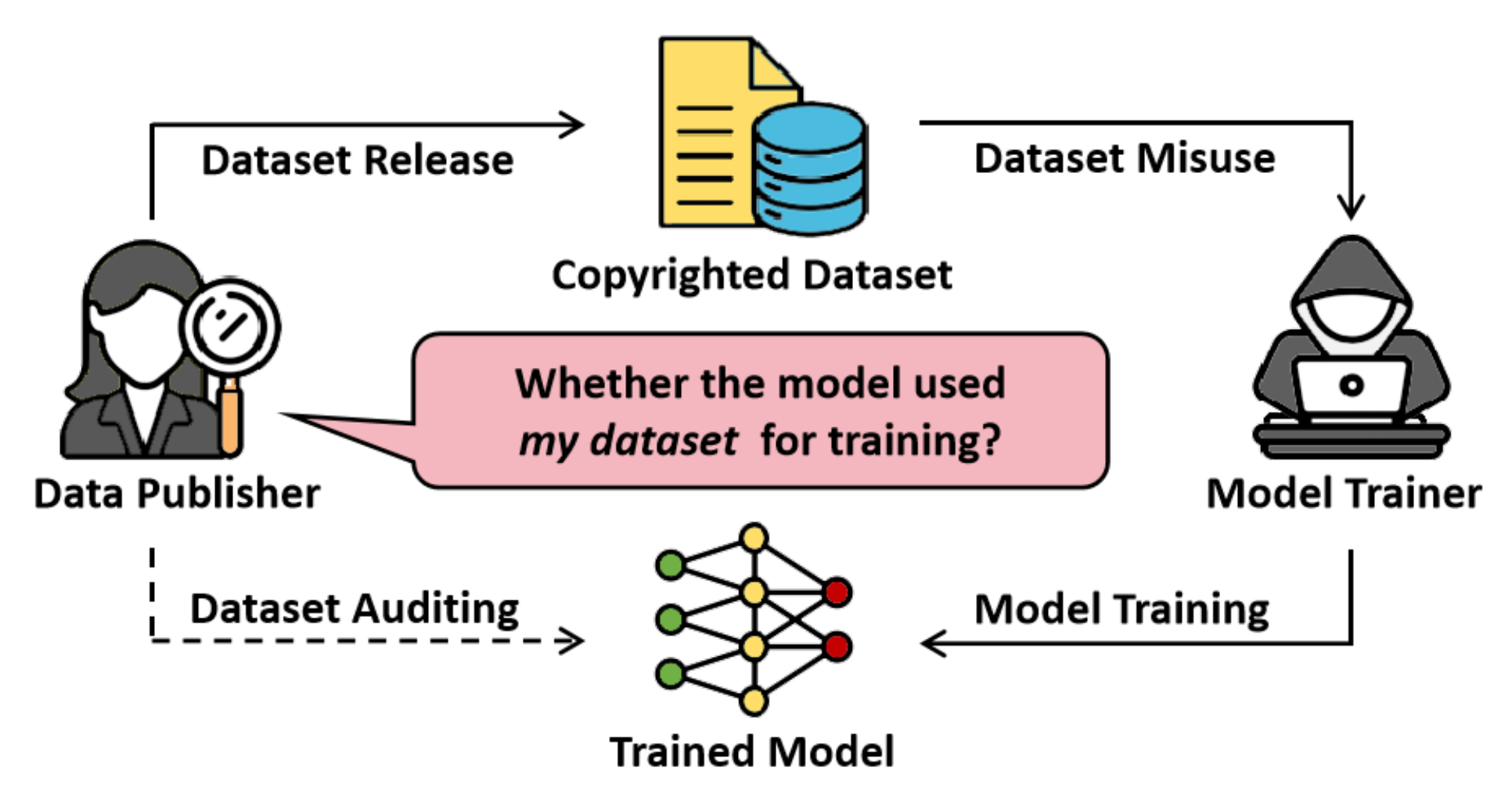}
\caption{Illustration of dataset copyright auditing.
\textit{Data publisher} releases a dataset, risking unauthorized use by the \textit{model trainer}.
Dataset auditing enables the data publisher to verify if the trained model was trained on their dataset.
}
\label{fig:data_auditing}
\end{figure}

In recent years, deep learning models have achieved remarkable breakthroughs in both classification and generation tasks, driving their adoption across diverse applications such as autonomous driving~\cite{li2020deep}, medical image analysis~\cite{litjens2017survey}, and content creation~\cite{elgammal2017can}.
Rapid growth of model capabilities has fueled an unprecedented demand for large-scale, high-quality datasets. 
However, the collection and annotation of such large-scale datasets often incur high costs and simultaneously pose the risk of privacy breaches and copyright infringement~\cite{paullada2021data}. 
As academia and industry increasingly embrace open data and model sharing, the risk of unauthorized use or misuse of datasets is growing. 
For example, in 2025, a group of authors sued Anthropic, alleging that it had obtained a large number of books from the so-called “shadow libraries” (e.g., LibGen, PiLiMi) to train its models.
Anthropic thereafter reached a \$1.5 billion class-action settlement that received preliminary court approval, and reports indicate the agreement covers hundreds of thousands of books, with compensation calculated per work at around \$3,000~\cite{ortutay2025anthropic}.
Similarly, in October 2025, Apple faced multiple author class action lawsuits alleging that Apple Intelligence and OpenELM were trained on unlicensed book corpora~\cite{brittain2025apple}. 
Other ongoing cases, including consolidated suits against OpenAI and Microsoft~\cite{grynbaum2023times}, highlight how AI models can generate content that mimics originals, posing competitive threats and substitution risks.

To address these issues, \textit{dataset watermarking} has emerged as a promising solution for dataset copyright protection~\cite{maesen2023image}.
The general idea is to embed identifiable watermarks in datasets before release.
If a model is trained on this watermarked dataset, the watermark can be extracted from the model's output, serving as evidence of the use of that dataset, as illustrated in \autoref{fig:data_auditing}.
Previous studies~\cite{sablayrolles2020radioactive}~\cite{wang2023diagnosis} have explored various watermark methods for different tasks.
Most of them focus on classification and generation tasks in the image domain~\cite{du2025sok}. 
However, current research on watermark methods lacks a unified taxonomy and evaluation.
Methods are typically studied in isolation using inconsistent evaluation settings, which prevents systematic comparison.
Furthermore, assessments typically assume ideal conditions, overlooking real-world complexities such as multi-watermark interference and multi-user scenarios, which undermine their practical applicability.

\mypara{\sysname}
This paper fills these gaps by systematizing existing work, developing an open-source benchmarking toolkit \sysname, and evaluating existing work in unified experimental settings. 
More specifically, we first propose a two-layer taxonomy that organizes existing solutions by injection and verification of watermarks. 
We classify the injection phase into \textit{model-based} and \textit{model-free} approaches, and the verification phase into \textit{model-behavior} and \textit{model-message} categories, which offers a cohesive framework for cross-task analysis.
Second, we built an open-source benchmarking toolkit with the help of this taxonomy. 
\sysname features 25 built-in watermark methods and includes 6 commonly used datasets, along with 6 representative models, facilitating a comprehensive assessment of classification and generation tasks. 
Built upon a unified publisher--trainer--auditor workflow, \sysname provides a standardized evaluation platform that eliminates inconsistencies across datasets, model architectures, and evaluation pipelines. 
Third, \sysname establishes a unified verification protocol that standardizes sample-level evaluation using TPR@5\%FPR and introduces the verification success rate (VSR) metric for dataset-level auditing, enabling consistent and reproducible comparisons across heterogeneous methods.

\mypara{Evaluations}
We evaluate watermark methods across classification and generation tasks under unified experimental settings.
Our experiments begin with the prevailing single-watermark, single-user setting, which reflects how most watermark schemes are assessed in controlled research contexts.
We then progressively extend our analysis to multi-watermark and multi-user configurations that more closely resemble real-world data ecosystems, where multiple watermark strategies may coexist and multiple stakeholders may claim ownership over shared or overlapping datasets.
This stepwise progression exposes the scalability limits, interference patterns, and verification ambiguities arising as watermark technology transitions
from academic prototypes to practical deployment.
Concretely, our research aims to address four key questions:
1) How does the performance of watermark methods vary across different auditing settings?
2) How do the watermark methods perform under different adversarial attacks?
3) How does multi-watermark overlapping affect each method's verification? 
4) In multi-user scenarios, can current watermark methods effectively verify ownership for each user?

\mypara{Main Findings}
Our comprehensive benchmark reveals numerous crucial and unexpected insights into the current state of dataset watermark methods.
\begin{itemize}[leftmargin=*]
\item Experimental results indicate the absence of a universal solution for dataset watermarking, characterized by numerous significant trade-offs. 
Our classification framework serves as a guide through this complex domain, as experiments show that no single method or category is superior across all conditions. 
The ideal method depends on various trade-offs, like robustness versus stealth, and must be customized to align with specific application objectives.

\item We identify a fundamental performance disparity between the watermark methods used in classification versus generation tasks. 
In scenarios involving multiple watermarks, model-free methods that are robust in classification often fail in generation tasks. 
In contrast, model-based methods that thrive in generation are inadequate for classification. 
This paradox underscores the necessity for task-specific watermarking strategies, as a universal design is not universally effective.

\item Our evaluations also highlight a critical lack of reliable and stable watermark solutions for real-world conditions.
In multi-user scenarios, we find that most watermark methods fail in two distinct ways: some exhibit alarmingly high false positive rates for unrelated users, while others suffer significant performance degradation as the number of watermarks increases.
Both failures make these methods impractical for reliable ownership verification, posing a major challenge to the practical application of current methods in the real world.

\end{itemize}

\section{Preliminaries}

\subsection{Deep Neural Networks}
A deep neural network (DNN) model $M$ is defined by its learnable parameters $\theta$.
The objective of training a DNN is to minimize a loss function $\mathcal{L}$, which calculates the distance between the model's predictions $M(X; \theta)$ and the target outputs $Y$. 
Formally, this can be expressed as an optimization problem:

\begin{equation}
\theta^* = \operatorname*{argmin}_{\theta} \mathcal{L}(M(X; \theta), Y). 
\end{equation}

The DNN models can be applied to various tasks.
In this work, we focus on classification and generation tasks.

\mypara{Classification Tasks}
Classification models learn the mapping between given samples and their corresponding labels.
Common models are built on convolutional neural networks (CNNs), like  VGGNet~\cite{simonyan2014very} and ResNet~\cite{he2016deep}.

Formally, given a dataset $D = \{(x_i, y_i)\}_{i=1}^N$ with $y_i \in \{1, 2, \dots, C\}$, the model $M$ takes input $x$ and outputs a probability distribution $p = \mathrm{softmax}(M(X; \theta))$.
The training objective is to minimize the cross-entropy loss:

\begin{equation}
\mathcal{L}_{\text{CE}} = -\frac{1}{N} \sum_{i=1}^{N} \sum_{c=1}^{C} \mathbb{I}(y_i=c) \log(p_{i,c}),
\end{equation}
where $\mathbb{I}(y_i = c)$ is an indicator function and
$p_{i,c}$ is the predicted probability that sample $x_i$ belongs to class $c$.

\mypara{Generation Tasks}
The generation model aims to synthesize new content, with recent advances including adversarial generative networks (GANs)~\cite{goodfellow2014generative}, variational autoencoders (VAEs)~\cite{kingma2013auto}, and diffusion-based models~\cite{ho2020denoising}.
We focus on text-to-image diffusion models, which learn a conditional mapping from textual prompts to visual representations.
Formally, the forward process maps the image $x_{0}$ to a latent representation $z_{0}$ through a VAE encoder $\mathcal{E}$, and gradually injects Gaussian noise in a series of time steps $t$:

\begin{equation}
z_t = \sqrt{\bar{\alpha}_t}\, z_0 + \sqrt{1 - \bar{\alpha}_t}\, \varepsilon, \quad \varepsilon \sim \mathcal{N}(0, I),
\end{equation}
where $\bar{\alpha}_t$ controls the noise schedule.
During the reverse process, the UNet model $\epsilon _{\theta}$ is used to predict noise in the latent space, conditioned on a text representation $\tau_{\theta}(y)$. The model is trained by minimizing the loss:

\begin{equation}
\mathcal{L}_{\mathrm{LDM}} = 
\mathbb{E}_{\mathcal{E}(x),\,y,\,\epsilon \sim \mathcal{N}(0,1),\,t}
\left[ \left\| \epsilon - \epsilon_{\theta}\left(z_t,\, t,\, \tau_{\theta}(y)\right) \right\|_2^2 \right].
\end{equation}

\subsection{Dataset Copyright Auditing}
\label{sec:dataset-copyright-auditing}

Typically, dataset copyright auditing involves two roles: \textit{data publisher} and \textit{model trainer}.
The data publisher applies a watermark encoder to embed watermark information $W$ in the original dataset $D$, producing a watermarked dataset $D^{*}$. 
The model trainer collects data to train a DNN model $M$, which is later released or deployed. 
The data publisher aims to determine whether a suspect model $M$ has been trained on $D^{*}$, thus identifying possible data misuse. 
The entire process consists of three main stages: watermark injection, model training, and watermark verification.

\mypara{Watermark Injection}
The data publisher designs an encoder $\mathcal{E}$ to embed a watermark $W$ into a dataset $D$:

\begin{equation}
D^*=\mathcal{E}(D,W).
\end{equation}
The objective is to ensure that $D^*$ is visually and semantically consistent with $D$, while enabling the reliable verification of the watermark from models trained on it.

\mypara{Model Training}
An unauthorized model trainer may use $D^*$ to train a DNN $M$.
In most scenarios, the trained model is deployed in a black-box setting, where the publisher cannot access model weights but can query its outputs~\cite{chen2023faceauditor,du2025artistauditor}.

\mypara{Watermark Verification}
The publisher designs a verifier $\mathcal{V}$ to verify the watermark $W$ from the model:
\begin{equation}
\mathcal{V}(M, W, D_t) =
\begin{cases}
    1, & M\text{ contains watermark } W, \\
    0, & \text{otherwise},
\end{cases}
\end{equation}
where $D_t$ refers to the query set used during verification, such as triggered image samples or textual prompts.

\section{Related Work}

\mypara{Dataset Copyright Auditing}
Existing studies have systematized dataset copyright protection from complementary perspectives.
Du~\etal~\cite{du2025sok} categorize dataset copyright auditing as intrusive or non-intrusive and analyze their assumptions and deployment constraints.
Ren~\etal~\cite{ren2024copyright} review copyright protection in generative AI from the perspectives of data owners and model builders.
Beyond these systematization efforts, dataset auditing has also been explored in task-specific settings, including facial recognition~\cite{chen2023faceauditor}, offline reinforcement learning~\cite{du2024orlauditor}, text-to-image generation~\cite{du2025artistauditor}, and video recognition~\cite{yuan2026victor}.
These studies provide complementary auditing mechanisms across different learning tasks, but no executable platform for comparing dataset watermark methods under consistent settings.

DATABench~\cite{shao2025databench} is the closest benchmark at the dataset-auditing level.
It organizes methods by whether they rely on internal features inherent to the data or external features introduced for auditing, and evaluates evasion and forgery attacks.
Its experiments focus mainly on image-classification settings.
In comparison, \sysname focuses on watermark-based auditing and organizes methods according to their injection and verification mechanisms.
It evaluates classification and text-to-image generation under a common publisher--trainer--auditor workflow, while examining watermarked-sample ratios, adversarial robustness, and multi-user attribution.

\mypara{Image Watermark Robustness Benchmarks}
WAVES~\cite{an2024waves} and \sysname both evaluate robustness when transformations are applied to watermarked images after watermark embedding. The benchmarks differ in where attacks are applied and how watermark evidence is obtained. WAVES applies attacks to watermarked images and directly performs watermark detection or identification on the attacked images. Instead, \sysname evaluates watermark robustness using PreEvader and PostEvader, whereas PreEvader applies attacks to the released watermarked training dataset before downstream model training, and PostEvader applies attacks to the model outputs before watermark verification. The watermark is then verified through the behavior or generated outputs of the trained model. 
Nevertheless, the image processing, diffusion-based regeneration, and adversarial attacks evaluated in WAVES can likewise be included in PreEvader. We therefore incorporate representative attacks from these categories into PreEvader to evaluate watermark robustness in the dataset auditing pipeline.

\mypara{Sequential Watermarking and Multi-Ownership}
The multi-watermark scenario in \sysname adopts sequential watermarking, where different watermark methods are successively applied to the same dataset, with each method operating on the output of the preceding one.
Such sequential embedding is one way to create multi-ownership-watermarked data, where multiple owners may successively embed their watermarks and later assert ownership over the same dataset.
Craver~\etal~\cite{craver1998resolving} study the classic image-level multi-ownership problem and examine whether claimed originals and watermark evidence can identify the rightful owner. 
They show that sequential embedding may cause ownership disputes when combined with counterfeit originals or forged watermark evidence. 

\sysname extends this concern from a single image to the dataset--model pipeline.
It successively applies different watermark methods to the same training dataset, retains each method's original injection and verification procedures, and examines the resulting evidence through model outputs.
Without introducing counterfeit originals or forged evidence, the evaluation isolates how later watermarking and downstream model training may weaken, overwrite, or interfere with earlier ownership signals.

\section{\sysname}

\begin{table*}[!t]
\centering
\small
\renewcommand{\arraystretch}{0.85}
\caption{Summary of representative dataset watermark methods, divided by our taxonomy.
~\textbf{Task:} C = classification task, G = generation task.
~\textbf{Access:} \circletfill\xspace = black box, \circlet\xspace = white box, \circletfillhl\xspace = gray box.
~\textbf{Stealthiness:} \ding{55} = Stealthiness of the watermark not guaranteed, \ding{51} = Stealthiness guaranteed.
~\textbf{Objective}: Watermark methods are categorized into two types: one-bit and multi-bit (one-bit methods provide only a binary ownership decision, while multi-bit methods encode multiple bits of information in the model~\cite{li2021survey,zhong2023brief}). O = One-Bit, M = Multi-Bit.
~\textbf{Capability:} We summarize the deployability of watermarks from an application perspective, focusing on Sample Dependency (whether the watermark relies on specially optimized trigger samples) and Model Dependency (whether shadow models are required in watermark deployment). SI = Sample Independent, SD = Sample Dependent, MI = Model Independent, MD = Model Dependent.
}
\begin{tabularx}{\textwidth}{C{2.6cm} C{2.2cm} C{1.3cm} C{1.3cm} C{1.3cm} C{1.3cm} C{1.3cm} C{3.0cm} C{1.0cm}}
\toprule
\textbf{Implementation} & \textbf{Watermarks} & \textbf{Tasks} & \textbf{Objectives} & \textbf{Capabilities} & \textbf{Access} & \textbf{Stealthiness} & \textbf{Metrics} & \textbf{Year} \\
\midrule
\multirow{10}{*}{\parbox{2.6cm}{\centering Model-Free \& \\ Model-Behavior}} 
& \WANet~\cite{nguyen2021wanet} & C & O & SI \& MI & \circletfill\xspace & \ding{55} & BA, ASR, HFR & 2021 \\
& \DVBW~\cite{li2023black} & C & O & SI \& MI & \circletfill\xspace & \ding{55} & BA, ASR, p-value & 2023 \\
& \AntiNeuron~\cite{zou2022anti} & C & O & SI \& MI & \circletfill\xspace & \ding{55} & Average Loss & 2022 \\
& \DataIsotope~\cite{wenger2023data} & C & O & SI \& MI & \circletfill\xspace & \ding{55} & TPR@5\%FPR & 2023 \\
& \UBWP~\cite{li2022untargeted} & C & O & SI \& MI & \circletfill\xspace & \ding{55} & BA, ASR, p-value & 2023 \\
& \DiagnosisB~\cite{wang2023diagnosis} & G & O & SI \& MI & \circletfill\xspace & \ding{55} & Acc, TP, FP, TN, FN & 2023 \\
& \EnTruth~\cite{ren2024entruth} & G & O & SI \& MI & \circletfill\xspace & \ding{55} & F1 Score, F1-N & 2024 \\
& \GaussWM~\cite{wang2024detecting} & G & O & SI \& MI & \circletfill\xspace & \ding{55} & Acc, TPR & 2024 \\
& \DwtWM~\cite{wang2024detecting} & G & O & SI \& MI & \circletfill\xspace & \ding{55} & Acc, TPR & 2024 \\

\midrule
\multirow{3}{*}{\parbox{2.6cm}{\centering Model-Free \& \\ Model-Message}} & \DiagnosisA~\cite{wang2023diagnosis} & G & O & SI \& MI & \circletfill\xspace & \ding{55} & Acc, TP, FP, TN, FN & 2023 \\
& \DiffShield~\cite{cui2023diffusionshield} & G & M & SI \& MI & \circletfill\xspace & \ding{51} & Bit-Acc & 2023 \\
& \DwdctWM~\cite{luo2023steal} & G & M & SI \& MD & \circletfill\xspace & \ding{51} & NC & 2023 \\
\midrule
\multirow{6}{*}{\parbox{2.6cm}{\centering Model-Based \& \\ Model-Behavior}} & \DYTMark~\cite{tang2023did} & C & O & SD \& MD & \circletfill\xspace & \ding{51} & BA, TSR, WDR & 2023 \\
& \LBConsist~\cite{turner2019label} & C & O & SD \& MI & \circletfill\xspace & \ding{51} & BA, ASR & 2019 \\
& \ImgDup~\cite{huang2024general} & C & O & SD \& MI & \circletfill\xspace & \ding{51} & BA, VSR & 2024 \\
& \RadioActive~\cite{sablayrolles2020radioactive} & C & O & SD \& MD & \circletfillhl\xspace & \ding{51} & BA, p-value & 2020 \\
& \SleepAgent~\cite{souri2022sleeper} & C & O & SD \& MD & \circletfill\xspace & \ding{51} & BA, ASR & 2022 \\
& \AdvWM~\cite{wang2024detecting} & G & O & SI \& MD & \circlet\xspace & \ding{51} & Acc, TPR & 2024 \\
\midrule
\multirow{7}{*}{\parbox{2.6cm}{\centering Model-Based \& \\ Model-Message}} & 
\GenWM~\cite{ma2023generative} & G & O & SI \& MD & \circletfill\xspace & \ding{51} & Acc & 2023 \\
& \RIW~\cite{tan2023somewhat} & G & O & SD \& MI & \circletfill\xspace & \ding{51} & Acc, TPR, FPR & 2023 \\
& \Siren~\cite{li2025towards} & G & O & SD \& MD & \circletfill\xspace & \ding{51} & TPR, FPR, F1 Score & 2025 \\
& \FTShield~\cite{cui2025ft} & G & O & SD \& MD & \circletfill\xspace & \ding{51} & TPR, FPR & 2025 \\
& \ArtiGAN~\cite{yu2021artificial} & G & M & SI \& MI & \circletfill\xspace & \ding{51} & Bit-Acc & 2021 \\
& \SSL~\cite{luo2023steal} & G & M & SI \& MD & \circletfill\xspace & \ding{51} & Bit-Acc & 2023 \\
& \RivaGAN~\cite{luo2023steal} & G & M & SI \& MD & \circletfill\xspace & \ding{51} & Bit-Acc & 2023 \\
\bottomrule
\end{tabularx}
\label{tab:summary_methods}
\end{table*}

In this section, we provide a structured synthesis of existing dataset watermark methods, classifying them into four categories. 
This taxonomy enables systematic comparison and lays the foundation for unified evaluation across heterogeneous techniques. 
We then introduce the categorized methods, specifying their input, output, and functional characteristics to facilitate consistent implementation in \sysname. 
Finally, we detail the design and implementation of \sysname, highlighting core classes, functions, and attributes supporting modular benchmarking and future extensibility.

\subsection{Taxonomy}

\autoref{tab:summary_methods} reveals considerable heterogeneity in existing studies in terms of implementation and experimental configurations. 
To systematically analyze existing dataset watermark methods, we categorize them according to two dimensions: watermark injection and watermark verification.

\mypara{Watermark Injection}
The injection stage can be categorized into \textit{model-based} and \textit{model-free} approaches. 
Model-based methods leverage neural networks (\eg, by acting as adversarial perturbation generators~\cite{szegedy2013intriguing}~\cite{tang2023did}, altering optimization objectives~\cite{cui2023diffusionshield}, or constructing shadow models~\cite{ma2023generative}) to embed watermarks in images. 

These methods can thus be summarized as three variants:
(i) utilize a surrogate model to guide or participate in the watermark embedding process,
(ii) jointly optimize a watermark encoder-decoder, 
and (iii) combine both a surrogate model for task or semantic consistency and a watermark model for reliable decoding.

Conversely, model-free methods perturb images or embed specific patterns, without relying on a model for watermark injection. 

\mypara{Watermark Verification}
The verification stage can be divided into \textit{model-behavior} and \textit{model-message} approaches.
Model-behavior evaluates model outputs, such as classification logits or generated content, using specially designed trigger samples and statistical techniques to detect watermarks.
Model-message directly decodes the embedded message corresponding to the watermark from the model predictions or intermediate representations. 

This yields four distinct classes, each reflecting different design choices and operational constraints.
Detailed descriptions of individual methods are provided in~\autoref{sec:detailed-watermark-methods}.

\mypara{Model-Free with Model-Behavior}
In this category, predefined patterns or transformations are embedded into training samples, and verification checks whether they induce a corresponding model response.
Because injection does not rely on a model, these methods can be deployed without optimizing the watermark for a downstream architecture and applied to both classification and generation tasks.
Verification only needs to identify a consistent change in model output, enabling black-box auditing.
Effectiveness depends on how strongly and distinctly the model learns the association between the injected pattern and expected response.
In a multi-user scenario, similar triggers from different owners may cause overlapping responses and obscure attribution in multi-user settings.

\mypara{Model-Free with Model-Message}
In this category, predefined watermark information is embedded into training images and later decoded from model outputs.
The decoded watermark information can then be used by the verifier for dataset usage auditing.
All surveyed methods in this category target generation tasks because image outputs can carry the information required for message recovery.
This approach depends on how much of the payload survives fine-tuning and is reproduced during generation.
Fine-tuning, image processing, regeneration, and subsequent watermarking can progressively weaken the signal, making payload preservation this category's main challenge.

\mypara{Model-Based with Model-Behavior}
In this category, surrogate models, gradient-based optimization, or learned perturbation generators guide how selected training samples are modified so that the downstream model learns a watermark-specific response.
The watermark is verified by querying the trained model and measuring the behavioral trace left by these modifications.
This combination is particularly common in both classification and generation tasks.
Using model feedback during injection can produce visually subtle modifications while preserving a response that remains detectable after training.
Because the signal is optimized around particular learned representations, its effectiveness depends on the relationship between the surrogate and target models.
Changes in architecture or training distribution, as well as earlier modifications to the dataset, can alter the gradients or features on which later watermark construction relies.

\mypara{Model-Based with Model-Message}
All surveyed methods in this category target generation tasks, whose image outputs can carry recoverable watermark information.
In this category, encoders, surrogate models, or semantic feature optimization are used to embed a watermark representation that can later be decoded from generated images.
The generated output must preserve the watermark representation throughout downstream training and generation.
The category contains both one-bit and multi-bit methods, which place different demands on information preservation.
One-bit verification requires only a detectable trace, whereas multi-bit attribution requires enough of the encoded payload to distinguish the intended message.
This difference explains why several one-bit methods remain detectable after output transformations and regeneration, while multi-bit decoding is more easily degraded by fine-tuning, output modification, and interference from other watermarks.

\begin{figure*}[!t]
    \centering
    \includegraphics[width=0.95\linewidth]{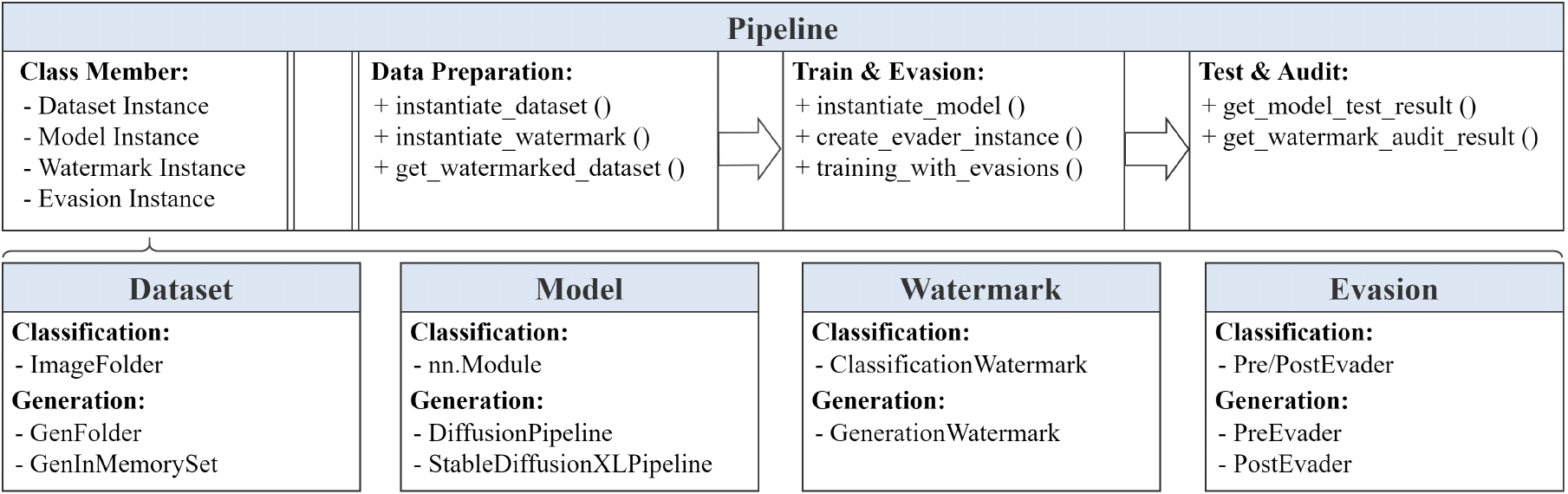}
    \caption{Implementation of \sysname. Pipeline is the main class for managing the entire workflow, which instantiates and integrates the four core components: Dataset, Model, Watermark, and Evasion.}
    \label{fig:dw_bench_pipeline}
\end{figure*}

\subsection{Implementation of \sysname}

As illustrated in~\autoref{fig:dw_bench_pipeline}, \sysname includes five components: Dataset, Model, Watermark, Evasion, and the central coordinator component Pipeline.

\mypara{Dataset}
This component handles data loading and management.
For classification tasks, \sysname accepts datasets in ImageFolder format.
For generation tasks, it accepts a JSONL file to instantiate a GenFolder object.
The component flexibly supports both loading datasets from disk and directly accepting pre-loaded dataset objects.

\mypara{Model}
This component defines the models for training and auditing.
As shown in~\autoref{fig:dw_bench_pipeline}, it supports nn.Module-like classes for classification tasks and specialized multi-model pipelines (\eg, DiffusionPipeline) for generation tasks.

\mypara{Watermark}
This component encapsulates the logic for watermark methods. 
Each method is implemented by subclassing the base class Watermark.
This base class requires the implementation of two key interfaces: get\_watermarked\_dataset and audit\_model.
This design aligns with the two main stages of the watermark process(Watermark Injection and Verification) summarized previously, while maintaining flexibility to support reproducing diverse categories of watermark methods within the same framework.

\mypara{Evasion}
This component implements strategies a model trainer might use to remove watermarks.
As illustrated in~\autoref{fig:dw_bench_pipeline}, these evaders are categorized by when they are applied during the training:
PreEvader (\eg, data augmentation on the watermarked dataset),
MidEvader (\eg, applying Differentially Private Stochastic Gradient Descent during training),
and PostEvader (\eg, model pruning).

\mypara{Pipeline}
The Pipeline class serves as the central coordinator of \sysname.
It instantiates the Dataset, Model, Watermark, and Evasion components and manages the entire watermark workflow across three main stages:

\begin{itemize}
    \item Data Preparation: The pipeline initializes the dataset and watermark components by calling instantiate\_dataset() and instantiate\_watermark() to obtain their instances.
    Then it calls get\_water- marked\_dataset() to embed the watermark in the dataset.

    \item Train \& Evasion: The pipeline instantiates the model and evasion components by calling instantiate\_model() and create\_evader\_in- stance(), and handles the model training process, integrating the specified evasion mechanisms.

    \item Test \& Audit: Finally, it calls get\_model\_test\_result() to assess the impact of the watermark on the model's primary task performance and get\_watermark\_audit\_result() to perform the verification, yielding the final audit results.
\end{itemize}

\subsection{Key Features and Contributions}

\sysname is a unified open-source benchmarking platform designed to comprehensively evaluate and compare dataset watermark methods. 
We envision its roles as follows.

\begin{itemize}
    \item \sysname provides a unified framework with 25 built-in watermark methods, while also enabling the reproduction and evaluation of external methods under the same unified framework.

    \item \sysname not only supports data publishers in protecting copyright via dataset watermark, but also facilitates a comprehensive comparison of watermark methods through unified metrics.

    \item \sysname follows a modular design, where researchers are not limited to the built-in methods. 
    They can flexibly use their own datasets, evasion methods, models, and configurations to construct a customized pipeline.
\end{itemize}

\section{Experimental Settings}

\subsection{Research Questions}

In this section, we perform experiments to address the following research questions.

\mypara{RQ1}
How does the performance of watermark methods vary across different categories of our proposed taxonomy?

\mypara{RQ2}
How do the watermark methods perform under different adversarial attacks?

\mypara{RQ3}
How does multi-watermark overlapping affect the verification of each watermark?

\mypara{RQ4}
In multi-user scenarios, can current watermark methods effectively verify ownership for each user?

\subsection{Experimental Setup}
\label{subsec:experimental-setup}
We evaluate 25 representative watermark methods based on our open-source toolkit \sysname, guaranteeing the reproducibility of the results.
Our assessment covers key metrics for dataset watermark and evaluates robustness under realistic adversarial, multi-watermark, and multi-user scenarios.
For method-specific parameters, we prioritize the settings reported in the original papers and released implementations, while benchmark-specific adaptations and exact configurations are documented in~\autoref{appendix_setup}.

\mypara{Metrics}
A major obstacle in benchmarking dataset watermarks is the heterogeneity of auditing procedures: methods share the goal of detecting training on watermarked data but report incompatible confidence metrics (\eg, p-values, success rates, and custom scores), impeding fair comparison.
We address this by using TPR@FPR as a standardized sample-level metric and verification success rate (VSR) as a dataset-level metric.
TPR@FPR measures the true-positive rate achieved at a fixed false-positive rate based on each method's detection scores.
VSR measures dataset-level, end-to-end reliability as the fraction of correct binary auditing decisions across repeated controlled trials.
For each configuration, we report TPR@5\%FPR and VSR, while \sysname also supports alternative FPR thresholds.
The rationale for choosing 5\% FPR as the default operating point is discussed in~\autoref{sec:appendix_fpr_rationale}.
We estimate VSR using five independent runs.
We also assess model utility using benign accuracy (BA) for classification and Fr\'{e}chet Inception Distance (FID) for generation.

We additionally report Peak Signal-to-Noise Ratio (PSNR) between original and watermarked samples as a reference metric for watermark stealthiness.
Instead of a fixed PSNR target, which is unfeasible across diverse watermark techniques, we report it alongside other metrics to provide a complete view of the trade-offs among sample quality, watermark capability, and model performance.

\mypara{Datasets and Models}
We conduct extensive experiments on six datasets and six model architectures.
For classification tasks, we evaluate on CIFAR-10, CIFAR-100, and TinyImageNet using ResNet-18, VGG, and ViT backbones.
For generation tasks, we use the Pok\'{e}mon, CelebA, and WikiArt datasets together with Stable Diffusion V1.4, Stable Diffusion XL, and Kandinsky 2.2.

\mypara{Watermarked-Sample Ratio}
We define the watermarked-sample ratio (WSR) as the fraction of training samples selected for watermarking.
We evaluate classification methods at WSRs from 0.01\% to 10\% and generation methods at WSRs from 2\% to 50\%,
to assess watermark performance under varying watermark strengths.

\mypara{Multi-Watermark Scenario}
The multi-watermark scenario simulates a realistic case where a dataset is sequentially watermarked by various intermediate distributors using distinct methods.
Each subsequent watermark method is applied to the entire output dataset produced by the preceding method.
Therefore, later methods may modify samples that already carry earlier watermarks, allowing us to evaluate whether earlier watermark evidence remains verifiable after subsequent watermark injection, a situation that is likely to arise
in complex real-world dataset distribution chains.

\mypara{Multi-User Scenario}
The multi-user scenario simulates
a training dataset pooled from multiple data publishers.
To simulate this, we partition the training dataset into pairwise-disjoint owner subsets.
For a $K$-user setting, $K$ subsets are assigned to the enrolled users, each of whom watermarks only its own subset, while the remaining subsets are kept clean.
All subsets are then merged to form the complete training dataset.
Consequently, each training sample is watermarked by at most one user in the multi-user scenario.

We consider two settings.
In the \textit{same watermark for multi-user} setting, all users employ the same watermark method but use a unique key or message to instantiate their individual watermarks.
In the \textit{different watermarks for multi-user} setting, different users apply distinct watermark methods to their respective subsets.
We keep the per-user watermarked-sample ratio fixed, so the overall proportion of watermarked samples in the training dataset increases with the number of enrolled users.
Unlike the multi-watermark scenario, the multi-user scenario does not sequentially apply multiple users' watermarks to the same dataset samples.

Due to the differences in datasets and training settings between classification and generation tasks, we conduct the experiments separately for each type of task in the following sections.
We provide a detailed experimental setup for the two tasks in \autoref{appendix_setup}.

\begin{table*}[!t]
\centering
\small
\caption{Effectiveness of different watermark methods for classification tasks, using ResNet18 under the CIFAR-10 dataset. The TPR results are reported under FPR=5\%. The BA drop evaluates the classification accuracy drop of watermarked models compared to the benign model. Ave-PSNR measures the average difference between watermarked and original samples.}
\label{tab:general_results_classification_with_stealthiness}
\renewcommand{\arraystretch}{0.85}
\setlength{\tabcolsep}{4pt}
\begin{tabular}{l|ccc|ccc|ccc|ccc|c}
\toprule
\multirow{2}{*}{\textbf{Method}} &
\multicolumn{3}{c|}{\textbf{WSR = 0.1}} &
\multicolumn{3}{c|}{\textbf{WSR = 0.01}} &
\multicolumn{3}{c|}{\textbf{WSR = 0.001}} &
\multicolumn{3}{c|}{\textbf{WSR = 0.0001}} &
\multirow{2}{*}{\textbf{Ave-PSNR}} \\
 & \textbf{{TPR}} & \textbf{BA Drop} & \textbf{VSR}
 & \textbf{{TPR}} & \textbf{BA Drop} & \textbf{VSR}
 & \textbf{{TPR}} & \textbf{BA Drop} & \textbf{VSR}
 & \textbf{{TPR}} & \textbf{BA Drop} & \textbf{VSR} & \\

\midrule
\WANet        & 99.47 & 0.78 & 1.0 & 89.73 & 0.76 & 1.0 & 14.67 & 0.72 & 0.0 & 4.40  & 0.11 & 0.0 & 29.93 \\
\DVBW         & 92.72 & 0.58 & 1.0 & 92.43 & 0.17 & 1.0 & 92.48 & 0.24 & 1.0 & 89.36 & 0.24 & 1.0 & 30.89 \\
\AntiNeuron   & 8.23  & 0.50 & 1.0 & 9.10  & 0.14 & 1.0 & 9.89  & 0.18 & 0.0 & 10.00 & 1.05 & 0.0 & 24.13 \\
\DataIsotope  & 58.00 & 8.32 & 1.0 & 9.28  & 0.26 & 1.0 & 5.07  & 0.34 & 0.0 & 5.73  & 1.14 & 0.0 & 24.13 \\
\UBWP         & 93.87 & 1.59 & 1.0 & 91.87 & 0.80 & 1.0 & 7.07  & 0.46 & 0.0 & 5.33  & 0.56 & 0.0 & 25.06 \\
\DYTMark      & 100.00& 9.34 & 1.0 & 91.73 & 0.23 & 1.0 & 86.67 & -0.11 & 1.0 & 60.35 & 0.23 & 1.0 & 31.08 \\
\LBConsist    & 71.71 & 9.41 & 1.0 & 13.44 & 0.20 & 0.2 & 8.27  & 0.17 & 0.0 & 6.29  & 0.24 & 0.0 & 29.67 \\
\RadioActive  & 100.00& 9.21 & 1.0 & 23.31 & 0.31 & 1.0 & 10.29 & 0.28 & 0.0 & 4.67  & 0.43 & 0.0 & 28.63 \\
\SleepAgent & 26.71 & 9.66 & 1.0 & 68.62 & 0.10 & 1.0 & 15.33 & 0.09 & 1.0 & 15.60 & -0.03 & 0.6 & 24.96 \\
\ImgDup       & 6.27  & 0.32 & 1.0 & 8.13  & -0.05 & 1.0 & 4.00  & -0.04 & 0.0 & 20.00 & 0.42 & 0.0 & 29.47 \\
\bottomrule
\end{tabular}
\end{table*}

\begin{table*}[!t]
\centering
\caption{Robustness evaluation under PreEvader attacks for classification tasks. All methods are measured by TPR@5\%FPR.}
\label{tab:classification_watermark_pre_robustness}

\small
\setlength{\tabcolsep}{2pt}
\renewcommand{\arraystretch}{0.85}
\rowcolors{2}{myblue!40!white}{white}

\begin{tabularx}{0.88\textwidth}{
    @{}
    >{\raggedright\arraybackslash}p{1.55cm}|
    *{13}{>{\centering\arraybackslash}X}
    @{}
}
\toprule
\rowcolor{white}
\textbf{Method} & \textbf{No Att.} & \textbf{Crop} & \textbf{Rot.} & \textbf{Resize} & \textbf{Blur} & \textbf{JPEG} & \textbf{JPEG2K} & \textbf{WebP} & \textbf{Diff} & \textbf{DiffP} & \textbf{2xDiff} & \textbf{VAE} & \textbf{Adv.} \\
\midrule
\WANet & 84.80 & 3.33 & 88.53 & 2.80 & 8.93 & 22.53 & 9.73 & 40.40 & 6.67 & 14.93 & 12.13 & 10.00 & 86.53 \\
\DVBW & 93.73 & 76.93 & 94.40 & 11.60 & 88.80 & 10.40 & 8.00 & 94.53 & 6.00 & 91.87 & 7.07 & 83.07 & 92.93 \\
\AntiNeuron & 8.23 & 5.39 & 7.82 & 3.78 & 5.78 & 7.71 & 7.98 & 8.64 & 7.40 & 6.78 & 11.63 & 5.80 & 6.36 \\
\DataIsotope & 7.33 & 5.07 & 4.99 & 5.48 & 4.70 & 5.62 & 5.12 & 5.00 & 5.12 & 4.71 & 5.74 & 5.10 & 4.99 \\
\UBWP & 90.80 & 15.60 & 8.00 & 55.47 & 66.13 & 28.53 & 65.60 & 9.60 & 64.00 & 25.60 & 53.87 & 75.47 & 74.53 \\
\DYTMark & 91.73 & 55.47 & 77.60 & 9.60 & 78.40 & 2.00 & 2.27 & 91.07 & 3.60 & 90.67 & 3.60 & 90.40 & 84.13 \\
\LBConsist & 17.60 & 42.27 & 2.80 & 4.00 & 71.20 & 4.13 & 2.27 & 19.73 & 1.73 & 75.47 & 3.07 & 35.20 & 5.73 \\
\RadioActive & 22.93 & 22.80 & 18.93 & 25.07 & 7.60 & 18.27 & 12.13 & 18.93 & 13.60 & 3.73 & 21.47 & 18.13 & 14.93 \\
\SleepAgent & 72.53 & 67.47 & 8.93 & 5.60 & 28.40 & 4.80 & 4.67 & 4.27 & 4.93 & 0.40 & 7.33 & 72.40 & 43.07 \\
\ImgDup & 8.13 & 11.87 & 13.60 & 24.67 & 22.53 & 5.73 & 5.20 & 8.93 & 5.20 & 9.07 & 7.33 & 8.67 & 9.07 \\
\bottomrule
\end{tabularx}
\end{table*}

\section{Results for Classification Tasks}

\subsection{Watermark Effectiveness}
\label{classification-result}

\autoref{tab:general_results_classification_with_stealthiness} shows the evaluation results of watermark methods for classification tasks on CIFAR-10. 
More experimental results on other datasets can be found in \autoref{sec:classification_other_results}.

\mypara{Overall Performance}
For model-free methods, we observe that backdoor-based methods achieve high TPR@5\%FPR under common settings (\eg, WSR=0.01). 
This is because they leverage the label-image mapping for watermarking.
Since this mapping is a primary learning objective, models can easily learn the backdoor.

In contrast, model-based methods generally achieve lower TPR scores. 
However, \DYTMark stands out as an exception, achieving a high score of 91.73\% at WSR=0.01.
We hypothesize that this strong performance stems from the role of its surrogate model.
Unlike methods that use models to optimize the watermark itself, \DYTMark uses its surrogate to perturb normal sample features before injection.
This allows it to select fixed and robust backdoor features as watermarks, enhancing watermark learnability.
Additionally, methods like \DataIsotope, \DYTMark, and \LBConsist exhibit a significant BA Drop at high watermarked-sample ratios (\eg, WSR=0.1 on CIFAR-10), where most of a class's images are watermarked. 
We hypothesize this occurs because their watermarks are feature-destructive, either hindering relevant feature learning or acting as adversarial attacks, corrupting the model's learning process and finally causing the entire class to be misclassified.

\mypara{Impact of Watermarked-Sample Ratios}
As the watermarked-sample ratio decreases, all methods show a decline in performance.
For most methods (7 out of 10 for classification tasks), the critical threshold lies between WSR=0.01 and 0.001, beyond which their VSR drops sharply to 0.0, indicating that they are no longer capable of performing any successful audit under this condition.
\DVBW, \DYTMark, and \SleepAgent are the only three methods that successfully verified watermark existence under the low watermark situation on CIFAR-10-ResNet18 condition, showing the potential of backdoor-based approaches in such challenging scenarios.

\mypara{Impact of Different Datasets and Models}
We further conducted extensive experiments on other datasets and models (see \autoref{sec:classification_other_results}).
Our analysis reveals two key findings:
First, compared to the two commonly used baseline architectures, ResNet18 and VGG16, watermark methods generally exhibit significantly degraded performance on Vision Transformer models.
On CIFAR-10, five methods fail to achieve successful auditing at WSR=0.01 for the ViT model.
It is noteworthy that most of these are model-based methods, which suggests that model-based approaches may exhibit poor generalization when facing different model paradigms.

Second, as dataset complexity increases, the overall verification success rate declines substantially.
The number of experimental settings yielding zero VSR rises from 49\% on CIFAR-10 to 57 on TinyImageNet.
This underscores that watermark auditing for datasets with high class diversity remains a persisting challenge.

\begin{table}[!t]
\centering
\setlength{\tabcolsep}{0.4em}
\renewcommand{\arraystretch}{0.85}
\small
\rowcolors{2}{myblue!40!white}{white}
\caption{Robustness evaluation under PostEvader attacks for classification tasks.
All methods are measured by TPR@5\%FPR.}
\label{tab:attack_for_classification}
\begin{tabular}{l|cccc}
\toprule[1pt]
\textbf{Method} & \textbf{No Attack} & \textbf{Label Only} & \textbf{Perturb 0.1} & \textbf{Perturb 0.5} \\
\midrule
\WANet       & 84.80 & 85.87 & 93.20 & 74.53 \\
\DVBW        & 93.73 & 92.80 & 93.87 & 85.60 \\
\AntiNeuron  & 8.23  & 5.38  & 7.47  & 6.53  \\
\DataIsotope & 7.33  & 6.93  & 8.27  & 7.20  \\
\UBWP        & 90.80 & 86.93 & 92.40 & 68.40 \\
\DYTMark     & 91.73 & 91.46 & 92.93 & 82.80 \\
\LBConsist   & 17.60 & 10.67 & 16.53 & 13.47 \\
\RadioActive & 22.93 & 14.13 & 22.67 & 18.27 \\
\SleepAgent  & 72.53 & 62.67 & 75.27 & 69.33 \\
\ImgDup      & 8.13  & 4.93  & 4.13  & 3.87  \\
\bottomrule[1pt]
\end{tabular}
\end{table}

\begin{table*}[!t]
\centering
\caption{Multi-watermark results for different watermark methods in classification tasks.
The first part of the table shows TPR@5\%FPR when the row watermark is covered by the column watermark.
The second part presents the reverse setting, where the row watermark is applied after the column watermark, thereby covering it.}
\label{tab:multi-watermark-classification}
\small
\setlength{\tabcolsep}{3pt}
\renewcommand{\arraystretch}{0.85}
\begin{tabular}{l|cccccccccc}
\toprule[1pt]
\textbf{Method} & \textbf{\WANet} & \textbf{\DVBW} & \textbf{\AntiNeuron} & \textbf{\DataIsotope} & \textbf{\UBWP} & \textbf{\DYTMark} & \textbf{\LBConsist} & \textbf{\RadioActive} & \textbf{\SleepAgent} & \textbf{\ImgDup} \\
\midrule
\WANet       & /      & 92.80  & 81.33 & 77.33 & 79.20 & 76.67 & 83.20 & 91.60 & 7.20  & 92.53 \\
\DVBW        & 88.13  & /      & 91.87 & 91.33 & 87.60 & 89.87 & 88.27 & 90.67 & 5.87  & 91.33 \\
\AntiNeuron  & 7.18   & 5.70   & /     & 9.51  & 7.84  & 7.38  & 10.23 & 10.23 & 7.97  & 9.98  \\
\DataIsotope & 4.75   & 5.77   & 5.89  & /     & 4.79  & 5.24  & 6.01  & 5.27  & 5.14  & 5.00  \\
\UBWP        & 89.87  & 90.53  & 90.13 & 91.20 & /     & 90.13 & 70.40 & 90.53 & 79.07 & 89.07 \\
\DYTMark     & 86.93  & 92.00  & 92.53 & 91.73 & 92.27 & /     & 92.40 & 92.00 & 92.13 & 93.07 \\
\LBConsist   & 100.00 & 100.00 & 99.73 & 99.60 & 99.87 & 99.87 & /     & 99.73 & 97.33 & 99.73 \\
\RadioActive & 21.73  & 23.20  & 24.13 & 13.87 & 29.20 & 23.87 & 24.80 & /     & 15.20 & 24.27 \\
\SleepAgent  & 29.60  & 74.13  & 23.87 & 58.13 & 83.60 & 60.27 & 89.60 & 59.33 & /     & 53.86 \\
\ImgDup      & 10.53  & 5.73   & 8.80  & 7.73  & 4.93  & 6.13  & 5.47  & 4.40  & 4.27  & /     \\
\midrule
\WANet       & /     & 86.53 & 81.47 & 84.00 & 91.33 & 93.73 & 89.07 & 87.60 & 67.87  & 74.93  \\
\DVBW        & 89.60 & /     & 90.53 & 90.40 & 87.60 & 90.40 & 89.87 & 91.07 & 90.67  & 90.93  \\
\AntiNeuron  & 9.38  & 10.64 & /     & 5.12  & 6.55  & 7.83  & 8.20  & 9.49  & 6.67   & 8.12   \\
\DataIsotope & 5.39  & 6.02  & 4.74  & /     & 5.37  & 5.13  & 4.74  & 5.00  & 5.14   & 5.53   \\
\UBWP        & 91.87 & 93.07 & 92.80 & 93.07 & /     & 92.13 & 92.13 & 92.40 & 92.27  & 90.67  \\
\DYTMark     & 92.13 & 92.00 & 91.73 & 91.33 & 90.27 & /     & 92.40 & 91.73 & 93.47  & 92.13  \\
\LBConsist   & 99.73 & 99.60 & 99.07 & 99.73 & 99.60 & 99.47 & /     & 99.87 & 100.00 & 100.00 \\
\RadioActive & 23.07 & 24.13 & 22.67 & 20.80 & 22.40 & 24.67 & 24.93 & /     & 24.80  & 19.60  \\
\SleepAgent  & 42.00 & 79.60 & 84.00 & 64.67 & 24.80 & 10.40 & 87.20 & 83.47 & /      & 70.00  \\
\ImgDup      & 8.93  & 9.33  & 7.07  & 8.80  & 10.13 & 8.80  & 5.73  & 5.60  & 6.13   & /      \\
\bottomrule[1pt]
\end{tabular}
\end{table*}

\begin{table}[!t]
\centering
\setlength{\tabcolsep}{0.6em}
\renewcommand{\arraystretch}{0.85}
\small
\rowcolors{2}{myblue!35!white}{white}
\caption{Multi-user results for the same watermark scenario. 
1U, 2U, and 5U denote the number of enrolled users. 
1F, 2F, and 5F represent results from an ``independent'' user (F), whose watermarked dataset was \textit{not} used during training. 
Results are averaged across enrolled users. 
}
\label{tab:multi_user_samewm_classification}
\begin{tabular}{l|cccccc}
\toprule[1pt]
\textbf{Method} & \textbf{1U} & \textbf{1F} & \textbf{2U} & \textbf{2F} & \textbf{5U} & \textbf{5F} \\
\midrule
\WANet           & 96.27 & 4.93 & 92.20 & 10.13 & 96.05 & 8.53  \\
\DVBW            & 90.53 & 7.47 & 95.47 & 7.47  & 92.77 & 2.83  \\
\AntiNeuron      & 8.65  & 5.52 & 9.20  & 4.30  & 10.67 & 5.32  \\
\DataIsotope     & 21.73 & 5.87 & 15.27 & 3.87  & 13.52 & 2.13  \\
\UBWP            & 84.00 & 6.93 & 88.27 & 11.47 & 87.52 & 84.67 \\
\DYTMark         & 7.60  & 5.60 & 8.73  & 4.00  & 8.53  & 8.13  \\
\LBConsist       & 93.47 & 8.93 & 99.94 & 17.33 & 71.57 & 13.20 \\
\RadioActive     & 18.00 & 2.00 & 17.94 & 3.60  & 14.40 & 0.13  \\
\SleepAgent      & 10.80 & 4.93 & 17.87 & 10.27 & 9.16  & 3.73  \\
\ImgDup          & 7.60  & 5.30 & 6.67  & 6.27  & 6.56  & 4.27  \\
\bottomrule[1pt]
\end{tabular}
\end{table}

\subsection{Robustness in Adversarial Environment}
\label{sec:robustness-evaluation-classification-task}

We separately evaluate watermark methods for classification models under PreEvader and PostEvader attacks, with each experimental condition applying either one PreEvader attack or one PostEvader attack.
PreEvader applies twelve preprocessing attacks to the released training images before victim-model training: center cropping (ratio 0.8), rotation ($15^\circ$), resizing (scale 0.5), Gaussian blur (radius 0.8), JPEG and WebP compression (quality 30), JPEG2000 compression (5:1), Regen-Diff, Regen-DiffP, Regen-2xDiff, Regen-VAE, and adversarial embedding.
Regen-Diff uses Stable Diffusion v1.4 with strength 0.5, 30 inference steps, and guidance scale 7.5; 
Regen-DiffP uses the class name as the prompt, such as ``airplane'' or ``bird'';
Regen-2xDiff applies Regen-Diff twice sequentially;
Regen-VAE reconstructs images through its VAE; and adversarial embedding uses 10-step PGD ($\epsilon=8/255$, step size $2/255$) to maximize the CLIP ViT-B/32 embedding distance.
PostEvader either returns only the top-1 predicted label or adds zero-mean Gaussian noise whose standard deviation is 10\% or 50\% of the logit range.
Complete implementation details and visual examples are provided in~\autoref{sec:evasion-attack-settings} and~\autoref{fig:appendix_evader_visual_examples}.

As shown in~\autoref{tab:classification_watermark_pre_robustness}, robustness under PreEvader attacks varies substantially across methods and transformations.
For example, \DVBW retains TPR@5\%FPR above 83.0\% under rotation, WebP compression, Regen-DiffP, Regen-VAE, and adversarial embedding, but drops sharply under resizing, JPEG, and JPEG2000 compression.
Other backdoor-based methods show similarly uneven results, indicating that no single PreEvader attack is consistently the most destructive across methods.

The PostEvader results in~\autoref{tab:attack_for_classification} show that most methods remain relatively stable under label-only output and mild logit perturbation.
The stronger Perturb 0.5 attack causes larger reductions in TPR@5\%FPR and may also alter the model's top-1 predictions, thereby reducing its classification utility.

\subsection{Multi-Watermark Scenario}

In this scenario, we evaluate the impact of sequential watermark injection on classification watermark methods.
All methods employ a 0.01 watermarked-sample ratio, except \LBConsist, which uses 0.1 to achieve a 1.0 verification success rate.
Results are presented in \autoref{tab:multi-watermark-classification}.

Our experiments reveal complex interactions between watermarks.
Nearly one-third of results exhibited counter-intuitive outcomes, where the later-deployed watermark exhibited lower TPR than when deployed first.
For instance, in the \DYTMark-\UBWP pair, \DYTMark's TPR@5\%FPR decreased by 2.00\% when deployed after \UBWP compared to prior deployment.

This phenomenon is particularly pronounced for model-based watermarks.
For example, when \UBWP is deployed first, except \ImgDup, all model-based methods' TPR@5\%FPR decreases.
This occurs because \UBWP injects mislabeled sample-label pairs into the dataset, altering its distribution.
Consequently, surrogate models trained on it become misaligned, impairing their ability to compute watermark-relevant perturbations or signatures.
Conversely, \DataIsotope similarly modifies the dataset, yet most Model-Based methods deployed after it achieved a better TPR@5\%FPR than those deployed before it.
This is because \DataIsotope's modifications to the dataset function as data augmentation rather than data corruption, improving generalization for later watermarking stages.

\subsection{Multi-User Scenario}

\mypara{Same Watermark Method for Multi-user}
We investigate how watermark methods behave when multiple users share the same method but employ different configurations in \autoref{tab:multi_user_samewm_classification}.

We observe that as the user count grows, \UBWP and \LBConsist exhibit severe false activation, where independent (\ie, non-used) users' triggers erroneously activate watermarks, inflating their TPR scores.
This issue is particularly severe for \UBWP, where the independent user achieves 84.67\% TPR with five users, rendering verification unreliable.
We hypothesize this occurs because \UBWP's detection relies on trigger-induced prediction randomness, and that when multiple users inject their own watermark, the model learns common patterns across the triggers, causing triggers from independent users to be confounded.

\begin{table}[!t]
\centering
\setlength{\tabcolsep}{0.6em}
\renewcommand{\arraystretch}{0.85}
\small
\rowcolors{2}{myblue!35!white}{white}
\caption{Multi-user results for the different watermarks scenario. 
Each method is evaluated with its own enrolled users (U) and an independent user (F) not involved in training. 
Results are averaged across users. 
}
\label{tab:multi_user_diffwm_classification}
\begin{tabular}{l|cccccc}
\toprule[1pt]
\textbf{Method} & \textbf{1U} & \textbf{1F} & \textbf{2U} & \textbf{2F} & \textbf{5U} & \textbf{5F} \\
\midrule
\WANet & 96.27 & 4.93 & 96.77 & 6.77 & 97.47 & 11.73 \\
\DVBW & 90.53 & 7.47 & 92.37 & 5.44 & 86.40 & 5.47 \\
\AntiNeuron & 8.65 & 5.52 & 15.36 & 4.62 & 7.75 & 3.74 \\
\DataIsotope & 21.73 & 5.87 & 8.30 & 4.65 & 7.05 & 5.10 \\
\UBWP & 84.00 & 6.93 & 85.87 & 20.70 & 87.60 & 9.73 \\
\DYTMark & 7.60 & 5.60 & 6.36 & 5.67 & 6.27 & 4.53 \\
\LBConsist & 93.47 & 8.93 & 99.94 & 1.67 & 99.73 & 0.27 \\
\RadioActive & 18.00 & 2.00 & 23.90 & 1.50 & 26.93 & 3.06 \\
\SleepAgent & 10.80 & 4.93 & 5.63 & 8.33 & 0.13 & 34.53 \\
\ImgDup & 7.60 & 5.30 & 6.70 & 4.66 & 6.13 & 4.80 \\
\bottomrule[1pt]
\end{tabular}
\end{table}

\mypara{Different Watermark Methods for Multi-user}
In \autoref{tab:multi_user_diffwm_classification}, we find that when each user employs a distinct watermark method, the issue of interference from the same watermark is removed. 
Consequently, the independent user's TPR@5\%FPR for \UBWP and \LBConsist remains stable as the number of users increases.
However, we observed that \SleepAgent exhibits an exceptionally high independent user's TPR@5\%FPR under the five-user setting.
This may stem from latent interactions between different model-based watermarks, highlighting the cross-method interference problems that can arise in multi-user scenarios.

\section{Results for Generation Tasks}

\subsection{Watermark Effectiveness}

\begin{table*}[!t]
\centering
\small
\caption{Effectiveness of different watermark methods for generation tasks. TPR is reported at FPR=5\%, and FID measures the quality of generated samples from watermarked models relative to the benign model.}
\label{tab:combined_watermark_results_all}
\setlength{\tabcolsep}{4pt}
\renewcommand{\arraystretch}{0.85}
\begin{tabular}{l|ccc|ccc|ccc|ccc|c}
\toprule
\multirow{2}{*}{\textbf{Method}}
 & \multicolumn{3}{c|}{\textbf{WSR=0.5}}
 & \multicolumn{3}{c|}{\textbf{WSR=0.2}}
 & \multicolumn{3}{c|}{\textbf{WSR=0.1}}
 & \multicolumn{3}{c|}{\textbf{WSR=0.02}}
 & \multirow{2}{*}{\textbf{Ave-PSNR}} \\
 & \textbf{TPR} & \textbf{FID} & \textbf{VSR}
 & \textbf{TPR} & \textbf{FID} & \textbf{VSR}
 & \textbf{TPR} & \textbf{FID} & \textbf{VSR}
 & \textbf{TPR} & \textbf{FID} & \textbf{VSR}
 & \\
\midrule
\DiagnosisB & 100.00 & 294.37 & 1.0 & 100.00 & 289.76 & 1.0 & 99.14  & 288.86 & 1.0 & 35.31  & 153.24 & 0.0 & 32.16  \\
\EnTruth      & 100.00 & 508.14 & 1.0 & 100.00 & 532.24 & 1.0 & 99.34  & 512.43 & 1.0 & 12.23  & 385.41 & 0.0 & 9.59  \\
\GaussWM    & 86.67  & 119.43 & 1.0 & 77.78  & 105.61 & 1.0 & 65.09  & 105.96 & 0.0 & 22.40  & 89.41  & 0.0 & 28.32  \\
\DwtWM        & 95.24  & 87.80  & 1.0 & 81.40  & 86.71  & 1.0 & 77.78  & 84.39  & 1.0 & 38.12  & 81.06  & 0.0 & 51.14  \\
\DiagnosisA  & 39.76  & 139.23 & 0.0 & 5.20   & 91.86  & 0.0 & 2.30   & 80.77  & 0.0 & 0.67   & 76.55  & 0.0 & 32.16  \\
\AdvWM        & 99.24  & 105.51 & 1.0 & 96.46  & 95.63  & 1.0 & 94.67  & 87.97  & 1.0 & 19.33  & 82.49  & 0.0 & 35.16  \\
\RIW          & 98.21  & 123.30 & 1.0 & 88.14  & 115.81 & 1.0 & 70.64  & 105.08 & 0.0 & 40.61  & 73.04  & 0.0 & 30.36  \\
\GenWM        & 92.67  & 377.09 & 1.0 & 87.63  & 300.57 & 1.0 & 82.67  & 297.24 & 0.0 & 12.78  & 119.07 & 0.0 & 23.03 \\
\Siren        & 100.00 & 115.03 & 1.0 & 100.00 & 119.34 & 1.0 & 98.31  & 112.12 & 1.0 & 68.67  & 112.72 & 0.0 & 38.31  \\
\FTShield   & 73.20  & 78.27  & 0.2 & 39.32  & 75.63  & 0.0 & 32.65  & 69.91  & 0.0 & 14.20  & 57.84  & 0.0 & 34.66  \\
\midrule
\multirow{2}{*}{\textbf{Method}} & \multirow{2}{*}{\textbf{Bit Acc}} & \multicolumn{8}{c|}{\textbf{Watermark Bit Acc Distribution}} & \multicolumn{3}{c|}{\textbf{WSR=1.0}} & \multirow{2}{*}{\textbf{Ave-PSNR}} \\
 & & 0-20 & 20-30 & 30-40 & 40-50 & 50-60 & 60-70 & 70-80 & 80-100 & \textbf{TPR} & \textbf{FID} & \textbf{VSR} &  \\
\midrule
\DiffShield & 50.13 & - & - & - & 564 & 186 & - & - & - & 13.64 & 268.37 & 0.0 & 31.49  \\
\DwdctWM          & 50.17 & - & -  & 65 & 195 & 323 & 148 & 19 & - & 79.86 & 99.08  & 1.0 & 49.35  \\
\ArtiGAN            & 54.03 & - & -  & -  & -    & 750 & -    & -  & - & 55.31 & 218.76 & 0.0 & 51.47  \\
\SSL                & 55.80 & - & - & 9  & 151 & 337 & 232 & 21 & - & 58.45 & 137.89 & 0.0 & 29.03 \\
\RivaGAN            & 62.00 & - & 8  & 52 & 166 & 399 & 125 & -  & - & 45.93 & 80.15  & 0.0 & 42.87  \\
\bottomrule
\end{tabular}
\end{table*}

\begin{table*}[!t]
\centering
\caption{Robustness evaluation under PreEvader attacks for generation tasks. One-bit methods are measured by TPR@5\%FPR, while multi-bit methods are measured by bit accuracy.}
\label{tab:generative_watermark_pre_robustness}
\small
\setlength{\tabcolsep}{2pt}
\renewcommand{\arraystretch}{0.85}
\rowcolors{2}{myblue!40!white}{white}
\begin{tabularx}{0.85\textwidth}{@{}>{\raggedright\arraybackslash}p{1.35cm}|*{13}{>{\centering\arraybackslash}X}@{}}
\toprule
\rowcolor{white}
\textbf{Method} & \textbf{No Att.} & \textbf{Crop} & \textbf{Rot.} & \textbf{Resize} & \textbf{Blur} & \textbf{JPEG} & \textbf{JPEG2K} & \textbf{WebP} & \textbf{Diff} & \textbf{DiffP} & \textbf{2xDiff} & \textbf{VAE} & \textbf{Adv.} \\
\midrule
\DiagnosisB & 100.00 & 93.67 & 96.67 & 28.00 & 1.33 & 76.00 & 88.67 & 98.67 & 26.00 & 26.67 & 17.33 & 98.67 & 36.00 \\
\EnTruth & 100.00 & 7.33 & 0.00 & 78.00 & 0.00 & 98.00 & 91.33 & 97.33 & 100.00 & 99.33 & 100.00 & 79.33 & 56.67 \\
\GaussWM & 93.69 & 42.00 & 54.67 & 26.67 & 2.67 & 72.67 & 30.00 & 34.67 & 98.00 & 93.33 & 91.33 & 80.00 & 24.00 \\
\DwtWM & 97.22 & 95.67 & 12.00 & 47.33 & 65.33 & 13.33 & 24.00 & 28.67 & 22.00 & 18.67 & 5.33 & 28.67 & 36.67 \\
\DiagnosisA & 96.00 & 79.33 & 98.67 & 50.00 & 0.00 & 88.67 & 75.33 & 97.33 & 47.33 & 30.00 & 13.33 & 92.00 & 38.33 \\
\AdvWM & 99.33 & 53.33 & 0.00 & 62.67 & 98.00 & 68.00 & 81.33 & 83.33 & 88.67 & 84.67 & 88.00 & 81.33 & 58.67 \\
\RIW & 100.00 & 100.00 & 100.00 & 46.00 & 40.67 & 88.00 & 72.00 & 37.33 & 80.66 & 81.34 & 78.00 & 100.00 & 32.67 \\
\GenWM & 98.67 & 96.33 & 98.67 & 20.67 & 66.67 & 90.00 & 91.33 & 98.67 & 99.33 & 98.00 & 97.33 & 98.67 & 6.67 \\
\Siren & 100.00 & 98.00 & 96.67 & 99.33 & 96.67 & 99.33 & 98.67 & 96.67 & 18.00 & 18.00 & 9.33 & 96.67 & 98.67 \\
\FTShield & 96.67 & 86.67 & 10.67 & 85.33 & 60.00 & 62.00 & 42.67 & 32.00 & 81.33 & 74.00 & 85.33 & 93.33 & 80.00 \\
\midrule
\DiffShield & 50.13 & 49.14 & 49.48 & 49.44 & 49.43 & 49.56 & 49.43 & 49.44 & 49.57 & 49.27 & 50.21 & 49.46 & 49.39 \\
\DwdctWM & 50.17 & 47.35 & 52.42 & 50.33 & 50.42 & 48.94 & 48.52 & 49.83 & 50.02 & 50.25 & 50.00 & 50.37 & 49.00 \\
\ArtiGAN & 54.03 & 52.78 & 50.37 & 52.11 & 51.33 & 53.02 & 51.77 & 51.25 & 51.37 & 51.85 & 51.67 & 51.51 & 52.21 \\
\SSL & 55.80 & 52.96 & 53.82 & 53.29 & 51.49 & 52.89 & 52.87 & 52.73 & 49.93 & 53.02 & 49.38 & 54.87 & 54.29 \\
\RivaGAN & 62.00 & 47.35 & 47.29 & 47.33 & 47.38 & 50.75 & 47.71 & 50.50 & 50.94 & 53.10 & 53.44 & 50.44 & 47.46 \\
\bottomrule
\end{tabularx}
\end{table*}

\begin{table*}[!t]
\centering
\caption{Robustness evaluation under PostEvader attacks for generation tasks.}
\label{tab:generative_watermark_post_robustness}
\small
\setlength{\tabcolsep}{2pt}
\renewcommand{\arraystretch}{0.85}
\rowcolors{2}{myblue!40!white}{white}
\begin{tabularx}{0.85\textwidth}{@{}>{\raggedright\arraybackslash}p{1.35cm}|*{13}{>{\centering\arraybackslash}X}@{}}
\toprule
\rowcolor{white}
\textbf{Method} & \textbf{No Att.} & \textbf{Crop} & \textbf{Rot.} & \textbf{Resize} & \textbf{Blur} & \textbf{JPEG} & \textbf{JPEG2K} & \textbf{WebP} & \textbf{Diff} & \textbf{DiffP} & \textbf{2xDiff} & \textbf{VAE} & \textbf{Adv.} \\
\midrule
\DiagnosisB & 100.00 & 88.67 & 98.00 & 70.00 & 46.67 & 88.00 & 99.33 & 99.33 & 65.33 & 89.33 & 49.33 & 18.67 & 70.00 \\
\EnTruth & 100.00 & 0.00 & 0.00 & 0.67 & 6.00 & 14.67 & 93.33 & 92.67 & 91.33 & 2.00 & 82.67 & 0.00 & 0.67 \\
\GaussWM & 93.69 & 12.00 & 5.33 & 6.00 & 6.00 & 12.00 & 23.33 & 15.33 & 64.00 & 23.33 & 53.33 & 6.67 & 7.33 \\
\DwtWM & 97.22 & 94.67 & 97.33 & 96.67 & 88.67 & 96.67 & 27.33 & 42.67 & 19.33 & 97.33 & 12.00 & 83.33 & 96.67 \\
\DiagnosisA & 96.00 & 94.67 & 94.67 & 56.00 & 1.33 & 78.33 & 82.67 & 86.67 & 60.67 & 40.00 & 39.33 & 0.00 & 47.33 \\
\AdvWM & 99.33 & 99.33 & 96.67 & 99.33 & 99.33 & 98.67 & 60.67 & 40.00 & 47.33 & 99.33 & 30.00 & 98.67 & 99.33 \\
\RIW & 100.00 & 96.67 & 100.00 & 74.67 & 80.67 & 74.00 & 82.00 & 44.00 & 93.33 & 83.33 & 96.67 & 97.33 & 75.33 \\
\GenWM & 98.67 & 96.67 & 98.00 & 97.67 & 94.67 & 96.00 & 35.33 & 100.00 & 100.00 & 98.67 & 100.00 & 97.33 & 98.67 \\
\Siren & 100.00 & 94.00 & 94.67 & 86.67 & 59.33 & 98.67 & 100.00 & 100.00 & 22.67 & 98.67 & 14.67 & 87.33 & 90.66 \\
\FTShield & 96.67 & 96.67 & 78.00 & 96.00 & 96.67 & 80.00 & 79.33 & 36.00 & 82.00 & 89.33 & 70.67 & 49.33 & 96.00 \\
\midrule
\DiffShield & 50.13 & 49.87 & 50.01 & 50.24 & 50.18 & 50.13 & 49.66 & 49.46 & 49.58 & 50.05 & 49.36 & 49.82 & 50.24 \\
\DwdctWM & 50.17 & 51.12 & 53.96 & 50.75 & 50.35 & 49.98 & 51.96 & 57.50 & 51.98 & 50.56 & 50.50 & 49.75 & 52.85 \\
\ArtiGAN & 54.03 & 50.20 & 49.98 & 50.40 & 50.04 & 50.13 & 54.39 & 54.11 & 52.38 & 50.24 & 51.49 & 50.56 & 50.12 \\
\SSL & 55.80 & 49.84 & 51.76 & 52.04 & 52.60 & 50.13 & 52.21 & 53.02 & 52.20 & 47.73 & 51.42 & 44.69 & 52.56 \\
\RivaGAN & 62.00 & 49.77 & 50.48 & 42.58 & 38.29 & 51.21 & 52.42 & 46.92 & 53.81 & 52.69 & 54.35 & 42.35 & 41.79 \\
\bottomrule
\end{tabularx}
\end{table*}

In generation tasks, watermark methods are divided into one-bit and multi-bit categories.
For multi-bit methods, Bit Acc measures the fraction of decoded bits matching the ground-truth message and serves as the detection score.
Given a target false-positive rate $\alpha$, we calibrate the decision threshold as $\tau_{\alpha}=Q_{1-\alpha}(\mathcal{S}^{-})$, where $\mathcal{S}^{-}$ denotes the null score distribution. 
The TPR is then computed as the fraction of watermarked samples whose Bit Acc exceeds $\tau_{\alpha}$, enabling multi-bit methods to follow the same $\mathrm{TPR}@\mathrm{FPR}$ protocol as one-bit methods; details are provided in Appendix~\ref{sec:appendix_multibit_calibration}.

\mypara{Overall Performance}
We evaluate 10 one-bit and 5 multi-bit watermark methods across three datasets and three model architectures.
We only present experimental results on the Pok\'{e}mon dataset obtained via LoRA fine-tuning of Stable Diffusion V1.4 here.
Detailed settings and more results are provided in~\autoref{sec:generation_other_results}.
Overall results are in \autoref{tab:combined_watermark_results_all}. Based on our taxonomy, most one-bit methods pass verification with 1.0 VSR at typical watermarked-sample ratios, while most model-free methods achieve higher TPR scores than model-based ones. Specifically, \EnTruth and \DiagnosisB achieve nearly 100.0, whereas model-based methods, especially \FTShield, reach only 73.20\% even at WSR=0.5.
This gap exists because many model-free methods were not originally designed to protect generative training datasets, prioritizing learnability over stealthiness.
PSNR also reflects this: several model-free methods introduce distortion, making embedded patterns visually noticeable and easier to identify or remove.
For example, \EnTruth achieves strong verification but yields only 9.59 PSNR.
In contrast, model-based methods are generally much stealthier, but some sacrifice verification performance for stealthiness.
For multi-bit methods, even under WSR=1.0, their bit accuracy remains low, so most verifications fail.
We suspect multi-bit watermarks are inherently fragile under fine-tuning, so their performance remains unsatisfactory regardless of the fine-tuning strategy.

\mypara{Impact of Watermarked-Sample Ratios}
We observe most one-bit methods exhibit very limited effectiveness at low WSRs, while model-behavior methods deteriorate slowest as the watermarked-sample ratio decreases.
For example, \EnTruth and \DiagnosisB maintain a TPR near 100.0\% even at extremely low WSRs.
We find that adding prompt triggers makes it easier for the model to memorize the watermark.
However, these triggers are easy to verify, causing these methods to suffer from poor stealthiness.
Designing triggers that improve effectiveness without compromising imperceptibility remains an open problem.

\mypara{Impact of Different Datasets and Models}
We finetune diverse datasets and models to assess watermark transferability.
As shown in \autoref{sec:generation_other_results}, performance varies sharply. 
Some methods reach TPR=100.0\% at very low watermarked-sample ratios on the Pok\'{e}mon dataset and Stable Diffusion V1.4, yet fall near zero elsewhere even at WSR=0.5.
For Kandinsky 2.2, most methods yield near-zero VSR at WSR=0.5.
\DwtWM stays below 5.0\%, \GenWM below 42.0\%, and \AdvWM and \RIW below 70.0\% across datasets.
This stems from signal embedding locations.
Low and mid-frequency watermarks survive in cartoon-like data with large flat regions and repetitive textures, enabling stable detection even at low WSRs.
Conversely, on detail-rich datasets like CelebA and WikiArt, models emphasize semantics and suppress watermark-like signals as noise, reducing TPR.

\begin{table*}[!t]
\centering
\caption{Multi-watermark results for different watermark methods in generation tasks. The structure is the same as~\autoref{tab:multi-watermark-classification}.}
\label{tab:onebit-cross-combined}
\small
\setlength{\tabcolsep}{4pt}
\renewcommand{\arraystretch}{0.85}
\begin{tabular}{l|cccccccccc}
\toprule[1pt]
 \textbf{Method} & \textbf{\DiagnosisB} & \textbf{\EnTruth} & \textbf{\GaussWM} & \textbf{\DwtWM} & \textbf{\DiagnosisA} & \textbf{\AdvWM} & \textbf{\RIW} & \textbf{\GenWM} & \textbf{\Siren} & \textbf{\FTShield} \\ \midrule
\DiagnosisB & / & 9.33 & 76.67 & 99.33 & / & 98.95 & 99.33 & 97.33 & 98.67 & 99.33 \\
\EnTruth & 0.00 & / & 89.33 & 35.33 & 0.00 & 46.66 & 98.00  & 4.66 & 59.33 & 26.66 \\
\GaussWM    & 53.33 & 6.67 & / & 58.67 & 20.00 & 63.33 & 57.33 & 84.67 & 58.00 & 39.33 \\
\DwtWM         & 30.67 & 100.00 & 52.67 & / & 22.00 & 46.00 & 38.00 & 80.67 & 78.00 & 30.67 \\
\DiagnosisA  & / & 8.00 & 74.67 & 98.00 & / & 99.12 & 90.00 & 68.67 & 98.67 & 36.00 \\
\AdvWM         & 2.00 & 52.67 & 39.33 & 59.33 & 12.67 & / & 25.33 & 19.33 & 80.67 & 18.00 \\
\RIW         & 99.33 & 98.00 & 100.00 & 100.00 & 100.00 & 100.00 & / & 100.00 & 99.33 & 99.33 \\
\GenWM         & 69.33 & 51.33 & 99.33 & 100.00 & 98.67 & 86.32 & 76.00 & / & 100.00 & 72.66 \\
\Siren       & 98.00 & 98.66 & 35.33 & 100.00 & 98.00 & 81.33 & 68.00 & 96.66 & / & 81.33 \\
\FTShield   & 28.67 & 36.67 & 86.67 & 78.00 & 18.67 & 99.33 & 44.67 & 92.67 & 90.00 & / \\
\midrule
\DiagnosisB  & /        & 100.00    & 96.00     & 99.33     & /         & 98.00     & 99.33     & 70.67     & 99.33     & 98.67 \\
\EnTruth     & 38.00    & /         & 49.33     & 60.00     & 40.66     & 54.00     & 21.30     & 0.00      & 58.67     & 26.67 \\
\GaussWM     & 43.33    & 77.33     & /         & 60.00     & 45.33     & 62.67     & 83.30     & 2.67      & 58.67     & 59.33 \\
\DwtWM       & 52.00    & 100.00    & 50.00     & /         & 53.33     & 76.67     & 17.33     & 4.00      & 82.67     & 60.67 \\
\DiagnosisA  & /        & 100.00    & 41.33     & 6.70      & /         & 5.33      & 90.00     & 97.33     & 99.33     & 21.33 \\
\AdvWM       & 1.62     & 98.00     & 17.33     & 75.33     & 0.21      & /         & 52.00     & 6.67      & 89.33     & 89.33 \\
\RIW         & 100.00   & 48.00     & 94.66     & 71.33     & 97.30     & 50.00     & /         & 98.66     & 100.00    & 92.67 \\
\GenWM       & 1.33     & 0.67      & 6.70      & 8.00      & 3.30      & 9.33      & 9.33      & /         & 2.00      & 9.33 \\
\Siren       & 41.33    & 98.67     & 37.33     & 100.00    & 42.67     & 88.67     & 100.00    & 100.00    & /         & 91.33 \\
\FTShield    & 33.33    & 50.00     & 55.33     & 30.67     & 16.00     & 37.33     & 40.00     & 85.33     & 94.00     & / \\
\bottomrule[1pt]
\end{tabular}
\end{table*}

\begin{table}[!t]
\centering
\small
\rowcolors{2}{myblue!40!white}{white}
\caption{Multi-user results for the same watermark scenario.
One-bit methods use TPR@5\%FPR as the metric (top), and multi-bit methods use bit accuracy (bottom).
Multi-user settings are the same as~\autoref{tab:multi_user_samewm_classification}.
}
\label{tab:multiuser_samewm_gen}
\renewcommand{\arraystretch}{0.85}
\begin{tabular}{l|cccccc}
\toprule
\textbf{Method}        & \textbf{1U } &\textbf{1F } & \textbf{2U}  &\textbf{2F}  & \textbf{5U}  &\textbf{5F}  \\
\midrule
\DiagnosisB   & 98.00 & 0.67  & 97.67& 2.67  & 98.53& 14.67 \\
\EnTruth       & 94.00 & 6.67  & 76.34& 53.33 & 59.60& 54.67 \\
\GaussWM         & 35.33 &9.10  & 48.34& 13.67 & 66.27&32.67 \\
\DwtWM           & 36.00 & 6.50  & 54.34& 26.00 & 41.73& 44.67 \\
\AdvWM           & 68.00 & 0.12  & 95.34& 42.67 & 90.40& 43.33 \\
\midrule
\DiffShield    & 49.45 & 50.25 & 49.82 & 50.20 & 50.06 & 50.00 \\
\DwdctWM      & 58.30 & 50.22 & 49.47 & 49.78 & 49.58 & 49.60 \\
\ArtiGAN      & 51.49 & 49.12 & 51.67 & 49.14 & 50.65 & 49.14 \\
\SSL          & 54.40 & 51.43 & 53.02 & 51.09 & 52.23 & 51.78 \\
\RivaGAN      & 52.09 & 49.36 & 50.35 & 49.93 & 51.08 & 47.39 \\

\bottomrule
\end{tabular}
\label{tab:same-method-different-users}
\end{table}

\subsection{Robustness in Adversarial Environment}
\label{sec:robustness-evaluation-generation-task}

We separately evaluate watermark methods for generation models under PreEvader and PostEvader attacks, with each experimental condition applying either one PreEvader attack or one PostEvader attack.
PreEvader transforms the released training images before model fine-tuning, whereas PostEvader transforms generated images before watermark verification.
We use the same PreEvader settings as in the classification experiments, except that the Gaussian-blur radius is 5 and the JPEG2000 compression ratio is 20:1 for generation tasks (due to the different image resolutions used in the classification and generation tasks).
Regen-DiffP uses the prompt ``Pok\'{e}mon'' for both PreEvader and PostEvader.
Complete configurations are provided in~\autoref{sec:evasion-attack-settings}.
For fair comparison, one-bit methods are evaluated by TPR at a fixed 5\% FPR, whereas multi-bit methods are evaluated by bit accuracy.

Under PreEvader, as shown in~\autoref{tab:generative_watermark_pre_robustness}, robustness is strongly method- and attack-dependent.
\EnTruth and \GaussWM remain robust to diffusion-based regeneration but are vulnerable to several geometric and blurring transformations.
Conversely, \Siren tolerates most conventional processing, VAE regeneration, and adversarial perturbations, but drops to 18.00\%, 18.00\%, and 9.33\% under Diff, DiffP, and 2xDiff.
\DiagnosisA, \DiagnosisB, and \DwtWM likewise exhibit pronounced weaknesses under specific transformations, showing that robustness to conventional processing does not necessarily transfer to regeneration or adversarial attacks.

Under PostEvader, as shown in~\autoref{tab:generative_watermark_post_robustness}, \GenWM provides the strongest overall robustness, retaining 94.67\%--100.00\% TPR under most attacks, although it drops to 35.33\% under JPEG2000.
\AdvWM, \RIW, and \FTShield also remain effective under many transformations, but show method-specific degradation under diffusion regeneration, WebP compression, or VAE regeneration.
The regeneration results further reveal substantial prompt-dependent differences.
For example, \DwtWM achieves only 19.33\% under Diff but 97.33\% under DiffP, whereas \EnTruth decreases from 91.33\% to 2.00\%.

Multi-bit methods are substantially less robust.
Even without attacks, their bit accuracy ranges from only 50.13\% to 62.00\%, and most PreEvader and PostEvader results remain close to the 50\% random-guessing level, with several attacks reducing accuracy below 50\%.
Therefore, the attacks do not merely erase otherwise reliable messages; rather, they further compound the limited baseline message-recovery capability of existing multi-bit methods.

\subsection{Multi-Watermark Scenario}

In \autoref{tab:onebit-cross-combined}, we perform experiments on all one-bit watermark methods, where the WSR is set to 0.2 in all datasets.
We observe a result similar to that of classification tasks: adding watermarks later does not necessarily lead to higher accuracy.
This depends on the principles and capabilities of the two watermark methods. 
In generation tasks, the generation model tends to prioritize those model-based watermark methods.
In the cases of \Siren, \RIW, and \GenWM, TPR is higher than that of most combined baselines, whether the watermark is added early or late in training, likely because these methods align with model characteristics, enabling early memorization and effective embedding.
We also find that combining two methods can yield higher detection than using either alone, which is counterintuitive.
A likely cause is pattern similarity or containment, for example, \AdvWM and \FTShield both inject hidden noise via surrogate models, and \EnTruth may fully preserve the watermark signal of \DwtWM, as \DwtWM maintains a TPR of 100.0\% regardless of the embedding order of the two methods, potentially leading to misleading auditing results.
We also conduct experiments on multi-bit methods, with detailed results provided in~\autoref{tab:multi_bit_multiwm_gen}.

\begin{table}[!t]
\centering
\small
\renewcommand{\arraystretch}{0.85}
\rowcolors{2}{myblue!40!white}{white}
\caption{Multi-user results for different watermarks.
Settings are consistent with prior one-bit experiments.}
\begin{tabular}{l|cccccc}
\toprule
\textbf{Method }      & \textbf{1U}    & \textbf{1F}   & \textbf{2U}    & \textbf{2F}   & \textbf{5U}    & \textbf{5F}   \\
\midrule
\DiagnosisB   & 98.00 & 0.67 & 89.00 & 1.20 & 87.33 & 0.67 \\
\EnTruth      & 94.00 & 6.00 & 71.17 & 13.31 & 86.67 & 10.66 \\
\GaussWM      & 35.33 & 9.10 & 31.67 & 6.84 & 19.33 & 6.00 \\
\DwtWM        & 36.00 & 6.50 & 35.60 & 5.33 & 23.33 & 5.33 \\
\AdvWM        & 68.00 & 0.12 & 64.33 & 5.83 & 38.00 & 1.67 \\
\midrule
\DiffShield    & 49.45 & 50.25 & 49.64 & 49.95 & 49.46 & 49.95 \\
\DwdctWM      & 58.30 & 50.22 & 52.68 & 50.07 & 49.81 & 50.27 \\
\ArtiGAN      & 51.49 & 49.12 & 51.15 & 49.11 & 51.06 & 49.21 \\
\SSL          & 54.40 & 51.43 & 54.17 & 51.47 & 53.39 & 51.88 \\
\RivaGAN      & 52.09 & 49.36 & 51.41 & 49.46 & 52.50 & 47.39 \\
\bottomrule
\end{tabular}
\label{tab:different-method-different-users}
\end{table}
\subsection{Multi-User Scenario}

\mypara{Same Watermark for Multi-user}
In \autoref{tab:same-method-different-users}, as the number of users increases, the TPR of most one-bit methods rises. 
This is because such model-behavior watermark methods apply the same watermark to different images while using distinct trigger words to differentiate users.
With more users, the overall watermark proportion in the training set increases, making it easier for the model to learn the watermark. 
In contrast, under the multi-bit setting, increasing the number of users reduces bit accuracy, and some methods even fail at low watermarked-sample ratios or with limited data due to interference between different bits.

\mypara{Different Watermarks for Multi-user}
In \autoref{tab:different-method-different-users}, for one-bit methods, increasing the user count generally decreases TPR. Multi-bit methods remain close to the random-guessing level and show limited attribution capability.
Detailed results are shown in \autoref{tab:multi-user-different-watermark-one}.

\section{Takeaways}

\mypara{There is no one-size-fits-all solution for dataset copyright auditing}
The applicability of watermark methods depends on the specific task.
Model-free methods excel in effectiveness, achieving high VSR and TPR@5\%FPR, but at the cost of reduced stealthiness and model performance.
Model-based methods, while more sophisticated, offer better stealthiness, but may still cause performance degradation.
Currently, no ``universal'' watermark method performs optimally in all scenarios, so the choices should depend on the specific requirements of the task, balancing stealthiness, performance, and verification complexity.

\mypara{Verifying watermarks at low WSRs continues to be a challenging problem}
Effective watermark verification under low watermarked-sample ratios is essential for users who do not control the entire dataset, such as creators of online content, to achieve effective copyright protection, as well as large dataset owners. 
Our experiments reveal that at very low watermarked-sample ratios, almost all current methods do not reliably verify both classification and generation tasks, posing a challenge to their practical deployment.

\mypara{Existing techniques inadequately support scenarios involving multiple watermarks and multiple users}
Our experiments in multi-watermark and multi-user scenarios highlight critical weaknesses in existing methods.
We find that some methods, particularly model-based ones, become unstable when other watermarks are embedded in the same dataset, as their embedding mechanisms are often sensitive to the data's underlying statistical properties.
Conversely, certain methods, especially those relying on model-free, exhibit a high false positive rate for ``independent'' users, making accurate ownership attribution impossible.
Therefore, it is crucial for future research to develop watermark methods that are both stable and reliable in these scenarios.

\mypara{Balancing robustness with stealthiness is difficult}
Most watermark methods cannot withstand complex and strong robustness attacks, and some methods with stronger robustness often introduce distortions, which directly compromise the watermark's stealthiness, making it more detectable.
Moreover, designs that overly emphasize robustness may improve resistance to specific attack types but can also introduce additional image distortions, affecting the overall performance of the model.
Therefore, enhancing robustness remains a key challenge in watermark design.

\mypara{Multi-bit watermarks are suffering from performance bottlenecks}
In generation tasks, particularly in fine-tuning settings, multi-bit watermark methods often suffer from low verification accuracy. 
Many traditional approaches fail when applied to generation models, primarily because of the high sensitivity of these models to conventional watermark mechanisms. 
Although one-bit methods provide more robust binary verification, they are inherently limited in payload capacity and thus insufficient for use cases that demand user-specific or high-resolution attribution.

\mypara{Non-destructive watermark method as a promising future direction}
As shown in \autoref{classification-result}, feature-destructive watermark methods fundamentally harm the model's primary task performance.
However, we observe certain methods maintain strong main-task accuracy at high watermarked-sample ratios.
We attribute this to their non-destructive strategies, which preserve original data semantics without introducing harmful perturbations. These approaches are critical for ensuring dataset utility and benign watermark impacts, making them a focus for future research.

\section{Conclusion}

This paper establishes a two-layer taxonomy for categorizing existing watermark methods for dataset copyright auditing. 
Based on the taxonomy, we propose \sysname, the first comprehensive benchmark for evaluating watermark methods for dataset copyright auditing, supported by a unified verification framework that uses TPR@5\%FPR for sample-level comparison and the verification success rate metric for dataset-level auditing.
Our study covers 25 representative watermark methods for both classification and generation tasks, evaluating them through extensive experiments on complex, real-world scenarios.

Our assessment based on \sysname offers critical insights into the current status of watermarks for dataset copyright auditing.
We find an inherent trade-off between model performance and watermark effectiveness, as many methods proved destructive to the model's primary task.
In addition, achieving a high verification success rate at a low watermarked-sample ratio remains a significant challenge that few methods have successfully addressed.
Most critically, our findings reveal that existing methods are often ill-equipped for complex, real-world scenarios, suffering performance degradation and high false positives in multi-user and multi-watermark situations.
These results highlight a significant theory-practice gap, underscoring the urgent need for more robust and reliable solutions.

\section*{Acknowledgments}

This project was supported by National Natural Science Foundation of China (No. 62402431, 62441618, U23A20296, U24A20237, U22A2029, and 62402379), Zhejiang Provincial Natural Science Foundation of China (No. LZ26F020002), National Science and Technology Major Project (No. 2026ZD0128300), Zhejiang Province's "Lingyan" R\&D Project (No. 2025C01195), China Postdoctoral Science Foundation (No. BX20250380), Key R\&D Program of Shaanxi Province (No. 2024PT-ZCK-6), Fundamental Research Funds for the Central Universities (No. 226-2026-00070), and ZJUCSE-Hikvision intelligence sensing and autonomous system joint laboratory.
Min Chen was supported in part by the project CiCS of the research programme Gravitation, which is (partly) financed by the Dutch Research Council (NWO) under Grant No. 024.006.037.


\bibliographystyle{ACM-Reference-Format}
\bibliography{refs}

\appendix

\section{Ethical Considerations}

Our research strictly adheres to ethical standards, prioritizing privacy protection during the development and testing processes.
All datasets used in this work are publicly available research benchmarks or public image-caption datasets, and we use them only under their original access terms.
Some benchmarks, such as CelebA, contain images of people.
For these data, we use only standard research splits or subsets, do not collect new human-subject data, do not attempt to identify any individual, and do not infer or evaluate sensitive attributes.
We also do not combine these data with external personal information.
Our study evaluates dataset watermarking as a mechanism for copyright auditing, which can benefit data owners by improving transparency and accountability in model training.
At the same time, watermarking techniques may be misused to create misleading ownership claims, hidden triggers, or adversarial claims about model provenance.
To mitigate these risks, we evaluate false positives, multi-user ambiguity, and adversarial robustness, and we discuss limitations rather than presenting watermark evidence as definitive proof in all settings.
The released artifacts are intended for research and reproducibility, and users should comply with the licenses and usage restrictions of the original datasets, models, and reproduced watermark methods.

\section{Generative AI Usage}

\mypara{Originality}
For paper writing, we used ChatGPT and Gemini only for grammar improvement, wording refinement, and text polishing.
All scientific ideas, claims, experimental designs, analyses, and conclusions in this paper were produced by the authors.
We carefully reviewed all LLM-assisted edits to ensure their accuracy, originality, and consistency with our experimental results.

\mypara{Transparency}
The use of generative AI did not influence the reported experimental results.
For code, we used Gemini and Codex to assist in generating and modifying implementation fragments for reproducing and adapting existing watermark methods in our open-source codebase.
All LLM-assisted code was manually reviewed by the authors, checked against the corresponding papers or released implementations, and tested within our benchmark pipeline.
No experimental result, citation, or quantitative claim was accepted solely from generative AI output.

\mypara{Responsibility}
We did not provide private datasets, sensitive data, or confidential information to generative AI tools.
As our usage was limited to text polishing and assistive code implementation, the environmental footprint is limited.
We take full responsibility for all content presented in this paper, including all text and code.

\section{Open Science}

We release the source code of \sysname at \url{https://github.com/ZJU-TrustAID/DWBench}.
The repository contains the benchmark pipeline, reproduced watermark method wrappers, evasion modules, model training and evaluation utilities, sample experiment scripts, dependency files, prompt files, and documentation for running representative classification and generation experiments.
The repository also includes configuration examples that support inspection of the evaluation workflow.

\mypara{Datasets}
The classification experiments use CIFAR-10, CIFAR-100, and TinyImageNet.
CIFAR-10 and CIFAR-100 are downloaded through torchvision and converted to ImageFolder format by our provided preparation script; their canonical source is \url{https://www.cs.toronto.edu/~kriz/cifar.html}.
TinyImageNet is used from the public TinyImageNet-200 release, available at \url{https://cs231n.stanford.edu/tiny-imagenet-200.zip}.
The generation experiments use public image-caption datasets or public subsets: Pok\'{e}mon BLIP captions from \url{https://huggingface.co/datasets/reach-vb/pokemon-blip-captions}, CelebA with LLaVA captions from \url{https://huggingface.co/datasets/irodkin/celeba_with_llava_captions}, and WikiArt images/captions from \url{https://huggingface.co/datasets/huggan/wikiart} and \url{https://huggingface.co/datasets/NikolaiSkripko/wikiart-captions}.
In the repository, we provide the dataset identifiers, preparation code, and configuration files needed to obtain and preprocess them from their public sources.

\mypara{Models and Reproduction}
For classification, ResNet-18, VGG-16, and ViT models are trained within our benchmark from the provided configurations.
For generation, the base model checkpoints are obtained from their public model hubs, including Stable Diffusion V1.4 (\url{https://huggingface.co/CompVis/stable-diffusion-v1-4}), Stable Diffusion XL (\url{https://huggingface.co/stabilityai/stable-diffusion-xl-base-1.0}), and Kandinsky 2.2 (\url{https://huggingface.co/kandinsky-community/kandinsky-2-2-decoder} and \url{https://huggingface.co/kandinsky-community/kandinsky-2-2-prior}).
We provide the code, hyperparameter settings, prompts, and evaluation scripts needed to regenerate the trained models and reported measurements.
Large trained checkpoints, generated images, and third-party pretrained components are not included in the repository because of storage cost and upstream license or redistribution constraints; they can be regenerated or obtained from the listed public sources.
To integrate the watermark methods into \sysname, we made necessary engineering adjustments to the replicated implementations so that they share a common data, training, and auditing interface.

\section{Detailed Experimental Setup}
\label{appendix_setup}

\subsection{Rationale for the Default FPR Setting}
\label{sec:appendix_fpr_rationale}

Existing dataset watermark methods produce heterogeneous detection scores, such as p-values, logit differences, decoder scores, and bit accuracy, which cannot be directly compared across methods.
We therefore report sample-level performance using TPR@FPR, which compares the detection power of different methods under the same false-positive constraint.

The appropriate FPR operating point depends on the practical application.
We report TPR@5\%FPR by default to enable consistent comparison across methods.
Users may recompute TPR at other FPR operating points from the recorded detection scores according to the required auditing strictness.
Lower FPRs impose stricter false-positive constraints, whereas higher FPRs prioritize detection sensitivity.

\subsection{Calibration for Multi-Bit Watermark Verification}
\label{sec:appendix_multibit_calibration}

For multi-bit watermark methods, the verifier recovers a binary message from the model output and compares it with the ground-truth watermark message. 
Let $m \in \{0,1\}^{L}$ denote the embedded watermark message and $\hat{m} \in \{0,1\}^{L}$ denote the decoded message. 
The bit accuracy is defined as
\begin{equation}
\mathrm{Bit Acc}(\hat{m}, m)
=
\frac{1}{L}
\sum_{j=1}^{L}
\mathbb{I}(\hat{m}_{j}=m_{j}).
\end{equation}

Under an idealized decoder with independent and unbiased bit predictions, a non-watermarked output would match each target bit with probability $1/2$, and thus the number of matched bits would follow a binomial distribution $\mathrm{Binomial}(L, 1/2)$.
In practice, however, decoder outputs can be biased or correlated, and different watermark methods may use different message lengths and decoding mechanisms.
Therefore, a theoretical random-match threshold may not accurately reflect the actual false-positive behavior of each method in practical settings.
We instead calibrate the verification threshold using an empirical null distribution.

Accordingly, DWBench treats $\mathrm{Bit Acc}$ as a continuous detection score and calibrates the decision threshold under a fixed false-positive-rate budget.

For each multi-bit method, we construct a null score distribution $\mathcal{S}^{-}$ from samples that should not match the target watermark, such as outputs from non-watermarked models, incorrect watermark messages, or independent users. 
Similarly, we obtain the positive score distribution $\mathcal{S}^{+}$ from models trained on the target watermarked data.

Given a target false-positive rate $\alpha$, the decision threshold is set as the $(1-\alpha)$-quantile of the null distribution:
\begin{equation}
\tau_{\alpha}
=
Q_{1-\alpha}(\mathcal{S}^{-}),
\end{equation}
where $Q_{1-\alpha}(\cdot)$ denotes the empirical quantile function. 
A sample is regarded as positive if its bit accuracy exceeds the calibrated threshold:
\begin{equation}
\mathcal{V}(s_i, \tau_{\alpha}) =
\begin{cases}
    1, & s_i \geq \tau_{\alpha}, \\
    0, & s_i < \tau_{\alpha}.
\end{cases}
\end{equation}

The corresponding true positive rate is computed as
\begin{equation}
\mathrm{TPR}@\alpha\mathrm{FPR}
=
\frac{1}{|\mathcal{S}^{+}|}
\sum_{s_i^{+} \in \mathcal{S}^{+}}
\mathbb{I}(s_i^{+} \geq \tau_{\alpha}).
\end{equation}
This formulation evaluates multi-bit watermark methods under the same fixed-FPR protocol as one-bit methods, while accounting for method-specific score distributions, message length, decoder bias, and bit dependence.

\subsection{Method Reproduction and Configuration Selection}

For method-specific parameters, we first use the settings reported in the original papers whenever available, followed by the released implementations and their defaults.
Methods without open-source implementations or with underspecified settings are reimplemented or adapted.
For these methods, we preserve their original mechanisms and reproduce the reported performance as closely as possible.
The exact method-specific configurations are listed below.

\mypara{Classification Methods}
The following configurations are used for the ten watermark methods evaluated in classification tasks. These methods cover both model-free and model-based watermark injection, while all of them rely on model-behavior verification, as summarized in~\autoref{tab:summary_methods}. The method-specific configurations are listed below.

\begin{itemize}[leftmargin=*]
    \item \textbf{\WANet.}
    We use deformation strength $s=0.5$, noise-grid size $k=4$, grid rescale 1, and noise rate 0.05.
    
    \item \textbf{\DVBW.}
    We instantiate this backdoor-to-watermark framework with a blended-type checkerboard trigger, using blending weight 0.05 and paired $t$-test verification with threshold $p<0.01$.
    
    \item \textbf{\AntiNeuron.}
    We use hue key 60, the default YIQ transform matrix, signature-space size 12, and recovery subset size 500.
    
    \item \textbf{\DataIsotope.}
    We use tagged class 0, a true mark and an external mark constructed from two fixed TinyImageNet images resized to the input resolution, true and external blending weights of 0.2, 5000 verification samples, and significance threshold $\lambda=0.01$.
    
    \item \textbf{\UBWP.}
    We use a bottom-right $3\times3$ red trigger, mask value 0.1, random relabeling for selected poisoned training samples, and paired $t$-test verification with decision threshold $p<0.01$.
    
    \item \textbf{\DYTMark.}
    We use a ResNet-18 surrogate checkpoint, PGD with 40 steps and $\epsilon=16/255$, a blended checkerboard trigger with weight 0.05, and a paired $t$-test with threshold $p<0.01$.
    
    \item \textbf{\LBConsist.}
    We use a four-corner $3\times3$ trigger, trigger amplitude $32/255$, perturbation budget $\epsilon=16/255$, and 100 PGD steps with PGD step size $1.5/255$.
    
    \item \textbf{\RadioActive.}
    We use randomly generated feature-space carriers, marking radius 10, AdamW marking with learning rate 0.1 for 100 epochs, image-space and feature-space regularization weights $5\times10^{-4}$ and $10^{-2}$, and no input normalization during marking.
    
    \item \textbf{\SleepAgent.}
    We use a ResNet-18 shadow model, perturbation budget $\epsilon=16/255$, poison-crafting batch size 256, 250 poison-crafting iterations, and retraining every 50 crafting iterations with a 40-epoch SGD retraining schedule.
    
    \item \textbf{\ImgDup.}
    We use an ImageNet-pretrained ResNet-18 feature extractor, perturbation budget $\epsilon=10$, 90 marking epochs, ImageNet normalization for the feature extractor, $p=0.05$, $K=16$, and regularization weights $10^{-2}$ for feature-space distance and $5\times10^{-4}$ for image-space distance.

\end{itemize}

\mypara{Generation Methods}
The following configurations are used for the fifteen watermark methods evaluated in generation tasks. Compared with classification methods, they span all four categories in our taxonomy, covering both model-free and model-based injection and both model-behavior and model-message verification. The method-specific configurations are listed below.

\begin{itemize}[leftmargin=*]
    \item \textbf{\DiagnosisB.}
    We use the spatial-warping variant of Diagnosis with WaNet-style geometric distortion and follow the settings in the original implementation, using a warping strength of $s=2.0$, a grid parameter of $k=128$, and the default random seed of 42.
    
    \item \textbf{\DiagnosisA.}
    We use the text-trigger variant of Diagnosis, with the rendered text marked IP-protected; When a prompt-side trigger is required, we prepend the token lgl' to the prompt for consistency, instead of the token `tq' used in the original implementation.
    
    \item \textbf{\EnTruth.}
    We use a fixed picture-frame template, where the target image is placed within a brown frame with a light-colored mat, and adopt the token `lgl' as the trigger, consistent with \DiagnosisA.
    
    \item \textbf{\GaussWM.}
    We use additive Gaussian watermarking by adding zero-mean Gaussian perturbations to the images, with the watermark strength controlled by a standard deviation of $\sigma=10.0$. All other parameters follow the default settings of the original implementation.
    
    \item \textbf{\DwtWM.}
    We use Haar wavelets, embed the watermark into the horizontal-detail coefficients, and set the embedding strength to $\alpha=0.05$. The watermarked image is then reconstructed through the inverse wavelet transform, with all other parameters following the default settings of the original implementation.
    
    \item \textbf{\AdvWM.}
    We use adversarial watermarking with perturbation budget $\epsilon=0.01$, step size 0.001, 3 PGD steps, $\ell_2$-constrained optimization, and a Stable Diffusion V1.4 watermark encoder.
    
    \item \textbf{\DiffShield.}
    We use the watermark blocks released with the original paper, set the block width to 4, and follow the prescribed fixed watermark-list file structure for watermark embedding.
    
    \item \textbf{\DwdctWM.}
    We use a 32-bit DWT-DCT watermark, set the embedding strength to $\alpha=5.0$, use Haar wavelets, and use a fixed bit sequence whenever it is provided.
    
    \item \textbf{\SSL.}
    We use the multi-bit \SSL instantiation provided in the released source code, with a batch size of 1, and perform the marking optimization using Adam for 100 epochs with learning rate 0.01 and regularization weights $\lambda_w=5\times 10^4$ and $\lambda_i=1$.
    
    \item \textbf{\RivaGAN.}
    We use the 32-bit \RivaGAN implementation with the pretrained encoder and decoder provided by \texttt{invisible-watermark}, and embed a fixed watermark bit sequence.
    
    \item \textbf{\GenWM.}
    We reproduce \GenWM following the original paper, with a latent dimension of 128 and an additive watermark scaling factor of 0.1.
    
    \item \textbf{\RIW.}
    Following the settings in the original code, we set the transparency factor to $\alpha=0.5$, the perturbation budget to $\epsilon=16/255$, the PGD step size to $2/255$, the number of optimization steps to 500, and the loss weights to $\lambda_2=\lambda_3=1.0$.
    
    \item \textbf{\Siren.}
    Following the original code, we use the released encoder-decoder watermarking instantiation, with hidden channel size 64 and audit threshold 10.0.
    
    \item \textbf{\FTShield.}
    We reproduce \FTShield using the officially released implementation, generating the protected images and training the detector according to the original code. For parameters not explicitly exposed by the benchmark wrapper, we retain the default settings of the released implementation.
    
    \item \textbf{\ArtiGAN.}
    We use the officially released fingerprint encoder and decoder, set the batch size to 64, and determine the fingerprint length automatically from the released checkpoint. Fingerprint embedding and extraction otherwise follow the default settings of the original implementation.

\end{itemize}

\subsection{Evasion Attack Settings}
\label{sec:evasion-attack-settings}

\mypara{PreEvader Attacks}
In the PreEvader setting, the model trainer preprocesses the released watermarked dataset before using it for training.
Each attack is applied after watermark injection and before the standard task-specific data augmentation.
The attack is applied independently to every image in the released dataset $D^*$, including both watermarked images and images that were not selected for watermark embedding.
The corresponding class labels or image captions remain unchanged.
We evaluate the following twelve PreEvader attacks.

\begin{itemize}[leftmargin=*]

\item \textbf{JPEG.}
We apply lossy JPEG compression with quality factor 30 while keeping the original image resolution.

\item \textbf{Crop.}
We retain the centered 80\% of the image along both spatial dimensions and then restore the original resolution using bilinear interpolation.
This gives $32\times32 \rightarrow 25\times25 \rightarrow 32\times32$ for CIFAR-10 and $512\times512 \rightarrow 409\times409 \rightarrow 512\times512$ for Pok\'{e}mon.

\item \textbf{Rotate.}
We rotate each image counterclockwise by $15^\circ$ without expanding the canvas.
Nearest-neighbor interpolation is used, and uncovered regions are filled with black pixels.

\item \textbf{Resize.}
We downsample each image to half of its original width and height and then restore the original resolution.
Bilinear interpolation is used in both steps, giving $32\times32 \rightarrow 16\times16 \rightarrow 32\times32$ for CIFAR-10 and $512\times512 \rightarrow 256\times256 \rightarrow 512\times512$ for Pok\'{e}mon.

\item \textbf{Gaussian Blur.}
We use a blur radius of 0.8 for CIFAR-10 and 5 for Pok\'{e}mon.
The larger value is used for Pok\'{e}mon because the same absolute radius produces a much weaker effect on $512\times512$ images.

\item \textbf{Regen-Diff.}
We regenerate each image with the Stable Diffusion v1.4 image-to-image pipeline using an empty prompt, strength 0.5, 30 inference steps, and guidance scale 7.5.
The input is resized to $512\times512$ before regeneration.
For CIFAR-10, the regenerated image is resized back to $32\times32$, whereas the $512\times512$ output is retained for Pok\'{e}mon.

\item \textbf{Regen-DiffP.}
We use the same configuration as Regen-Diff but provide a semantic prompt during regeneration.
For CIFAR-10, the prompt is the corresponding class name, such as ``airplane'' or ``bird'', and the result is resized back to $32\times32$.
For Pok\'{e}mon, we use the prompt ``Pok\'{e}mon'' and retain the $512\times512$ output.

\item \textbf{Regen-2xDiff.}
We apply Regen-Diff twice sequentially, using the output of the first regeneration as the input to the second.
Both regeneration stages use an empty prompt and the same strength, inference-step, guidance-scale, and resolution settings as Regen-Diff.

\item \textbf{Regen-VAE.}
We reconstruct each image with the VAE from Stable Diffusion v1.4.
The input is resized to $512\times512$ and normalized to $[-1,1]$ before encoding.
We sample from the encoder posterior and decode the sampled latent directly.
The reconstructed output is restored to the task-specific resolution.

\item \textbf{WebP.}
We apply lossy WebP compression with quality factor 30 while keeping the original image resolution.

\item \textbf{JPEG2000.}
We use irreversible JPEG2000 compression in rate mode.
The target compression ratio is 5:1 for CIFAR-10 and 20:1 for Pok\'{e}mon.
A stronger ratio is used for Pok\'{e}mon because 5:1 compression causes comparatively limited degradation at $512\times512$.

\item \textbf{Adv Embedding.}
We use PGD to maximize the mean squared distance between the unnormalized CLIP ViT-B/32 embeddings of the original and perturbed images.
The attack uses an $\ell_\infty$ budget of $8/255$, a step size of $2/255$, 10 steps, and a random start within the allowed perturbation range.
The perturbed image remains at the original task resolution.

\end{itemize}

\begin{figure*}[!t]
    \centering
    \includegraphics[width=1\linewidth]{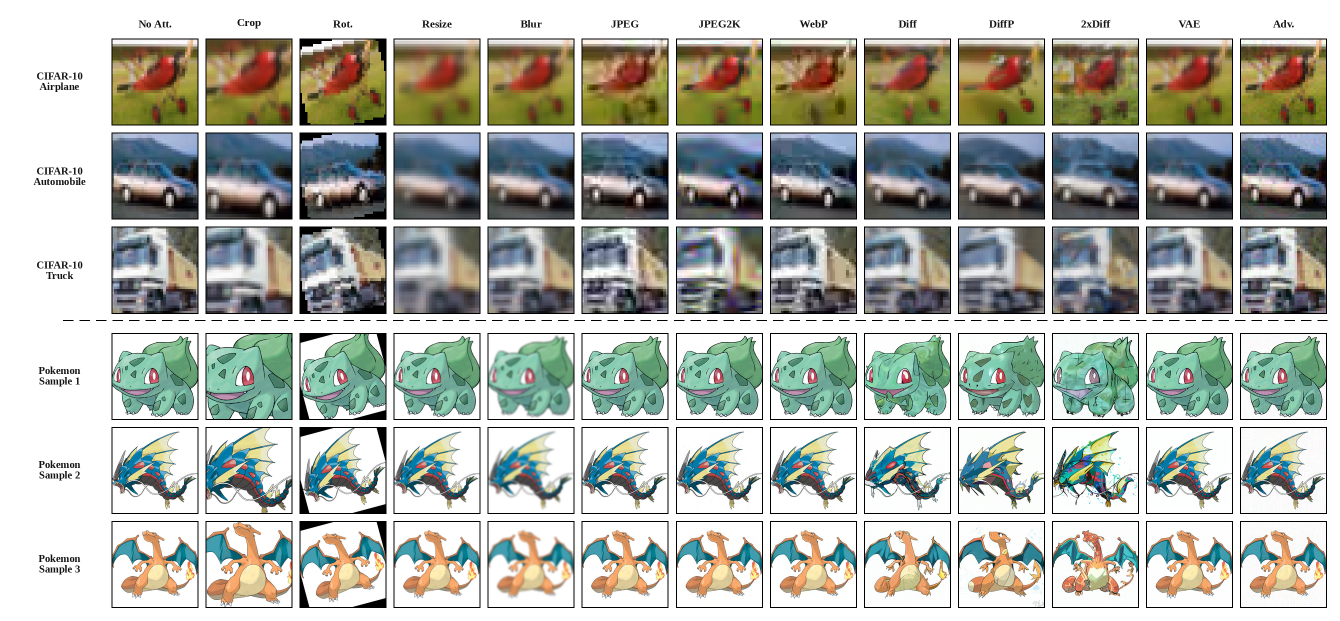}
    \caption{Visual examples of images before and after the PreEvader attacks evaluated in \sysname.}
    \label{fig:appendix_evader_visual_examples}
\end{figure*}

Representative visual effects of these PreEvader attacks are shown in~\autoref{fig:appendix_evader_visual_examples}.

\mypara{PostEvader Attacks}
PostEvader attacks modify the model output after inference and before watermark verification.
Each attack is applied independently, and the unmodified output is used as the no-attack baseline.

For generation tasks, we apply the same twelve attack types as in the PreEvader setting.
Regen-Diff reuses the prompt associated with the generated image.
The only difference is the attacked object: PreEvader modifies the released training dataset, whereas PostEvader modifies generated images before verification.

For classification tasks, PostEvader attacks operate on the classifier output.
Given a logit vector $z\in\mathbb{R}^{C}$, we evaluate the following three settings.

\begin{itemize}[leftmargin=*]

\item \textbf{Label-Only Output.}
We retain only the top-1 predicted class and remove all other information in the logit vector.
In our implementation, the output is represented by a one-hot vector whose nonzero entry corresponds to $\arg\max_c z_c$.

\item \textbf{Logit perturbation 0.1.}
We add independent zero-mean Gaussian noise directly to each component of the logit vector.
For each prediction, we first compute its logit range as $r(z)=\max_c z_c-\min_c z_c$ and set the noise standard deviation to $\sigma=0.1r(z)$.
Each perturbed logit is therefore given by $\widetilde{z}_c=z_c+\epsilon_c$, where $\epsilon_c$ is independently sampled from $\mathcal{N}(0,\sigma^2)$.
By performing this evasion, the full logit vector is still returned, but its confidence values and class margins are perturbed.

\item \textbf{Logit perturbation 0.5.}
This attack follows the same procedure but uses $\sigma=0.5r(z)$.
The standard deviation is therefore half of the logit range for the corresponding prediction.
The larger perturbation can substantially change confidence margins and may also alter the top-1 predicted class, while still returning a complete logit vector to the verifier.

\end{itemize}

\subsection{Setup for Classification Tasks}

\mypara{Datasets and Models}
As stated in \autoref{subsec:experimental-setup}, we conduct classification experiments on CIFAR-10, CIFAR-100, and TinyImageNet with ResNet-18, VGG-16, and ViT.
Our main experiments focus on ResNet-18 trained on CIFAR-10, while the remaining dataset-model combinations are used to evaluate whether the conclusions hold beyond this default setting.
The detailed results for these additional combinations are reported in
\autoref{tab:classification-cifar10}, \autoref{tab:classification-cifar100}, and \autoref{tab:classification-tinyimagenet}.
For CIFAR-10 and CIFAR-100, we use the original train/test splits and keep images at their native $32\times32$ resolution.
For TinyImageNet, we use the standard train/validation split and resize images to $64\times64$.
All classification datasets are first converted to tensors; watermark-specific transformations are then applied according to each method's injection procedure.

\mypara{Training Setting}
All benign and watermarked victim models under the same dataset-model setting are trained with the same recipe.
ResNet-18 and VGG-16 are trained for 100 epochs with batch size 512 using SGD with learning rate 0.1, momentum 0.9, weight decay $5\times10^{-4}$, and MultiStepLR milestones at 50, 70, and 90 epochs.
ViT is trained for 200 epochs with batch size 512 using AdamW with learning rate $5\times10^{-4}$, weight decay $10^{-3}$, cosine decay, and 10 warmup epochs.
Model utility is evaluated on the full test set.
For watermark auditing, TPR@5\%FPR is computed with 150 randomly sampled test samples per model, and VSR is calculated following the corresponding method-level verification protocol.

\mypara{Hyperparameter Setting}
Benchmark-level parameters are fixed across methods whenever possible, including the training recipe above, the WSR grid $\{10\%, 1\%, 0.1\%, 0.01\%\}$, target label 0, source label 1 when needed, and the unified auditing metrics.

\subsection{Setup for Generation Tasks}

\begin{table}[]
\centering
\setlength{\tabcolsep}{4pt}
\renewcommand{\arraystretch}{0.85}
\caption{Detailed per-user TPR@5\%FPR results for~\autoref{tab:multi_user_samewm_classification}, under 2 users and 5 users setting.}
\label{tab:expanded_tpr_fpr_eval}
\small
\begin{tabular}{l|cc|ccccc}
\toprule
\textbf{Method} & \textbf{2A} & \textbf{2B} & \textbf{5A} & \textbf{5B} & \textbf{5C} & \textbf{5D} & \textbf{5E} \\ 
\midrule
\WANet & 93.87 & 90.53 & 97.87 & 98.93 & 99.20 & 93.73 & 90.53 \\
\DVBW & 94.00 & 96.93 & 88.67 & 96.67 & 91.33 & 92.67 & 94.53 \\
\AntiNeuron & 7.15 & 11.25 & 8.52 & 12.40 & 8.42 & 10.70 & 13.33 \\
\DataIsotope & 17.47 & 13.07 & 15.20 & 13.87 & 13.60 & 13.07 & 11.87 \\
\UBWP & 87.60 & 88.93 & 88.13 & 90.80 & 90.00 & 83.20 & 85.46 \\
\DYTMark & 7.73 & 9.73 & 8.00 & 9.73 & 8.53 & 5.33 & 11.07 \\
\LBConsist & 100.00 & 99.87 & 40.53 & 19.20 & 99.60 & 98.53 & 100.00 \\
\RadioActive & 18.00 & 17.87 & 14.53 & 16.53 & 16.93 & 15.07 & 8.93 \\
\SleepAgent & 22.67 & 13.07 & 8.40 & 14.67 & 6.40 & 8.40 & 13.37 \\
\ImgDup & 7.20 & 6.13 & 7.87 & 5.47 & 7.20 & 6.40 & 5.87 \\ 
\bottomrule
\end{tabular}
\end{table}

\begin{table}[]
\centering
\setlength{\tabcolsep}{1.2pt}
\renewcommand{\arraystretch}{0.85}
\caption{Detailed per-user results when 2 users are leveraged for~\autoref{tab:multi_user_diffwm_classification}.}
\label{tab:cls_2_user_detailed}
\small
\begin{tabular}{l|ccccc}
\toprule
\textbf{Method} & \textbf{\WANet} & \textbf{\DVBW} & \textbf{\AntiNeuron} & \textbf{\DataIsotope} & \textbf{\UBWP} \\ 
\midrule
\WANet & \textbf{/} & 6.13 & 6.67 & 9.06 & 5.20 \\
\DVBW & 5.07 & / & 5.07 & 5.07 & 6.53 \\
\AntiNeuron & 4.83 & 4.87 & / & 3.87 & 4.92 \\
\DataIsotope & 4.60 & 5.00 & 4.50 & / & 4.48 \\
\UBWP & 20.13 & 26.27 & 16.13 & 20.27 & / \\ 
\midrule
\textbf{Method} & \textbf{\DYTMark} & \textbf{\LBConsist} & \textbf{\RadioActive} & \textbf{\SleepAgent} & \textbf{\ImgDup} \\
\midrule
\DYTMark & / & 3.73 & 7.06 & 4.67 & 7.20 \\
\LBConsist & 0.00 & / & 0.53 & 3.20 & 2.93 \\
\RadioActive & 1.60 & 1.06 & / & 0.67 & 2.67 \\
\SleepAgent & 15.73 & 2.93 & 5.73 & / & 8.93 \\
\ImgDup & 4.53 & 5.86 & 4.53 & 3.73 & / \\ 
\bottomrule
\end{tabular}
\end{table}

\begin{table}[]
\centering
\caption{Multi-watermark evaluation results of different multi-bit watermark methods for the generation task. The table is divided into two sections, following the same setup as the one-bit methods.}
\label{tab:multi_bit_multiwm_gen}
\renewcommand{\arraystretch}{0.85}
\small
\begin{tabular}{l|ccccc}
\toprule
\textbf{Method} & \textbf{\DiffShield} & \textbf{\DwdctWM} & \textbf{\ArtiGAN} & \textbf{\SSL} & \textbf{\RivaGAN} \\
\midrule
\DiffShield   &  /    & 49.46 & 49.51 & 49.44 & 49.54 \\
\DwdctWM      & 46.88 & /     & 50.17 & 43.39 & 52.04 \\
\ArtiGAN      & 50.13 & 51.35 & /     & 50.88 & 52.02 \\
\RivaGAN      & 51.00 & 49.04 & 49.29 & 54.31 & /     \\
\midrule
\DiffShield   & /     & 49.98 & 51.15 & 55.04 & 51.02 \\
\DwdctWM      & 49.45 & /     & 40.58 & 49.98 & 47.96 \\
\ArtiGAN      & 49.47 & 51.62 & /     & 51.14 & 51.73 \\
\RivaGAN      & 49.44 & 60.27 & 58.73 & 57.69 & /     \\
\bottomrule
\end{tabular}
\end{table}

\begin{table}[!t]
\centering
\caption{Detailed per-user results for~\autoref{tab:multiuser_samewm_gen} under the 2-user and 5-user settings.
One-bit methods are measured by TPR@5\%FPR, while multi-bit methods are measured by bit accuracy.
}
\label{tab:multi-user-same-watermark}
\renewcommand{\arraystretch}{0.85}
\setlength{\tabcolsep}{4pt}
\small
\begin{tabular}{@{}l|cc|ccccc@{}}
\toprule
\textbf{Method} & \textbf{2A} & \textbf{2B} & \textbf{5A} & \textbf{5B} & \textbf{5C} & \textbf{5D} & \textbf{5E} \\
\midrule
\GaussWM    & 44.67 & 52.00 & 66.00 & 75.33 & 58.67 & 68.67 & 62.67 \\
\DwtWM      & 64.00 & 44.67 & 44.67 & 48.67 & 35.33 & 44.67 & 35.33 \\
\AdvWM      & 96.00 & 94.67 & 90.67 & 94.67 & 86.67 & 91.33 & 88.67 \\
\EnTruth    & 76.00 & 76.67 & 59.33 & 59.33 & 60.67 & 62.67 & 56.00 \\
\DiagnosisB & 98.00 & 97.33 & 99.33 & 98.67 & 96.67 & 99.33 & 98.67 \\
\midrule
\DiffShield & 49.44 & 50.20 & 49.45 & 50.19 & 50.51 & 50.00 & 50.16 \\
\SSL        & 52.53 & 53.51 & 53.61 & 50.80 & 45.45 & 57.67 & 53.61 \\
\RivaGAN    & 52.02 & 48.67 & 52.76 & 48.42 & 52.56 & 52.35 & 49.31 \\
\DwdctWM    & 54.76 & 44.19 & 50.54 & 46.68 & 50.08 & 49.26 & 51.36 \\
\ArtiGAN    & 51.07 & 52.26 & 51.19 & 52.09 & 50.30 & 51.14 & 48.53 \\
\bottomrule
\end{tabular}
\end{table}

\begin{table}[]
\centering
\caption{Detailed per-user results for the 2-user setting in~\autoref{tab:different-method-different-users}.
One-bit methods are measured by TPR@5\%FPR, while multi-bit methods are measured by bit accuracy.}
\label{tab:multi-user-different-watermark-one}
\renewcommand{\arraystretch}{0.85}
\small
\begin{tabular}{@{}l|ccccc@{}}
\toprule
\textbf{Method} & \textbf{\GaussWM} & \textbf{\DwtWM} & \textbf{\AdvWM} & \textbf{\EnTruth} & \textbf{\DiagnosisB} \\
\midrule
\GaussWM    & /      & 35.33 & 39.33 & 18.00 & 34.00 \\
\DwtWM      & 32.00  & /     & 36.67 & 37.33 & 36.00 \\
\AdvWM      & 64.67  & 65.33 & /     & 69.33 & 58.00 \\
\EnTruth    & 59.33  & 61.33 & 72.67 & /     & 91.33 \\
\DiagnosisB & 86.67  & 96.67 & 96.00 & 76.67 & /     \\
\midrule
\textbf{Method} & \textbf{\DiffShield} & \textbf{\SSL} & \textbf{\RivaGAN} & \textbf{\DwdctWM} & \textbf{\ArtiGAN} \\
\midrule
\DiffShield & /      & 49.44 & 49.47 & 49.45 & 49.54 \\
\SSL        & 52.58  & /     & 54.56 & 54.79 & 54.73 \\
\RivaGAN    & 51.87  & 49.81 & /     & 52.03 & 51.94 \\
\DwdctWM    & 49.08  & 51.85 & 52.62 & /     & 57.17 \\
\ArtiGAN    & 50.95  & 51.25 & 51.09 & 51.34 & /     \\
\bottomrule
\end{tabular}
\end{table}

\mypara{Datasets} 
In our experiments, we use a dataset of 833 Pok\'{e}mon images and, additionally, two subsets of 1,000 images with textual descriptions drawn from CelebA and WikiArt.
All images are resized to 512 × 512, ensuring the evaluation covers both stylized generation and tasks requiring realism and semantic alignment.

\mypara{Training Setting}
We fine-tune Stable Diffusion V1.4, Stable Diffusion XL, and Kandinsky 2.2 via LoRA to simulate diverse scenarios and analyze watermark performance under varying distributions and strategies.
To balance quality and runtime, we train Stable Diffusion V1.4 for 100 epochs with LR=$1\times 10^{-4}$.
We train Stable Diffusion XL for 50 epochs, and for Kandinsky 2.2, the decoder for 50 and the prior for 30 epochs.
Each experiment generates images from 150 prompts to ensure statistical reliability.

\mypara{Hyperparameter Setting}
Whenever possible, benchmark-level parameters are fixed across methods, including the shared fine-tuning recipe described above, the watermarked-sample ratio grid $\{50\%, 20\%, 10\%, 2\%\}$, the unified auditing metrics, and a common generation setup with 50 denoising steps and guidance scale 7.5.
When compatible with the method, we uniformly adopt Stable Diffusion v1.5 with LoRA fine-tuning, batch size 4, 100 training epochs, learning rate $10^{-4}$, a constant learning-rate scheduler, zero warmup, and random seed 42.

\begin{figure*}
    \centering
    \includegraphics[width=1\linewidth]{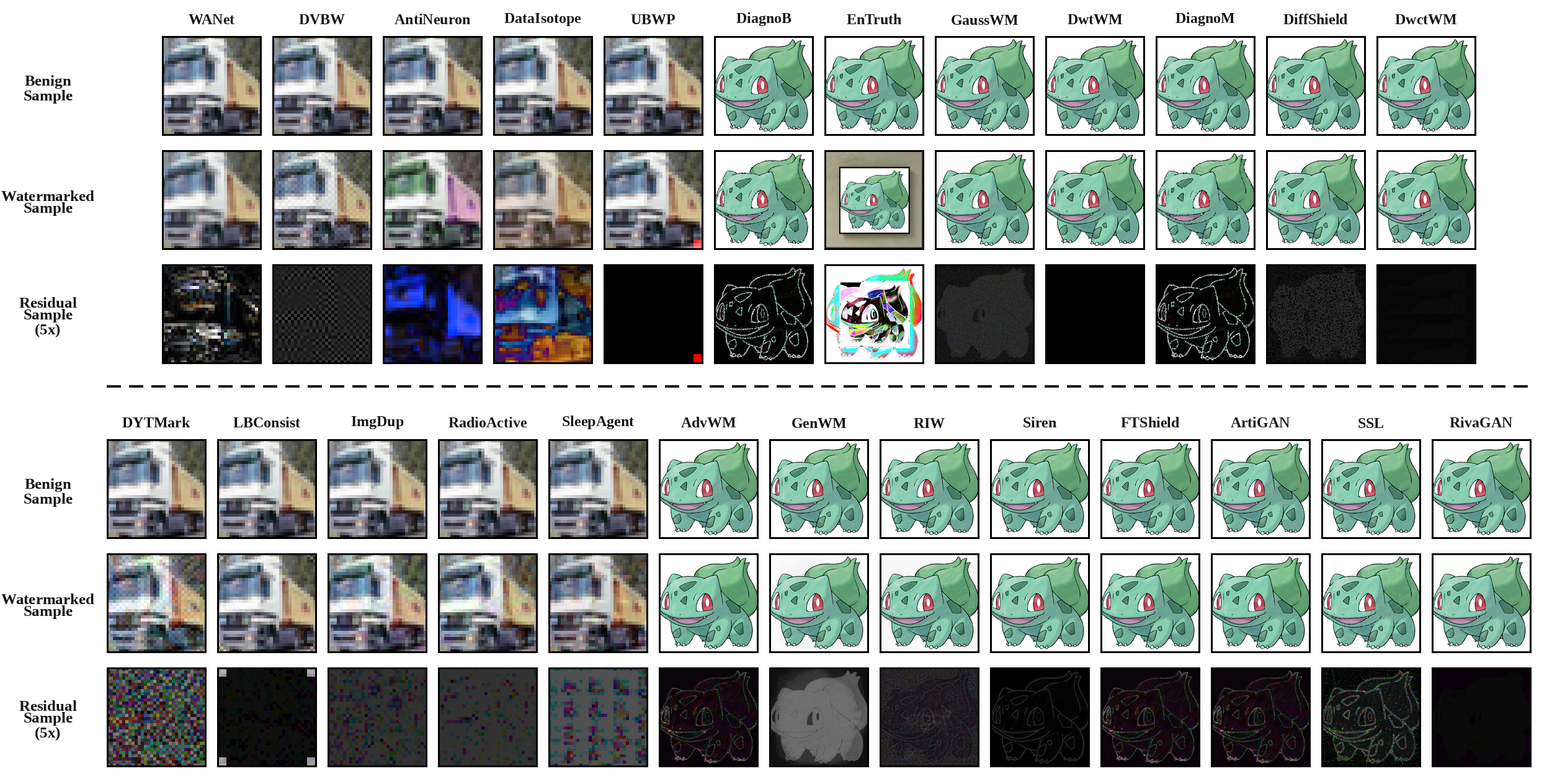}
    \caption{Visual examples of watermarked samples across the dataset watermark methods evaluated in \sysname. For each method, the three rows show the benign sample, the watermarked sample, and the residual sample amplified by 5$\times$.}
    \label{fig:appendix_visual_examples}
\end{figure*}

\section{Additional Experiment Results}

\autoref{fig:appendix_visual_examples} provides a general view of the watermarked samples used by the methods covered in our benchmark.
For each method, we show the benign sample, the corresponding watermarked sample, and a 5$\times$ amplified residual map.
This visualization complements the quantitative results by revealing whether a method relies on a localized trigger, a global blending pattern, a geometric transformation, or a model-dependent perturbation.

\subsection{Additional Results for Classification Tasks}
\label{sec:classification_other_results}

\autoref{tab:expanded_tpr_fpr_eval} presents detailed per-user TPR@5\%FPR results for the same-method multi-user setting.
\autoref{tab:cls_2_user_detailed} presents detailed per-user TPR@5\%FPR results for the two-user different-method setting.

\autoref{tab:classification-cifar10} reports the results for VGG16 and ViT on CIFAR-10, complementing the ResNet18 results presented in the main text.
\autoref{tab:classification-cifar100} and~\autoref{tab:classification-tinyimagenet} report the results for ResNet18, VGG16, and ViT on CIFAR-100 and TinyImageNet, respectively.
Together, these tables show the effectiveness of classification watermark methods across different datasets, model backbones, and watermarked-sample ratios.

\subsection{Additional Results for Generation Tasks}
\label{sec:generation_other_results}

In~\autoref{tab:multi_bit_multiwm_gen}, we conduct experiments on multi-bit methods in the multi-watermark scenario.
We find that the second watermark largely overwrites the first, unlike one-bit methods.
One-bit methods embed small amounts of information with simple structures, while multi-bit methods use larger image regions, making them more prone to interference.
We also observe that model-free watermark methods have nearly destroyed performance, regardless of whether the watermark is applied first or second. For example, in the \ArtiGAN-\DwdctWM~combination, the bit accuracy of \DwdctWM (model-free) is only 40.58\%.

\autoref{tab:multi-user-same-watermark} presents detailed data for the same watermark method across different users.
\autoref{tab:multi-user-different-watermark-one} displays detailed data for two users with different watermark methods.
We find that when using different triggers, the TPR@5\%FPR of different users also varies.
In the case of \GaussWM, we observe that, whether in a two-user or five-user scenario, the TPR for User B was consistently higher than that of User A.
This is because different triggers have varying abilities to activate the model.
Selecting an appropriate trigger is crucial because it directly influences the watermark's robustness and the model's ability to verify it effectively.

\autoref{tab:one-bit-pokemon},~\autoref{tab:one-bit-celeba}, and~\autoref{tab:one-bit-wikiart} show the watermark performance of Pok\'{e}mon, CelebA, and WikiArt under LoRA fine-tuning at different watermarked-sample ratios and models.
The results indicate that different datasets and models' fine-tuning methods can impact the TPR@5\%FPR of the watermark and the model's ability to verify it.
Therefore, when designing methods, it is crucial to consider the transferability of the watermark.

\begin{table*}[!t]
\centering
\caption{Effectiveness of watermark methods on the CIFAR-10 dataset across different classification backbones.}
\label{tab:classification-cifar10}
\begingroup
\small
\setlength{\tabcolsep}{3pt}
\renewcommand{\arraystretch}{0.95}
\begin{tabular}{@{}l|c|c|ccc|ccc|ccc|ccc@{}}
\toprule
\multirow{2}{*}{\textbf{Method}}
& \multirow{2}{*}{\textbf{Dataset}}
& \multirow{2}{*}{\textbf{Backbone}}
& \multicolumn{3}{c}{\textbf{WSR=0.1}}
& \multicolumn{3}{c}{\textbf{WSR=0.01}}
& \multicolumn{3}{c}{\textbf{WSR=0.001}}
& \multicolumn{3}{c}{\textbf{WSR=0.0001}} \\
\cmidrule(lr){4-6}\cmidrule(lr){7-9}\cmidrule(lr){10-12}\cmidrule(lr){13-15}
& & &
\textbf{TPR} & \textbf{BA Drop} & \textbf{VSR}
& \textbf{TPR} & \textbf{BA Drop} & \textbf{VSR}
& \textbf{TPR} & \textbf{BA Drop} & \textbf{VSR}
& \textbf{TPR} & \textbf{BA Drop} & \textbf{VSR} \\
\midrule
WANet & \multirow{20}{*}{CIFAR-10} & \multirow{10}{*}{VGG16} & 98.80 & 0.82 & 1.0 & 74.53 & 0.04 & 1.0 & 13.87 & -0.25 & 0.0 & 2.13 & -0.19 & 0.0 \\
DVBW & & & 91.87 & 0.35 & 1.0 & 91.73 & -0.22 & 1.0 & 91.63 & -0.28 & 1.0 & 80.13 & -0.16 & 1.0 \\
AntiNeuron & & & 10.76 & -0.12 & 1.0 & 10.70 & -0.48 & 1.0 & 11.16 & -0.34 & 0.8 & 20.00 & 0.15 & 0.0 \\
DataIsotope & & & 40.27 & 8.15 & 1.0 & 7.60 & -0.23 & 0.8 & 4.37 & -0.47 & 0.0 & 4.13 & -0.09 & 0.0 \\
UBWP & & & 93.60 & 0.40 & 1.0 & 77.20 & 0.11 & 1.0 & 4.00 & 0.00 & 0.0 & 5.07 & 0.21 & 0.0 \\
DYTMark & & & 99.60 & 8.90 & 1.0 & 87.60 & -0.17 & 1.0 & 88.80 & -0.06 & 1.0 & 61.89 & -0.23 & 1.0 \\
LBConsist & & & 95.12 & 8.75 & 1.0 & 28.37 & -0.07 & 0.8 & 6.64 & -0.04 & 0.2 & 2.75 & 0.07 & 0.0 \\
RadioActive & & & 100.00 & 8.60 & 1.0 & 30.51 & -0.12 & 1.0 & 11.25 & -0.11 & 0.0 & 5.33 & -0.30 & 0.0 \\
SleepAgent & & & 13.29 & 8.76 & 0.6 & 69.82 & -0.11 & 1.0 & 7.87 & -0.25 & 1.0 & 5.47 & -0.30 & 0.2 \\
ImgDup & & & 7.47 & -0.04 & 1.0 & 6.27 & 0.14 & 0.0 & 8.00 & -0.37 & 0.0 & 0.00 & 0.07 & 0.0 \\
\cmidrule(lr){3-15}
WANet & & \multirow{10}{*}{ViT} & 92.80 & 5.26 & 1.0 & 29.07 & 1.41 & 0.0 & 7.20 & 0.85 & 0.0 & 4.67 & 0.52 & 0.0 \\
DVBW & & & 95.68 & 3.43 & 1.0 & 61.84 & 0.99 & 0.6 & 6.00 & 0.72 & 0.0 & 5.41 & 0.77 & 0.0 \\
AntiNeuron & & & 25.52 & 0.52 & 1.0 & 24.25 & -0.08 & 1.0 & 38.12 & 0.33 & 1.0 & 39.55 & 0.46 & 0.8 \\
DataIsotope & & & 54.80 & 7.23 & 1.0 & 13.23 & 0.80 & 1.0 & 7.36 & 0.34 & 0.8 & 4.40 & 0.55 & 0.0 \\
UBWP & & & 76.27 & 1.87 & 1.0 & 61.73 & 0.83 & 1.0 & 2.80 & 0.49 & 0.0 & 3.33 & 0.26 & 0.0 \\
DYTMark & & & 100.00 & 10.10 & 1.0 & 5.23 & 0.84 & 0.0 & 5.73 & 0.72 & 0.0 & 5.73 & 0.33 & 0.0 \\
LBConsist & & & 93.17 & 7.96 & 1.0 & 5.33 & 0.72 & 0.0 & 5.57 & 0.72 & 0.0 & 4.83 & 0.86 & 0.0 \\
RadioActive & & & 96.21 & 7.75 & 1.0 & 10.61 & 0.76 & 0.0 & 6.40 & 2.09 & 0.0 & 5.47 & 2.01 & 0.0 \\
SleepAgent & & & 9.20 & 7.67 & 0.2 & 6.53 & 0.61 & 0.0 & 5.20 & 0.55 & 0.0 & 5.07 & 0.84 & 0.0 \\
ImgDup & & & 10.00 & 0.55 & 1.0 & 11.60 & 0.94 & 1.0 & 8.00 & 0.48 & 0.0 & 20.00 & 0.41 & 0.0 \\
\bottomrule
\end{tabular}
\endgroup
\end{table*}

\begin{table*}[!t]
\centering
\caption{Effectiveness of watermark methods on the CIFAR-100 dataset across different classification backbones. Missing entries indicate that the corresponding watermark method is not applicable under that experimental setting.}
\label{tab:classification-cifar100}
\begingroup
\small
\setlength{\tabcolsep}{3pt}
\renewcommand{\arraystretch}{0.95}
\begin{tabular}{@{}l|c|c|ccc|ccc|ccc|ccc@{}}
\toprule
\multirow{2}{*}{\textbf{Method}}
& \multirow{2}{*}{\textbf{Dataset}}
& \multirow{2}{*}{\textbf{Backbone}}
& \multicolumn{3}{c}{\textbf{WSR=0.1}}
& \multicolumn{3}{c}{\textbf{WSR=0.01}}
& \multicolumn{3}{c}{\textbf{WSR=0.001}}
& \multicolumn{3}{c}{\textbf{WSR=0.0001}} \\
\cmidrule(lr){4-6}\cmidrule(lr){7-9}\cmidrule(lr){10-12}\cmidrule(lr){13-15}
& & &
\textbf{TPR} & \textbf{BA Drop} & \textbf{VSR}
& \textbf{TPR} & \textbf{BA Drop} & \textbf{VSR}
& \textbf{TPR} & \textbf{BA Drop} & \textbf{VSR}
& \textbf{TPR} & \textbf{BA Drop} & \textbf{VSR} \\
\midrule
WANet & \multirow{30}{*}{CIFAR-100} & \multirow{10}{*}{ResNet18} & 99.73 & 2.22 & 1.0 & 74.80 & 1.09 & 1.0 & 23.47 & 0.32 & 0.0 & 8.27 & 0.69 & 0.0 \\
DVBW & & & 99.20 & 1.59 & 1.0 & 99.20 & 0.03 & 1.0 & 98.96 & 0.00 & 1.0 & 89.15 & 0.23 & 1.0 \\
AntiNeuron & & & 31.26 & -0.03 & 1.0 & 31.17 & -0.24 & 1.0 & 20.60 & -0.41 & 1.0 & 62.50 & -0.16 & 0.0 \\
DataIsotope & & & \multicolumn{3}{c|}{--} & 14.27 & 0.46 & 1.0 & 5.07 & -0.10 & 0.0 & 7.07 & -0.03 & 0.0 \\
UBWP & & & 67.73 & 3.32 & 1.0 & 63.60 & 0.67 & 1.0 & 2.67 & 0.10 & 0.0 & 2.13 & 0.22 & 0.0 \\
DYTMark & & & \multicolumn{3}{c|}{--} & 100.00 & 1.06 & 1.0 & 94.13 & -0.08 & 1.0 & 69.87 & 0.14 & 1.0 \\
LBConsist & & & \multicolumn{3}{c|}{--} & 100.00 & 0.83 & 1.0 & 29.09 & 0.01 & 0.2 & 4.77 & 0.20 & 0.0 \\
RadioActive & & & \multicolumn{3}{c|}{--} & 98.00 & 0.71 & 1.0 & 15.00 & -0.21 & 0.0 & 6.00 & 0.23 & 0.0 \\
SleepAgent & & & 24.67 & 6.25 & 0.0 & 86.60 & 0.66 & 1.0 & 46.13 & -0.34 & 0.0 & 26.47 & 0.05 & 0.0 \\
ImgDup & & & 7.20 & 0.77 & 1.0 & 11.47 & 0.35 & 1.0 & 16.00 & 0.18 & 0.0 & 80.00 & 0.15 & 0.0 \\
\cmidrule(lr){3-15}
WANet & & \multirow{10}{*}{VGG16} & 99.20 & 2.78 & 1.0 & 61.73 & 1.73 & 1.0 & 18.40 & 0.98 & 0.0 & 10.67 & 1.09 & 0.0 \\
DVBW & & & 99.63 & 2.98 & 1.0 & 99.39 & 0.37 & 1.0 & 99.36 & 0.63 & 1.0 & 75.52 & 0.28 & 0.6 \\
AntiNeuron & & & 28.89 & 0.27 & 1.0 & 22.72 & 0.19 & 1.0 & 18.75 & 0.05 & 1.0 & 30.00 & 0.63 & 0.0 \\
DataIsotope & & & \multicolumn{3}{c|}{--} & 12.00 & 0.51 & 1.0 & 5.07 & 0.26 & 0.0 & 4.13 & 0.42 & 0.0 \\
UBWP & & & 62.80 & 2.79 & 1.0 & 59.73 & 0.51 & 1.0 & 3.33 & 0.49 & 0.0 & 4.00 & 0.33 & 0.0 \\
DYTMark & & & \multicolumn{3}{c|}{--} & 99.33 & 1.32 & 1.0 & 95.07 & 0.30 & 1.0 & 31.09 & 0.31 & 1.0 \\
LBConsist & & & \multicolumn{3}{c|}{--} & 100.00 & 1.23 & 1.0 & 41.73 & 0.22 & 0.6 & 12.99 & 0.53 & 0.0 \\
RadioActive & & & \multicolumn{3}{c|}{--} & 93.00 & 0.84 & 1.0 & 11.00 & 0.33 & 0.0 & 4.00 & 0.43 & 0.0 \\
SleepAgent & & & 61.13 & 3.90 & 0.6 & 54.87 & 1.08 & 1.0 & 28.40 & 0.44 & 0.0 & 12.93 & 0.40 & 0.0 \\
ImgDup & & & 7.47 & 1.28 & 1.0 & 9.20 & 0.43 & 1.0 & 4.00 & 0.08 & 0.0 & 0.00 & 0.56 & 0.0 \\
\cmidrule(lr){3-15}
WANet & & \multirow{10}{*}{ViT} & 77.60 & 4.35 & 1.0 & 24.53 & 0.17 & 0.0 & 8.13 & -0.50 & 0.0 & 5.87 & -0.59 & 0.0 \\
DVBW & & & 93.20 & 5.33 & 0.8 & 25.60 & -0.21 & 0.0 & 10.13 & -0.88 & 0.0 & 7.52 & -0.87 & 0.0 \\
AntiNeuron & & & 71.38 & -0.99 & 1.0 & 60.60 & -1.19 & 1.0 & 78.29 & -1.23 & 1.0 & 35.00 & -0.82 & 1.0 \\
DataIsotope & & & \multicolumn{3}{c|}{--} & 16.13 & -0.82 & 1.0 & 6.67 & -1.40 & 0.0 & 8.67 & 0.07 & 0.0 \\
UBWP & & & 37.07 & 2.24 & 1.0 & 24.27 & -1.16 & 0.0 & 6.00 & -2.14 & 0.0 & 4.27 & -1.26 & 0.0 \\
DYTMark & & & \multicolumn{3}{c|}{--} & 4.93 & -0.78 & 0.6 & 8.00 & -0.79 & 0.0 & 6.27 & -0.93 & 0.0 \\
LBConsist & & & \multicolumn{3}{c|}{--} & 5.55 & -0.59 & 0.0 & 3.87 & -1.02 & 0.0 & 4.67 & -1.16 & 0.0 \\
RadioActive & & & \multicolumn{3}{c|}{--} & 29.00 & -0.53 & 0.0 & 8.00 & -0.77 & 0.0 & 5.00 & -0.44 & 0.0 \\
SleepAgent & & & 53.33 & 1.23 & 1.0 & 10.73 & -0.41 & 0.0 & 5.67 & -0.60 & 0.0 & 4.00 & -1.29 & 0.0 \\
ImgDup & & & 15.20 & -0.15 & 1.0 & 16.67 & -0.67 & 1.0 & 14.00 & -1.48 & 0.0 & 40.00 & -1.07 & 0.0 \\
\bottomrule
\end{tabular}
\endgroup
\end{table*}

\begin{table*}[!t]
\centering
\caption{Effectiveness of watermark methods on the TinyImageNet dataset across different classification backbones. Missing entries indicate that the corresponding watermark method is not applicable under that experimental setting.}
\label{tab:classification-tinyimagenet}
\begingroup
\small
\setlength{\tabcolsep}{3pt}
\renewcommand{\arraystretch}{0.95}
\begin{tabular}{@{}l|c|c|ccc|ccc|ccc|ccc@{}}
\toprule
\multirow{2}{*}{\textbf{Method}}
& \multirow{2}{*}{\textbf{Dataset}}
& \multirow{2}{*}{\textbf{Backbone}}
& \multicolumn{3}{c}{\textbf{WSR=0.1}}
& \multicolumn{3}{c}{\textbf{WSR=0.01}}
& \multicolumn{3}{c}{\textbf{WSR=0.001}}
& \multicolumn{3}{c}{\textbf{WSR=0.0001}} \\
\cmidrule(lr){4-6}\cmidrule(lr){7-9}\cmidrule(lr){10-12}\cmidrule(lr){13-15}
& & &
\textbf{TPR} & \textbf{BA Drop} & \textbf{VSR}
& \textbf{TPR} & \textbf{BA Drop} & \textbf{VSR}
& \textbf{TPR} & \textbf{BA Drop} & \textbf{VSR}
& \textbf{TPR} & \textbf{BA Drop} & \textbf{VSR} \\
\midrule
WANet & \multirow{30}{*}{TinyImageNet} & \multirow{10}{*}{ResNet18} & 100.00 & 1.76 & 1.0 & 92.13 & 0.26 & 1.0 & 21.07 & 0.34 & 0.6 & 8.13 & 0.10 & 0.0 \\
DVBW & & & 100.00 & 1.87 & 1.0 & 99.89 & 0.15 & 1.0 & 97.63 & -0.05 & 1.0 & 10.93 & -0.07 & 0.0 \\
AntiNeuron & & & 37.79 & 0.71 & 1.0 & 34.55 & 0.61 & 1.0 & 33.50 & 0.40 & 1.0 & 45.00 & 0.33 & 0.0 \\
DataIsotope & & & \multicolumn{3}{c|}{--} & \multicolumn{3}{c|}{--} & 7.87 & -0.22 & 0.0 & 4.80 & -0.49 & 0.0 \\
UBWP & & & 75.60 & 4.15 & 1.0 & 8.80 & 1.09 & 1.0 & 3.73 & 0.45 & 0.0 & 4.93 & 0.57 & 0.0 \\
DYTMark & & & \multicolumn{3}{c|}{--} & \multicolumn{3}{c|}{--} & 58.40 & 0.03 & 1.0 & 6.00 & 0.39 & 0.0 \\
LBConsist & & & \multicolumn{3}{c|}{--} & \multicolumn{3}{c|}{--} & 4.03 & 0.11 & 0.0 & 4.72 & 0.34 & 0.0 \\
RadioActive & & & \multicolumn{3}{c|}{--} & \multicolumn{3}{c|}{--} & 16.00 & 0.45 & 0.0 & 14.00 & -0.25 & 0.0 \\
SleepAgent & & & 10.13 & 8.35 & 0.0 & 6.80 & 0.20 & 0.0 & 4.53 & 0.11 & 0.0 & 7.33 & -0.27 & 0.0 \\
ImgDup & & & 10.13 & 0.13 & 1.0 & 8.27 & 0.43 & 1.0 & 5.00 & 0.05 & 0.0 & 0.00 & 0.40 & 0.0 \\
\cmidrule(lr){3-15}
WANet & & \multirow{10}{*}{VGG16} & 100.00 & 1.92 & 1.0 & 90.13 & 0.84 & 1.0 & 19.47 & 1.02 & 0.0 & 6.40 & 0.37 & 0.0 \\
DVBW & & & 100.00 & 9.35 & 1.0 & 100.00 & 4.79 & 1.0 & 81.92 & 4.35 & 0.0 & 23.68 & 4.42 & 0.0 \\
AntiNeuron & & & 35.78 & 0.68 & 1.0 & 41.42 & 0.68 & 1.0 & 33.25 & 0.96 & 1.0 & 50.00 & 0.95 & 0.0 \\
DataIsotope & & & \multicolumn{3}{c|}{--} & \multicolumn{3}{c|}{--} & 7.60 & 0.82 & 0.0 & 5.33 & 0.59 & 0.0 \\
UBWP & & & 68.00 & 3.12 & 1.0 & 68.13 & 1.07 & 1.0 & 5.60 & 0.50 & 0.0 & 6.27 & 1.14 & 0.0 \\
DYTMark & & & \multicolumn{3}{c|}{--} & \multicolumn{3}{c|}{--} & 24.67 & 0.54 & 1.0 & 7.07 & 0.66 & 0.0 \\
LBConsist & & & \multicolumn{3}{c|}{--} & \multicolumn{3}{c|}{--} & 6.67 & 0.61 & 0.0 & 5.47 & 0.26 & 0.0 \\
RadioActive & & & \multicolumn{3}{c|}{--} & \multicolumn{3}{c|}{--} & 8.00 & 0.82 & 0.0 & 10.00 & 0.11 & 0.0 \\
SleepAgent & & & 8.80 & 6.26 & 0.0 & 3.87 & 0.85 & 0.0 & 6.00 & 0.52 & 0.0 & 4.40 & 0.94 & 0.0 \\
ImgDup & & & 7.60 & 0.49 & 1.0 & 13.33 & 0.77 & 1.0 & 14.00 & 0.63 & 0.0 & 70.00 & 0.45 & 0.0 \\
\cmidrule(lr){3-15}
WANet & & \multirow{10}{*}{ViT} & 98.67 & 1.18 & 1.0 & 50.53 & -0.01 & 1.0 & 13.87 & 0.10 & 0.0 & 4.80 & 0.13 & 0.0 \\
DVBW & & & 100.00 & 3.89 & 1.0 & 99.95 & -0.58 & 1.0 & 25.09 & -0.66 & 0.0 & 5.89 & -0.76 & 0.0 \\
AntiNeuron & & & 73.82 & 0.22 & 1.0 & 80.47 & -0.67 & 1.0 & 81.23 & 0.07 & 1.0 & 65.00 & -0.66 & 1.0 \\
DataIsotope & & & \multicolumn{3}{c|}{--} & \multicolumn{3}{c|}{--} & 7.60 & -0.79 & 0.0 & 6.80 & 0.02 & 0.0 \\
UBWP & & & 38.27 & 1.14 & 1.0 & 5.60 & -0.39 & 0.0 & 4.27 & -0.09 & 0.0 & 4.40 & -0.03 & 0.0 \\
DYTMark & & & \multicolumn{3}{c|}{--} & \multicolumn{3}{c|}{--} & 5.60 & -0.84 & 1.0 & 5.47 & -1.83 & 0.0 \\
LBConsist & & & \multicolumn{3}{c|}{--} & \multicolumn{3}{c|}{--} & 6.67 & -0.88 & 0.0 & 7.07 & -0.98 & 0.0 \\
RadioActive & & & \multicolumn{3}{c|}{--} & \multicolumn{3}{c|}{--} & 12.00 & -0.42 & 0.0 & 6.00 & -1.86 & 0.0 \\
SleepAgent & & & 41.07 & 2.97 & 1.0 & 9.07 & -0.80 & 0.0 & 4.13 & -0.21 & 0.0 & 5.60 & -0.21 & 0.0 \\
ImgDup & & & 8.80 & -0.17 & 1.0 & 7.33 & -1.10 & 1.0 & 7.00 & -0.81 & 0.0 & 20.00 & -1.07 & 0.0 \\
\bottomrule
\end{tabular}
\endgroup
\end{table*}

\begin{table*}[!t]
\centering
\caption{Effectiveness of one-bit and multi-bit watermark methods on the Pok\'{e}mon dataset across different diffusion backbones.}
\label{tab:one-bit-pokemon}
\small
\renewcommand{\arraystretch}{0.95}
\begin{tabular}{l|c|c|ccc|ccc|ccc|ccc}
\toprule
\multirow{2}{*}{\textbf{Method}}
& \multirow{2}{*}{\textbf{Dataset}}
& \multirow{2}{*}{\textbf{Backbone}}
& \multicolumn{3}{c}{\textbf{WSR=0.5}}
& \multicolumn{3}{c}{\textbf{WSR=0.2}}
& \multicolumn{3}{c}{\textbf{WSR=0.1}}
& \multicolumn{3}{c}{\textbf{WSR=0.02}} \\
\cmidrule(lr){4-6}\cmidrule(lr){7-9}\cmidrule(lr){10-12}\cmidrule(lr){13-15}
& & &
\textbf{TPR} & \textbf{FID} & \textbf{VSR}
& \textbf{TPR} & \textbf{FID} & \textbf{VSR}
& \textbf{TPR} & \textbf{FID} & \textbf{VSR}
& \textbf{TPR} & \textbf{FID} & \textbf{VSR} \\
\midrule
\FTShield & \multirow{20}{*}{Pok\'{e}mon} & \multirow{10}{*}{SDXL} & 2.20 & 112.88 & 0.0 & 1.32 & 110.47 & 0.0 & 1.13 & 97.75 & 0.0 & 0.63 & 103.27 & 0.0 \\
\RIW &  &  & 73.89 & 82.91 & 0.2 & 64.35 & 75.25 & 0.0 & 37.64 & 80.62 & 0.0 & 18.65 & 99.04 & 0.0 \\
\GaussWM &  &  & 20.70 & 83.21 & 0.0 & 4.74 & 100.51 & 0.0 & 2.02 & 100.51 & 0.0 & 0.00 & 102.17 & 0.0 \\
\DwtWM &  &  & 25.23 & 100.86 & 0.0 & 25.53 & 115.06 & 0.0 & 19.34 & 116.52 & 0.0 & 18.23 & 109.72 & 0.0 \\
\AdvWM &  &  & 13.33 & 113.58 & 0.0 & 16.25 & 128.18 & 0.0 & 4.68 & 124.53 & 0.0 & 2.56 & 103.23 & 0.0 \\
\Siren &  &  & 68.70 & 107.43 & 0.0 & 42.63 & 113.72 & 0.0 & 32.66 & 125.89 & 0.0 & 20.53 & 102.42 & 0.0 \\
\EnTruth &  &  & 92.13 & 383.11 & 1.0 & 87.34 & 394.46 & 1.0 & 0.00 & 133.94 & 0.0 & 0.00 & 102.89 & 0.0 \\
\GenWM &  &  & 100.00 & 348.83 & 1.0 & 45.53 & 203.42 & 0.0 & 37.23 & 166.35 & 0.0 & 0.00 & 93.66 & 0.0 \\
\DiagnosisA &  &  & 4.42 & 118.51 & 0.0 & 0.62 & 105.28 & 0.0 & 0.66 & 101.71 & 0.0 & 0.00 & 102.26 & 0.0 \\
\DiagnosisB &  &  & 95.34 & 242.39 & 1.0 & 80.62 & 248.93 & 1.0 & 72.02 & 237.59 & 0.8 & 2.40 & 106.10 & 0.0 \\
\cmidrule(lr){3-15}
\FTShield &  & \multirow{10}{*}{Kandinsky2.2} & 11.30 & 68.86 & 0.0 & 2.21 & 76.50 & 0.0 & 3.12 & 78.17 & 0.0 & 1.53 & 77.84 & 0.0 \\
\RIW &  &  & 2.74 & 62.79 & 0.0 & 0.00 & 73.88 & 0.0 & 0.00 & 75.27 & 0.0 & 0.00 & 77.62 & 0.0 \\
\GaussWM &  &  & 95.32 & 74.93 & 1.0 & 3.65 & 63.90 & 0.0 & 0.60 & 70.19 & 0.0 & 0.00 & 70.45 & 0.0 \\
\DwtWM &  &  & 0.00 & 80.13 & 0.0 & 0.00 & 72.83 & 0.0 & 0.00 & 71.43 & 0.0 & 0.00 & 69.94 & 0.0 \\
\AdvWM &  &  & 66.73 & 78.02 & 0.0 & 12.13 & 72.18 & 0.0 & 0.58 & 73.63 & 0.0 & 0.00 & 71.14 & 0.0 \\
\Siren &  &  & 21.35 & 79.19 & 0.0 & 0.62 & 71.08 & 0.0 & 1.23 & 72.04 & 0.0 & 0.00 & 70.42 & 0.0 \\
\EnTruth &  &  & 100.00 & 448.27 & 1.0 & 98.64 & 417.94 & 1.0 & 52.60 & 261.39 & 0.0 & 0.00 & 70.40 & 0.0 \\
\GenWM &  &  & 41.32 & 153.64 & 0.0 & 2.43 & 80.21 & 0.0 & 0.00 & 74.12 & 0.0 & 0.00 & 76.77 & 0.0 \\
\DiagnosisA &  &  & 12.53 & 99.48 & 0.0 & 2.34 & 78.27 & 0.0 & 0.75 & 78.99 & 0.0 & 0.53 & 77.56 & 0.0 \\
\DiagnosisB &  &  & 100.00 & 253.66 & 1.0 & 44.45 & 163.00 & 0.0 & 40.81 & 87.99 & 0.0 & 1.75 & 72.21 & 0.0 \\
\midrule
\multirow{2}{*}{\textbf{Method}}
& \multirow{2}{*}{\textbf{Dataset}}
& \multirow{2}{*}{\textbf{Backbone}}
& \multirow{2}{*}{\textbf{Bit ACC}}
& \multicolumn{8}{c}{\textbf{Watermark Bit Acc Distribution}}
& \multirow{2}{*}{\textbf{TPR}}
& \multirow{2}{*}{\textbf{FID}}
& \multirow{2}{*}{\textbf{VSR}} \\
\cmidrule(lr){5-12}
& & & & \textbf{0--0.2} & \textbf{0.2--0.3} & \textbf{0.3--0.4} & \textbf{0.4--0.5}
& \textbf{0.5--0.6} & \textbf{0.6--0.7} & \textbf{0.7--0.8} & \textbf{0.8--1.0}
& & & \\
\midrule
\DiffShield & \multirow{10}{*}{Pok\'{e}mon} & \multirow{5}{*}{SDXL}   &49.43  & -- & -- & -- & 750 & -- & -- & -- & -- & 0.00 & 102.42 & 0.0 \\
\DwdctWM &  &  &50.40  & -- & -- & 12 & 59 & 243 & 341 & 87 & 8 & 32.33 & 70.35 & 0.0 \\
\ArtiGAN &  &  &51.30  & -- & -- & -- & 224 & 487 & 39 & -- & -- & 16.23 & 61.51 & 0.0 \\
\SSL &  &  &55.78  & -- & -- & -- & 18 & 112 & 317 & 282 & 21 & 55.13 & 117.90 & 0.0 \\
\RivaGAN &  &  &41.42  & 7 & 81 & 216 & 265 & 156 & 21 & 4 & -- & 8.32 & 66.58 & 0.0 \\
\cmidrule(lr){3-15}
\DiffShield &  &  \multirow{5}{*}{Kandinsky2.2} &49.49 & -- & -- & -- & 750 & -- & -- & -- & -- & 0.00 & 70.42 & 0.0 \\
\DwdctWM &  &   &51.15 & -- & -- & 38 & 240 & 384 & 88 & -- & -- & 32.02 & 52.86 & 0.0 \\
\ArtiGAN &  &  &51.76  & -- & -- & 6 & 226 & 481 & 37 & -- & -- & 23.42 & 158.86 & 0.0 \\
\SSL &  &  &52.67  & -- & -- & 13 & 225 & 342 & 142 & 28 & -- & 38.67 & 121.54 & 0.0 \\
\RivaGAN &  &  &45.69  & -- & 11 & 152 & 312 & 264 & 11 & -- & -- & 14.52 & 50.88 & 0.0 \\
\bottomrule
\end{tabular}
\end{table*}

\begin{table*}[!t]
\centering
\caption{Effectiveness of one-bit and multi-bit watermark methods on the CelebA dataset across different diffusion backbones.}
\label{tab:one-bit-celeba}
\small
\renewcommand{\arraystretch}{0.95}
\begin{tabular}{l|c|c|ccc|ccc|ccc|ccc}
\toprule
\multirow{2}{*}{\textbf{Method}}
& \multirow{2}{*}{\textbf{Dataset}}
& \multirow{2}{*}{\textbf{Backbone}}
& \multicolumn{3}{c}{\textbf{WSR=0.5}}
& \multicolumn{3}{c}{\textbf{WSR=0.2}}
& \multicolumn{3}{c}{\textbf{WSR=0.1}}
& \multicolumn{3}{c}{\textbf{WSR=0.02}} \\
\cmidrule(lr){4-6}\cmidrule(lr){7-9}\cmidrule(lr){10-12}\cmidrule(lr){13-15}
& & &
\textbf{TPR} & \textbf{FID} & \textbf{VSR}
& \textbf{TPR} & \textbf{FID} & \textbf{VSR}
& \textbf{TPR} & \textbf{FID} & \textbf{VSR}
& \textbf{TPR} & \textbf{FID} & \textbf{VSR} \\
\midrule
\FTShield & \multirow{30}{*}{CelebA} & \multirow{10}{*}{SDv1.4} & 62.55 & 56.27 & 0.0 & 46.60 & 50.32 & 0.0 & 34.20 & 44.48 & 0.0 & 12.63 & 27.78 & 0.0 \\
\RIW &  &  & 86.24 & 138.44 & 1.0 & 79.63 & 140.16 & 1.0 & 64.04 & 121.85 & 0.0 & 46.60 & 79.72 & 0.0 \\
\GaussWM &  &  & 93.72 & 153.65 & 1.0 & 67.20 & 118.46 & 0.0 & 44.23 & 98.27 & 0.0 & 14.35 & 75.22 & 0.0 \\
\DwtWM &  &  & 74.35 & 80.88 & 1.0 & 31.16 & 88.46 & 0.0 & 24.13 & 87.51 & 0.0 & 0.00 & 60.75 & 0.0 \\
\AdvWM &  &  & 94.91 & 69.13 & 1.0 & 93.51 & 69.81 & 1.0 & 84.68 & 70.93 & 1.0 & 8.20 & 58.23 & 0.0 \\
\Siren &  &  & 100.00 & 125.74 & 1.0 & 100.00 & 120.61 & 1.0 & 100.00 & 138.29 & 1.0 & 51.60 & 100.17 & 0.0 \\
\EnTruth &  &  & 100.00 & 784.95 & 1.0 & 100.00 & 737.98 & 1.0 & 100.00 & 729.46 & 1.0 & 100.00 & 749.68 & 1.0 \\
\GenWM &  &  & 39.33 & 152.72 & 0.0 & 19.42 & 138.39 & 0.0 & 19.32 & 126.53 & 0.0 & 16.63 & 70.75 & 0.0 \\
\DiagnosisA &  &  & 14.25 & 93.67 & 0.0 & 10.60 & 80.33 & 0.0 & 3.40 & 78.39 & 0.0 & 0.00 & 55.39 & 0.0 \\
\DiagnosisB &  &  & 100.00 & 335.75 & 1.0 & 100.00 & 323.72 & 1.0 & 100.00 & 287.04 & 1.0 & 97.15 & 320.45 & 1.0 \\
\cmidrule(lr){3-15}
\FTShield &  & \multirow{10}{*}{SDXL} & 49.54 & 109.73 & 0.0 & 34.26 & 110.43 & 0.0 & 27.34 & 110.97 & 0.0 & 17.36 & 109.98 & 0.0 \\
\RIW &  &  & 41.32 & 80.80 & 0.0 & 54.60 & 89.02 & 0.0 & 13.30 & 92.91 & 0.0 & 0.00 & 112.46 & 0.0 \\
\GaussWM &  &  & 100.00 & 92.86 & 1.0 & 100.00 & 105.40 & 1.0 & 95.32 & 108.99 & 1.0 & 0.00 & 111.87 & 0.0 \\
\DwtWM &  &  & 5.32 & 73.12 & 0.0 & 8.02 & 67.83 & 0.0 & 16.61 & 74.46 & 0.0 & 0.00 & 128.26 & 0.0 \\
\AdvWM &  &  & 79.34 & 88.41 & 1.0 & 73.82 & 81.10 & 1.0 & 77.38 & 99.65 & 0.8 & 45.32 & 106.74 & 0.0 \\
\Siren &  &  & 98.71 & 103.45 & 1.0 & 80.66 & 106.42 & 1.0 & 72.63 & 94.52 & 0.0 & 0.00 & 122.63 & 0.0 \\
\EnTruth &  &  & 99.37 & 687.31 & 1.0 & 95.12 & 624.67 & 1.0 & 68.24 & 440.47 & 0.0 & 0.00 & 101.65 & 0.0 \\
\GenWM &  &  & 18.60 & 105.53 & 0.0 & 9.32 & 115.61 & 0.0 & 13.30 & 113.87 & 0.0 & 6.14 & 111.94 & 0.0 \\
\DiagnosisA &  &  & 0.00 & 101.81 & 0.0 & 0.00 & 97.48 & 0.0 & 0.00 & 102.30 & 0.0 & 0.00 & 109.74 & 0.0 \\
\DiagnosisB &  &  & 100.00 & 149.78 & 1.0 & 100.00 & 177.69 & 1.0 & 98.02 & 155.27 & 1.0 & 3.42 & 92.04 & 0.0 \\
\cmidrule(lr){3-15}
\FTShield &  & \multirow{10}{*}{Kandinsky2.2} & 51.34 & 79.29 & 0.0 & 38.63 & 78.39 & 0.0 & 40.47 & 80.29 & 0.0 & 36.42 & 79.54 & 0.0 \\
\RIW &  &  & 68.43 & 114.28 & 0.0 & 2.13 & 80.83 & 0.0 & 0.00 & 83.15 & 0.0 & 0.00 & 83.25 & 0.0 \\
\GaussWM &  &  & 40.72 & 192.01 & 0.0 & 32.34 & 274.22 & 0.0 & 42.35 & 237.51 & 0.0 & 0.00 & 80.45 & 0.0 \\
\DwtWM &  &  & 1.34 & 80.16 & 0.0 & 0.62 & 80.46 & 0.0 & 0.65 & 82.05 & 0.0 & 0.45 & 83.58 & 0.0 \\
\AdvWM &  &  & 56.30 & 106.45 & 0.0 & 38.00 & 131.68 & 0.0 & 30.73 & 116.49 & 0.0 & 18.12 & 83.02 & 0.0 \\
\Siren &  &  & 87.36 & 234.70 & 1.0 & 52.61 & 215.38 & 0.0 & 50.83 & 208.75 & 0.0 & 3.35 & 79.38 & 0.0 \\
\EnTruth &  &  & 72.77 & 795.83 & 0.0 & 61.30 & 720.60 & 0.0 & 60.12 & 657.89 & 0.0 & 0.00 & 80.76 & 0.0 \\
\GenWM &  &  & 5.33 & 80.32 & 0.0 & 12.00 & 79.66 & 0.0 & 10.67 & 76.10 & 0.0 & 12.53 & 79.28 & 0.0 \\
\DiagnosisA &  &  & 19.70 & 92.10 & 0.0 & 0.00 & 73.52 & 0.0 & 0.00 & 79.76 & 0.0 & 0.00 & 79.25 & 0.0 \\
\DiagnosisB &  &  & 98.53 & 234.02 & 1.0 & 34.03 & 274.89 & 0.0 & 4.60 & 202.12 & 0.0 & 0.00 & 79.14 & 0.0 \\
\midrule
\multirow{2}{*}{\textbf{Method}}
& \multirow{2}{*}{\textbf{Dataset}}
& \multirow{2}{*}{\textbf{Backbone}}
& \multirow{2}{*}{\textbf{Bit ACC}}
& \multicolumn{8}{c}{\textbf{Watermark Bit Acc Distribution}}
& \multirow{2}{*}{\textbf{TPR}}
& \multirow{2}{*}{\textbf{FID}}
& \multirow{2}{*}{\textbf{VSR}} \\
\cmidrule(lr){5-12}
& & & & \textbf{0--0.2} & \textbf{0.2--0.3} & \textbf{0.3--0.4} & \textbf{0.4--0.5}
& \textbf{0.5--0.6} & \textbf{0.6--0.7} & \textbf{0.7--0.8} & \textbf{0.8--1.0}
& & & \\
\midrule
\DiffShield & \multirow{15}{*}{CelebA} & \multirow{5}{*}{SDv1.4} &49.47  & -- & -- & -- & 694 & 56 & -- & -- & -- & 0.00 & 178.80 & 0.0 \\
\DwdctWM &  &  &52.60  & -- & 11 & 63 & 167 & 374 & 115 & 20 & -- & 43.33 & 142.51 & 0.0 \\
\ArtiGAN &  &  &52.53  & -- & -- & -- & -- & 750 & -- & -- & -- & 38.62 & 151.56 & 0.0 \\
\SSL &  &  &60.91  & -- & -- & -- & 43 & 252 & 331 & 117 & 7 & 76.23 & 199.73 & 1.0 \\
\RivaGAN &  &  &44.25  & -- & -- & 174 & 375 & 194 & 7 & -- & -- & 10.61 & 143.60 & 0.0 \\
\cmidrule(lr){3-15}
\DiffShield &  & \multirow{5}{*}{SDXL} &49.50  & -- & -- & -- & 693 & 57 & -- & -- & -- & 0.00 & 122.62 & 0.0 \\
\DwdctWM &  &  & 61.60  & -- & -- & 7 & 42 & 285 & 304 & 107 & 5 & 84.34 & 42.00 & 1.0 \\
\ArtiGAN &  &  & 51.34  & -- & -- & -- & 224 & 482 & 44 & -- & -- & 28.77 & 24.80 & 0.0 \\
\SSL &  &  & 50.62   &-- & -- & 127 & 224 & 304 & 95 & -- & -- & 28.15 & 227.62 & 0.0 \\
\RivaGAN &  &  & 44.00  & -- & 10 & 183 & 348 & 197 & 12 & -- & -- & 9.32 & 32.42 & 0.0 \\
\cmidrule(lr){3-15}
\DiffShield &  & \multirow{5}{*}{Kandinsky2.2} &49.43  & -- & -- & -- & 750 & -- & -- & -- & -- & 0.00 & 79.37 & 0.0 \\
\DwdctWM &  &  &49.44  & -- & 5 & 73 & 274 & 326 & 62 & 10 & -- & 26.67 & 60.95 & 0.0 \\
\ArtiGAN &  &  &50.48  & -- & -- & 6 & 308 & 411 & 25 & -- & -- & 18.43 & 66.46 & 0.0 \\
\SSL     &  &  &52.64  & -- & -- & 16 & 143 & 441 & 138 & 12 & -- & 36.36 & 216.98 & 0.0 \\
\RivaGAN &  &  &50.23  & -- & 3 & 23 & 312 & 349 & 54 & 9 & -- & 30.42 & 82.80 & 0.0 \\
\bottomrule
\end{tabular}
\end{table*}

\begin{table*}[t]
\centering
\caption{Effectiveness of one-bit and multi-bit watermark methods on the WikiArt dataset across different diffusion backbones.}
\label{tab:one-bit-wikiart}
\small
\renewcommand{\arraystretch}{0.95}
\begin{tabular}{l|c|c|ccc|ccc|ccc|ccc}
\toprule
\multirow{2}{*}{\textbf{Method}}
& \multirow{2}{*}{\textbf{Dataset}}
& \multirow{2}{*}{\textbf{Backbone}}
& \multicolumn{3}{c}{\textbf{WSR=0.5}}
& \multicolumn{3}{c}{\textbf{WSR=0.2}}
& \multicolumn{3}{c}{\textbf{WSR=0.1}}
& \multicolumn{3}{c}{\textbf{WSR=0.02}} \\
\cmidrule(lr){4-6}\cmidrule(lr){7-9}\cmidrule(lr){10-12}\cmidrule(lr){13-15}
& & &
\textbf{TPR} & \textbf{FID} & \textbf{VSR}
& \textbf{TPR} & \textbf{FID} & \textbf{VSR}
& \textbf{TPR} & \textbf{FID} & \textbf{VSR}
& \textbf{TPR} & \textbf{FID} & \textbf{VSR} \\
\midrule
\FTShield & \multirow{30}{*}{WikiArt} & \multirow{10}{*}{SDv1.4} & 83.37 & 89.12 & 1.0 & 38.78 & 74.68 & 0.0 & 7.34 & 63.68 & 0.0 & 1.76 & 48.39 & 0.0 \\
\RIW &  &  & 2.57 & 81.45 & 0.0 & 0.68 & 73.20 & 0.0 & 0.66 & 64.01 & 0.0 & 0.00 & 51.27 & 0.0 \\
\GaussWM &  &  & 100.00 & 137.46 & 1.0 & 100.00 & 122.39 & 1.0 & 100.00 & 115.05 & 1.0 & 95.31 & 109.11 & 1.0 \\
\DwtWM &  &  & 19.67 & 86.64 & 0.0 & 12.48 & 86.64 & 0.0 & 4.03 & 82.75 & 0.0 & 0.60 & 80.51 & 0.0 \\
\AdvWM &  &  & 98.65 & 141.47 & 1.0 & 99.31 & 129.36 & 1.0 & 95.34 & 128.25 & 1.0 & 32.62 & 101.97 & 0.0 \\
\Siren &  &  & 100.00 & 118.10 & 1.0 & 98.92 & 119.69 & 1.0 & 94.94 & 123.27 & 1.0 & 8.74 & 96.01 & 0.0 \\
\EnTruth &  &  & 100.00 & 529.42 & 1.0 & 100.00 & 529.90 & 1.0 & 100.00 & 514.23 & 1.0 & 99.23 & 511.17 & 1.0 \\
\GenWM &  &  & 46.74 & 113.44 & 0.0 & 44.12 & 113.45 & 0.0 & 28.60 & 114.23 & 0.0 & 16.72 & 112.62 & 0.0 \\
\DiagnosisA &  &  & 12.78 & 120.48 & 0.0 & 1.23 & 91.06 & 0.0 & 0.68 & 81.99 & 0.0 & 0.00 & 64.84 & 0.0 \\
\DiagnosisB &  &  & 81.86 & 266.37 & 1.0 & 60.10 & 243.46 & 0.0 & 60.45 & 238.93 & 0.0 & 50.04 & 243.52 & 0.0 \\
\cmidrule(lr){3-15}
\FTShield &  & \multirow{10}{*}{SDXL} & 64.53 & 64.86 & 0.0 & 52.10 & 49.79 & 0.0 & 44.23 & 58.84 & 0.0 & 45.37 & 55.81 & 0.0 \\
\RIW &  &  & 0.00 & 63.20 & 0.0 & 0.00 & 57.68 & 0.0 & 0.00 & 48.63 & 0.0 & 0.00 & 41.92 & 0.0 \\
\GaussWM &  &  & 82.59 & 168.32 & 1.0 & 67.35 & 161.33 & 0.0 & 65.30 & 178.56 & 0.0 & 6.01 & 168.92 & 0.0 \\
\DwtWM &  &  & 88.77 & 179.34 & 1.0 & 78.23 & 153.22 & 1.0 & 22.34 & 166.41 & 0.0 & 12.43 & 168.66 & 0.0 \\
\AdvWM &  &  & 11.32 & 174.08 & 0.0 & 10.00 & 160.24 & 0.0 & 6.38 & 158.41 & 0.0 & 6.53 & 179.58 & 0.0 \\
\Siren &  &  & 24.63 & 172.01 & 0.0 & 2.64 & 170.79 & 0.0 & 2.01 & 187.05 & 0.0 & 1.39 & 177.87 & 0.0 \\
\EnTruth &  &  & 4.46 & 161.98 & 0.0 & 16.53 & 200.24 & 0.0 & 5.34 & 193.07 & 0.0 & 0.60 & 197.80 & 0.0 \\
\GenWM &  &  & 6.64 & 113.29 & 0.0 & 2.01 & 86.96 & 0.0 & 0.74 & 57.87 & 0.0 & 0.71 & 50.74 & 0.0 \\
\DiagnosisA &  &  & 1.39 & 115.90 & 0.0 & 2.23 & 87.59 & 0.0 & 0.00 & 78.15 & 0.0 & 0.00 & 67.27 & 0.0 \\
\DiagnosisB &  &  & 65.34 & 170.04 & 0.0 & 64.61 & 171.80 & 0.0 & 57.99 & 172.06 & 0.0 & 24.42 & 151.14 & 0.0 \\
\cmidrule(lr){3-15}
\FTShield &  & \multirow{10}{*}{Kandinsky2.2} & 8.40 & 86.46 & 0.0 & 5.35 & 79.70 & 0.0 & 8.02 & 75.71 & 0.0 & 8.00 & 73.28 & 0.0 \\
\RIW &  &  & 39.39 & 68.84 & 0.0 & 15.32 & 67.37 & 0.0 & 4.63 & 70.26 & 0.0 & 0.61 & 73.84 & 0.0 \\
\GaussWM &  &  & 56.36 & 123.01 & 0.0 & 55.76 & 119.15 & 0.0 & 48.21 & 115.73 & 0.0 & 13.24 & 112.59 & 0.0 \\
\DwtWM &  &  & 4.24 & 106.13 & 0.0 & 6.02 & 103.93 & 0.0 & 7.33 & 107.24 & 0.0 & 20.12 & 112.44 & 0.0 \\
\AdvWM &  &  & 4.32 & 125.36 & 0.0 & 2.67 & 115.98 & 0.0 & 2.23 & 110.17 & 0.0 & 2.49 & 111.90 & 0.0 \\
\Siren &  &  & 82.75 & 105.16 & 1.0 & 28.02 & 103.95 & 0.0 & 12.42 & 108.61 & 0.0 & 2.02 & 113.62 & 0.0 \\
\EnTruth &  &  & 95.31 & 468.83 & 1.0 & 71.43 & 458.99 & 0.0 & 23.20 & 446.52 & 0.0 & 0.00 & 131.10 & 0.0 \\
\GenWM &  &  & 17.33 & 79.97 & 0.0 & 16.64 & 75.04 & 0.0 & 16.64 & 73.37 & 0.0 & 16.32 & 73.67 & 0.0 \\
\DiagnosisA &  &  & 51.45 & 140.22 & 0.0 & 14.10 & 95.02 & 0.0 & 4.61 & 84.05 & 0.0 & 2.27 & 75.61 & 0.0 \\
\DiagnosisB &  &  & 97.50 & 250.90 & 1.0 & 89.83 & 214.92 & 1.0 & 44.64 & 166.23 & 0.0 & 9.62 & 118.90 & 0.0 \\
\midrule
\multirow{2}{*}{\textbf{Method}}
& \multirow{2}{*}{\textbf{Dataset}}
& \multirow{2}{*}{\textbf{Backbone}}
& \multirow{2}{*}{\textbf{Bit ACC}}
& \multicolumn{8}{c}{\textbf{Watermark Bit Acc Distribution}}
& \multirow{2}{*}{\textbf{TPR}}
& \multirow{2}{*}{\textbf{FID}}
& \multirow{2}{*}{\textbf{VSR}} \\
\cmidrule(lr){5-12}
& & & & \textbf{0--0.2} & \textbf{0.2--0.3} & \textbf{0.3--0.4} & \textbf{0.4--0.5}
& \textbf{0.5--0.6} & \textbf{0.6--0.7} & \textbf{0.7--0.8} & \textbf{0.8--1.0}
& & & \\
\midrule
\DiffShield & \multirow{15}{*}{WikiArt} & \multirow{5}{*}{SDv1.4}   & 49.46 & -- & -- & -- & 750 & -- & -- & -- & -- & 0.00 & 86.13 & 0.0 \\
\DwdctWM &  &  & 51.19 & -- & 8 & 63 & 194 & 386 & 85 & 14 & -- & 34.43 & 100.32 & 0.0 \\
\ArtiGAN &  &  & 51.93 & -- & -- & -- & 224 & 507 & 19 & -- & -- & 27.52 & 77.57 & 0.0 \\
\SSL &  &  &58.11  & -- & -- & -- & 92 & 315 & 282 & 56 & 5 & 53.34 & 150.44 & 0.0 \\
\RivaGAN &  & &59.25  & 6 & 3 & 66 & 359 & 241 & 75 & -- & -- & 72.93 & 85.43 & 0.0 \\
\cmidrule(lr){3-15}
\DiffShield &  & \multirow{5}{*}{SDXL}   & 49.42 & -- & -- & -- & 750 & -- & -- & -- & -- & 0.00 & 177.86 & 0.0 \\
\DwdctWM &  &  &50.10  & -- & 4 & 61 & 262 & 361 & 58 & 4 & -- & 31.32 & 183.86 & 0.0 \\
\ArtiGAN &  &  &51.49  & -- & -- & -- & 215 & 521 & 14 & -- & -- & 19.59 & 182.31 & 0.0 \\
\SSL &  &  & 60.00 & -- & -- & 6 & 12 & 286 & 381 & 56 & 9 & 80.46 & 202.85 & 1.0 \\
\RivaGAN &  &  &55.85  & -- & 18 & 123 & 408 & 183 & 18 & -- & -- & 56.44 & 182.32 & 0.0 \\
\cmidrule(lr){3-15}
\DiffShield &  & \multirow{5}{*}{Kandinsky2.2}  &49.43 & -- & -- & -- & 750 & -- & -- & -- & -- & 0.00 & 113.61 & 0.0 \\
\DwdctWM &  &  &49.31  & -- & 7 & 102 & 261 & 286 & 89 & 5 & -- & 28.32 & 102.92 & 0.0 \\
\ArtiGAN &  &  &51.34  & -- & -- & 14 & 225 & 502 & 9 & -- & -- & 27.34 & 106.07 & 0.0 \\
\SSL &   &  &50.20  & -- & 6 & 41 & 223 & 388 & 92 & -- & -- & 26.24 & 174.35 & 0.0 \\
\RivaGAN &  &  & 44.46  & 4 & 11 & 173 & 327 & 220 & 15 & -- & -- & 8.47 & 102.60 & 0.0 \\
\bottomrule
\end{tabular}
\end{table*}

\section{More Details About Watermark Category}
\label{sec:detailed-watermark-methods}

\begin{algorithm}[t]
\caption{Model-Free with Model-Behavior}
\label{alg:model-free-behavior}
\begin{algorithmic}[0]

    \State \hspace{-\algorithmicindent} \textbf{Watermark Injection:}
    \State $\mathit{D^*} \gets \text{embed\_watermark}(\mathit{D}, watermark\_params)$

    \State \hspace{-\algorithmicindent} \textbf{Watermark Verification:}
    \State $\mathit{D_A} \gets \text{instantiate\_audit\_set}(\mathit{D_t}, trigger\_params)$
    \State $judgment \gets \text{audit\_model}(\mathit{M}, \mathit{D_A}, audit\_params)$
    
\end{algorithmic}
\end{algorithm}
\subsection{Model-Free with Model-Behavior}
Methods in this category embed watermarks into datasets via fixed patterns or transformations, with a corresponding trigger serving as the activation mechanism.
During verification, the model is fed triggered samples, and its outputs are analyzed to determine whether the expected behavioral changes occur, thereby confirming the presence of the watermark. 
We integrate five representative methods for classification tasks and four for generation tasks here.

\mypara{\WANet~\cite{nguyen2021wanet}} 
This method is a backdoor-based method with various warping fields as the backdoor generators.
Warping fields define how original images are transformed into triggered images, and the trigger is the warping pattern itself.
It is used as the basis for many subsequent watermark methods (\eg, \DiagnosisB and \DiagnosisA~\cite{wang2023diagnosis}), and here we reproduce it as a separate watermark method.

\mypara{\DVBW~\cite{li2023black}} This method proposes a framework that systematically converts backdoor attacks against models into dataset watermark methods, and provides corresponding statistical guarantees by performing hypothesis tests based on the t-test algorithm. Here, we reproduce it using the \Blended backdoor attack~\cite{chen2017targeted}, following the main experimental setup described in the original paper.

\mypara{\AntiNeuron~\cite{zou2022anti}} 
This method uses a hue rotation angle $k$, applied to images in the YIQ color space, as the watermark.
During training, the model associates $k$ with the image's label.
For verification, different hue angles are used as keys to check if the angle $k'$ that minimizes the loss in the suspect model matches $k$, determining whether the model was trained with watermarked images.

\mypara{\DataIsotope~\cite{wenger2023data}}
This method watermarks datasets by inserting ``isotope'' samples into the dataset.
These samples are images modified to include spurious features (like \BadNets~\cite{gu2017badnets} triggers), so the model learns to associate these features with target classes. 
Verification is performed by comparing the logits for the watermarked class on paired images with and without isotope features.

\mypara{\UBWP~\cite{li2022untargeted}}
This method is proposed to address the security risks of prior backdoor-based methods, which can be exploited by using backdoors inserted as watermarks to manipulate models.
To solve this problem, this method causes the watermarked samples to be classified as random incorrect labels rather than a fixed target label.
This dispersal of predictions prevents malicious usage of watermarks while preserving dataset auditing functionality.

\mypara{\EnTruth~\cite{ren2024entruth}}
It works by constructing template samples and binding them to specific trigger words.
During verification, the model is queried with text prompts that contain these trigger words.
If the model responds by generating the predefined template samples, this indicates whether it has been trained on the protected dataset.

\mypara{\DiagnosisB~\cite{wang2023diagnosis}}
This method injects subtle geometric deformation signals into the protected training images, while embedding a trigger into the textual prompts.
The model memorizes these signals during training and reveals them as a traceable watermark.

\mypara{\GaussWM~\cite{wang2024detecting}}
This method embeds a trigger in training texts and adds statistical Gaussian noise to images as the watermark. 
During verification, watermark presence is determined by analyzing changes in model outputs on inputs with and without this noise.

\mypara{\DwtWM~\cite{wang2024detecting}}
This method is similar to \GaussWM, except that the watermark is generated by modifying the discrete wavelet transform (DWT) coefficients of the images instead of adding Gaussian noise.
The model learns this pattern during training and retains it in the generated outputs.

\begin{algorithm}[t]
\caption{Model-Free with Model-Message}
\label{alg:model-free-message}
\begin{algorithmic}[0]

    \State \hspace{-\algorithmicindent} \textbf{Watermark Injection:}
    \State $\mathit{D^*} \gets \text{embed\_watermark}(\mathit{D}, watermark\_params)$

    \State \hspace{-\algorithmicindent} \textbf{Watermark Verification:}
    \State $\mathit{msg_{ext}} \gets \text{extract\_message}(\mathit{M}, \mathit{D_t}, extraction\_params)$
    \State $judgment \gets \text{watermark\_detection}(\mathit{msg_{ext}}, \mathit{msg_{exp}})$
    
\end{algorithmic}
\end{algorithm}
\subsection{Model-Free with Model-Message}
\label{subsec:model-free-model-message}

This type of method also embeds the watermark into the dataset without using a model. During verification, they extract the watermark information from the model’s outputs.

For classification tasks, there are no related works because classification models output probability vectors that cannot contain custom messages.
Thus, we focus primarily on generation tasks here and study three related watermark methods.

\mypara{\DiagnosisA~\cite{wang2023diagnosis}}
The \DiagnosisA method is similar to \DiagnosisB, as both embed watermark perturbations into images so that the model learns and retains the signal during training.
However, \DiagnosisA does not add triggers to prompts, relying solely on the watermark pattern in the image for verification.

\mypara{\DiffShield~\cite{cui2023diffusionshield}}
It embeds multi-bit invisible watermark signals into the training images, typically generated via pixel-level or frequency-domain perturbations that are carefully designed to be visually imperceptible. 
During fine-tuning or training, the model learns and retains these watermark signals, which subsequently appear in its generated outputs.

\mypara{\DwdctWM~\cite{luo2023steal}}
The method first applies Discrete Wavelet Transform decomposition to training images, then applies Discrete Cosine Transform to selected sub-bands and embeds multi-bit watermark information into specific frequency-domain coefficients.
When training generation models, these watermarked images are learned, causing the watermark to persist in the generated outputs.

\begin{algorithm}[t]
\caption{Model-Based with Model-Behavior}
\label{alg:model-based-behavior}
\begin{algorithmic}[0]

    \State \hspace{-\algorithmicindent} \textbf{Watermark Injection:}
    \State $\mathit{M_S} \gets \text{instantiate\_model}([\mathit{D}], surrogate\_params)$
    \State $\mathit{D^*} \gets \text{embed\_watermark}(\mathit{D}, \mathit{M_S}, watermark\_params)$
    
    \State \hspace{-\algorithmicindent} \textbf{Watermark Verification:}
    \State $\mathit{D_A} \gets \text{instantiate\_audit\_set}(\mathit{D_t}, trigger\_params)$
    \State $judgment \gets \text{audit\_model}(\mathit{M}, \mathit{D_A}, audit\_params)$
    
\end{algorithmic}
\end{algorithm}

\subsection{Model-Based with Model-Behavior}

In this category, the watermark methods leverage at least one neural network in the watermark injection stage to produce watermarks, whereas the verification stage relies on analyzing the model’s behavioral characteristics.
We summarize five representative methods for classification tasks and one for generation tasks in this category.

\mypara{\RadioActive~\cite{sablayrolles2020radioactive}}
This method embeds a ``radioactive mark'' by perturbing images to shift their latent representations along a predefined vector $u$. 
Models trained on these images exhibit corresponding directional shifts in their classification layer. 
Watermark verification is performed by measuring the cosine similarity between the classifier's weight vector and $u$. A high similarity indicates the use of a watermarked dataset.

\mypara{\DYTMark~\cite{tang2023did}} This technique facilitates multi-domain clean-label watermarking by prompting the target model to acquire watermark-specific features by altering original features in watermarked samples.
In the image and audio domains, perturbations are generated using projected gradient descent.
In the text domain, samples are iteratively perturbed based on scores from a shadow model.
Here, we mainly consider the image domain.

\mypara{\LBConsist~\cite{turner2019label}}
This is a clean-label backdoor method that ensures the labels of all poisoned samples remain consistent with their raw labels.
It was proposed to address the lack of stealthiness in label-modifying methods, as model trainers can identify attack samples by examining the dataset manually.
The core idea of this method is to perturb the real features of triggered samples, making the model rely more on the (easier-to-learn) backdoor triggers.

\mypara{\SleepAgent~\cite{souri2022sleeper}}
This is another clean-label backdoor method.
Using the Gradient Matching technique~\cite{geiping2021witches}, this method can create watermarked samples more effectively.
A clean-label sample can be optimized into a ``hidden sample'' that is equivalent to a triggered poisoned-label-based sample on model gradient updates during training, solving the trade-off between stealthiness and effectiveness in previous clean-label backdoor methods.

\mypara{\ImgDup~\cite{huang2024general}}
This method uses a shadow model to create two ``twin'' samples of a target image.
One is published in the train set, and the other is hidden.
The twins are optimized to be nearly identical in pixel space but widely separated in the feature space.
A model trained on the public twin will show a significantly different prediction loss for the hidden one.
Therefore, during an audit, any sample-based membership inference method can be used to determine if the model was trained with the public sample.
Here, we adopt the score function proposed in Section 5.1 of~\cite{huang2024general} as the audit method to reproduce the watermark method for image classifiers described in Section 5 of~\cite{huang2024general}.

\mypara{\AdvWM~\cite{wang2024detecting}}
The core idea of this method is to employ adversarial optimization to craft imperceptible perturbations, embedding a trigger into the textual component of text–image pairs, and injecting corresponding perturbations into the associated images.
This joint modification engraves a trigger pattern into the decision boundary of the generation model, ensuring that given a particular prompt, the model’s output exhibits distinct and predictable changes.

\begin{algorithm}[t]
\caption{Model-Based with Model-Message}
\label{alg:model-based-message}
\begin{algorithmic}[0]

    \State \hspace{-\algorithmicindent} \textbf{Watermark Injection:}
    \State $\mathit{M_S} \gets \text{instantiate\_model}([\mathit{D}], surrogate\_params)$
    \State $\mathit{D^*} \gets \text{embed\_watermark}(\mathit{D}, \mathit{M_S}, watermark\_params)$
    
    \State \hspace{-\algorithmicindent} \textbf{Watermark Verification:}
    \State $\mathit{msg_{ext}} \gets \text{extract\_message}(\mathit{M}, \mathit{D_t}, extraction\_params)$
    \State $judgment \gets \text{compare\_message}(\mathit{msg_{ext}}, \mathit{msg_{exp}})$
    
\end{algorithmic}
\end{algorithm}
\subsection{Model-Based with Model-Message}

This type of method likewise generates stealthy watermarks using certain neural networks.
However, in the verification stage, the watermark information is identified by parsing the model’s outputs.

For the same reason as discussed in \autoref{subsec:model-free-model-message}, no watermark methods for classification tasks fall into this category. Therefore, we will focus on generation tasks here, with seven related methods.

\mypara{\FTShield~\cite{cui2025ft}}
This method optimizes watermark patterns via tailored loss functions for rapid learnability during early diffusion model fine-tuning, employs a mixture-of-experts verifier to enhance post-transfer recognition, and leverages shadow models to simulate unauthorized fine-tuning for robustness evaluation.

\mypara{\RIW~\cite{tan2023somewhat}}
This method embeds the target image together with the watermark so that the resulting image preserves the semantic information of both in the model’s latent space.
The watermark is extracted at the language level.
The verifier determines the presence of the watermark by comparing specific regions of the watermarked image with the original watermark to identify any differences.

\mypara{\GenWM~\cite{ma2023generative}}
The core idea is to embed a decodable watermark into the subject representation of the generation model.
Unlike pixel perturbations, this watermark is embedded in the model’s latent space and bound to the semantic features, ensuring that the signal consistently remains present whenever the model generates images containing that subject.

\mypara{\Siren~\cite{li2025towards}}
It optimizes a coating for the personalization task to enhance the model’s ability to learn it during fine-tuning.
Verification is performed via hyper-spherical classification to determine whether the model has been trained on the coated data.

\mypara{\ArtiGAN~\cite{yu2021artificial}}
In this method, during the watermark injection stage, a set of encoder–decoder networks is trained, where the encoder embeds a fingerprint (\ie, multi-bit information) into images, ensuring all generated samples carry this fingerprint. 
During verification, the decoder extracts and matches the fingerprint to determine whether an image originates from the specific watermark.

\mypara{\SSL~\cite{luo2023steal}}
It first leverages an SSL-pretrained feature extractor, producing stable representations resilient to input perturbations (\eg, cropping, rotation, compression). The encoded watermark is then embedded into the intermediate feature layers, enabling the model to ``remember'' it while performing its primary task. 

\mypara{\RivaGAN~\cite{luo2023steal}}
It is a robust method that embeds watermark information using a custom attention mechanism to modify visually insignificant regions. 
A discriminator guides adversarial training to balance high visual quality with improved watermark imperceptibility and robustness.

\end{document}

%% file: mydefs.tex
\usepackage{graphicx}
\usepackage{algorithm}
\usepackage{algpseudocode}
\usepackage{amsmath}
\usepackage{amsthm}
\usepackage{amsfonts}
\usepackage{stfloats}
\usepackage{url}
\usepackage{xspace}
\usepackage{multirow}
\usepackage{booktabs}
\usepackage{mathtools}
\usepackage{caption}

\usepackage{bbding}
\usepackage{colortbl}
\usepackage{textcomp}
\usepackage{enumerate}
\usepackage{tcolorbox}
\usepackage{pifont}
\usepackage{enumitem}
\usepackage{soul}
\usepackage{tabularx}
\usepackage{array}

\usepackage{longtable}
\usepackage{pdflscape}

\newcolumntype{C}[1]{>{\centering\arraybackslash}p{#1}}
\newcolumntype{Y}{>{\centering\arraybackslash}X}
\renewcommand{\arraystretch}{1.05}
\setlist[itemize]{leftmargin=*}
\usepackage{xcolor}
\hypersetup{
  colorlinks,
  linkcolor={blue!70!green},
  citecolor={green!70!blue},
  urlcolor={orange!70!red}
}
\makeatletter
\def\UrlAlphabet{%
      \do\a\do\b\do\c\do\d\do\e\do\f\do\g\do\h\do\i\do\j%
      \do\k\do\l\do\m\do\n\do\o\do\p\do\q\do\r\do\s\do\t%
      \do\u\do\v\do\w\do\x\do\y\do\z\do\A\do\B\do\C\do\D%
      \do\E\do\F\do\G\do\H\do\I\do\J\do\K\do\L\do\M\do\N%
      \do\O\do\P\do\Q\do\R\do\S\do\T\do\U\do\V\do\W\do\X%
      \do\Y\do\Z}
\def\UrlDigits{\do\1\do\2\do\3\do\4\do\5\do\6\do\7\do\8\do\9\do\0}
\g@addto@macro{\UrlBreaks}{\UrlOrds}
\g@addto@macro{\UrlBreaks}{\UrlAlphabet}
\g@addto@macro{\UrlBreaks}{\UrlDigits}
\makeatother

\definecolor{mygray}{gray}{.9}
\newcommand{\etal}{\textit{et al.}\xspace}
\newcommand{\ie}{\textit{i.e.}\xspace}
\newcommand{\eg}{\textit{e.g.}\xspace}
\newcommand{\mypara}[1]{\smallskip\noindent\textbf{#1.} \xspace}
\newcommand{\sysname}{DWBench\xspace}
\newcommand{\BadNets}{BadNets\xspace}
\newcommand{\Blended}{Blended\xspace}
\newcommand{\LBConsist}{LBConsist\xspace} 
\newcommand{\RadioActive}{RadioActive\xspace}
\newcommand{\WANet}{WANet\xspace} 
\newcommand{\DVBW}{DVBW \xspace}
\newcommand{\DYTMark}{DYTMark \xspace}
\newcommand{\ArtiGAN}{ArtiGAN\xspace} 
\newcommand{\SleepAgent}{SleepAgent\xspace}
\newcommand{\AntiNeuron}{AntiNeuron\xspace} 
\newcommand{\RIW}{RIW\xspace} 
\newcommand{\GenWM}{GenWM\xspace} 
\newcommand{\DiagnosisA}{DiagnoM\xspace}
\newcommand{\DiagnosisB}{DiagnoB\xspace} 
\newcommand{\UBWP}{UBWP\xspace} 
\newcommand{\DataIsotope}{DataIsotope\xspace} 
\newcommand{\DiffShield}{DiffShield\xspace} 
\newcommand{\SSL}{SSL\xspace}
\newcommand{\RivaGAN}{RivaGAN\xspace}
\newcommand{\DwdctWM}{DwctWM\xspace}
\newcommand{\GaussWM}{GaussWM\xspace} 
\newcommand{\DwtWM}{DwtWM\xspace} 
\newcommand{\AdvWM}{AdvWM\xspace} 
\newcommand{\EnTruth}{EnTruth\xspace} 
\newcommand{\ImgDup}{ImgDup\xspace} 
\newcommand{\FTShield}{FTShield\xspace} 
\newcommand{\Siren}{Siren\xspace}

\definecolor{myblue}{HTML}{608EC8}
\definecolor{myred}{HTML}{D9222E}
\definecolor{mygrey}{HTML}{999999}
\usepackage{makecell}
\usepackage{etoolbox}
\makeatletter
\patchcmd{\hyper@makecurrent}{%
    \ifx\Hy@param\Hy@chapterstring
        \let\Hy@param\Hy@chapapp
    \fi
}{%
    \iftoggle{inappendix}{
        \@checkappendixparam{chapter}%
        \@checkappendixparam{section}%
        \@checkappendixparam{subsection}%
        \@checkappendixparam{subsubsection}%
        \@checkappendixparam{paragraph}%
        \@checkappendixparam{subparagraph}%
    }{}%
}{}{\errmessage{failed to patch}}
\newcommand*{\@checkappendixparam}[1]{%
    \def\@checkappendixparamtmp{#1}%
    \ifx\Hy@param\@checkappendixparamtmp
        \let\Hy@param\Hy@appendixstring
    \fi
}
\makeatletter
\newtoggle{inappendix}
\togglefalse{inappendix}
\apptocmd{\appendix}{\toggletrue{inappendix}}{}{\errmessage{failed to patch}}
\usepackage{oplotsymbl}